\documentclass[%
a4paper,
preprint,
superscriptaddress,
groupedaddress,
runinaddress,
showpacs,
 amsmath,amssymb,
 aip,
 jcp,
showkeys
]{revtex4-1}

\usepackage[english]{babel}

\usepackage{etex}

\usepackage{graphicx}
\usepackage{color}
\usepackage{colortbl}
\usepackage{xcolor}
\usepackage{calc}

\usepackage{amsfonts}
\usepackage{amsmath}
\usepackage{amssymb}
\usepackage{wasysym}

\usepackage{float}
\usepackage{epstopdf}
\usepackage{nicefrac}

\usepackage{tcolorbox}

\usepackage[makeroom]{cancel}
\usepackage{threeparttable}

\usepackage[vcentermath]{youngtab} 

\usepackage{tikz}

\usetikzlibrary{trees}
\usetikzlibrary{calc}
\usetikzlibrary{positioning}
\usetikzlibrary{shapes.misc}
\usetikzlibrary{decorations.pathreplacing}

\usetikzlibrary{shapes.arrows} 
\usetikzlibrary{backgrounds}

\tikzset{
  solid node/.style = {circle, draw, inner sep = 3, fill = black},
  right angle/.style = {grow = 300},
  left angle/.style = {grow = 240},
  short/.style = {level distance = 1cm, },
  long/.style = {level distance = 2cm},
  left up/.style = {grow = 120},
  right up/.style = {grow = 60}
}

\usepackage{rotating}
\usepackage{mathtools}
\usepackage{hyperref}

\usepackage{units}
\usepackage{tabularx}
\usepackage{pgf}
\usepackage{pgfplots}
\usepackage{geometry}
\usetikzlibrary{calc,intersections}
\usepackage{relsize}
\usepackage{longtable}

\usepackage{dcolumn}
\usepackage{braket}
\usepackage{listings}
\usepackage{algorithm}
\usepackage{relsize}
\usepackage{nicefrac}
\usepackage{mathdots}

\usepackage{diagbox}

\usepackage{standalone}

\usepackage{bm}
\usepackage{multirow}

\usepackage{acronym}

\usepackage{tikzorbital}
\usepackage{sidecap}

\usepackage{setspace}
\usepackage{bbold}
\usepackage[inline]{enumitem}

\pgfplotsset{compat=newest}

\usepackage{geometry}

\usepackage{colortbl}

{%
   \end{list}%
}%

\newlength{\marginwidth}
\providecommand{\abs}[1]{\lvert#1\rvert}

\newcommand\commutator[2]{\ensuremath{\mathinner{%
    \mathopen[\,#1,#2\,\mathclose]}}}

\newcommand*{\dashfill}{\leavevmode\cleaders\hbox{-}\hfill\kern0pt}

\newcommand*{\midhrulefill}{
\leavevmode
\cleaders\hbox to 1ex{\raisebox{.5ex}{\rule{1ex}{.4pt}}}\hfill\kern0pt
}

\newcommand{\mustbe}{\stackrel{!}{=}}

\newcommand*{\braopket}[3]{\ensuremath{\langle{#1}|{#2}|{#3}\rangle}}

\renewcommand{\d}{\downarrow}
\renewcommand{\u}{\uparrow}
\newcommand{\s}{\sigma}

\newcommand{\mbf}[1]{\mathbf{#1}}

\newcommand{\floor}[1]{\lfloor #1 \rfloor}

\def\code#1{\texttt{\detokenize{#1}}}

\newcolumntype{L}{>{$}l<{$}} 
\newcolumntype{C}{>{$}c<{$}} 

\def\ul#1{\underline{#1}}
\def\ol#1{\overline{#1}}

\newcommand{\oL}{\ensuremath{\overline{L}}}
\newcommand{\oR}{\ensuremath{\overline{R}}}
\newcommand{\uL}{\ensuremath{\underline{L}}}
\newcommand{\uR}{\ensuremath{\underline{R}}}

\newcommand{\oLL}{\oL\oL}

\newcommand{\uLL}{\uL\uL}
\newcommand{\uRR}{\uR\uR}

\newcommand{\ra}{\ensuremath{\rightarrow}}

\newcommand{\D}{\Delta}

\def\bigO#1{\mathcal{O}(#1)}

\newcommand*{\citen}{}
\DeclareRobustCommand*{\citen}[1]{%
  \begingroup
    \romannumeral-`\x 
    \setcitestyle{numbers}%
    \cite{#1}%
  \endgroup
}

\definecolor{spawncolor}{RGB}{200 0 0}
\definecolor{deathcolor}{RGB}{0 200 0}
\definecolor{annihilcolor}{RGB}{255 99 71}
\definecolor{endcolor}{RGB}{255 99 71}

\tikzset{state node/.style={circle,thick,inner sep=1pt}}

\tikzset{state conn/.style={very thick}}
\tikzset{state dash/.style={very thick, dashed}}

\tikzset{walker/.style={very thick}}
\tikzset{walker new/.style={very thick, red}}
\tikzset{spawn/.style={very thick, color=spawncolor}}
\tikzset{death/.style={very thick, color=deathcolor}}
\tikzset{annihil/.style={very thick, color=annihilcolor}}
\newlength{\shortlength}
\newlength{\longlength}
\newlength{\afourwidth}
\newlength{\jctcsingle}
\newlength{\jctcdouble}
\newlength{\walkerlength}
\newlength{\tablesep}
\newlength{\gridsep}
\tikzset{branch/.style={circle,draw,thick,inner sep=1pt}}
\tikzset{growd/.style={-,very thick, red}}
\tikzset{grid/.style={-}}

\tikzset{lattline/.style={-,thick}}

\newlength{\orbsep}
\newlength{\sepone}
\newlength{\septwo}
\newlength{\septhree}
\tikzset{graph/.style={circle,draw,thick,inner sep=2pt}}
\tikzset{graphline/.style={-,thick}}

\tikzset{site/.style={circle,very thick,draw,opacity=0.8,inner sep=1pt}}

\newlength{\latticesep}
\newlength{\indicator}
\tikzset{orbline/.style={-,very thick}}

\tikzset{doubspin/.style={-,very thick}}
\tikzset{posspin/.style={-,very thick,blue}}
\tikzset{negspin/.style={-,very thick,red}}

\newlength{\picsep}
\newlength{\stepzero}
\tikzset{loop-tail/.style={circle,thick,inner sep=1pt,draw,fill}}
\tikzset{tree-out/.style={circle,thick,inner sep=1pt,draw,fill}}

\tikzset{mline/.style={thick}}

\newlength{\onedist}
\newlength{\twodist}
\newlength{\threedist}
\tikzset{inter/.style={circle,thick,inner sep=1pt,draw}}
\tikzset{mprime/.style={thick,dashed}}

\newlength{\taildist}
\newlength{\headdist}
\tikzset{line-out/.style={thick,dotted}}

\newlength{\gridlen}
\tikzset{drt-vert/.style={circle,very thick,draw}}

\tikzset{tree-start/.style={rectangle,very thick,draw,rounded corners=2pt,text width = {width("1")}, text height = {height("1")}, outer sep = 0,inner sep = 4pt}}
\tikzset{tree-mid/.style={circle,very thick, draw,  text width = {width("1")}, text height = {height("1")}, outer sep = 0, inner sep = 2pt,align=center}}

\definecolor{csf-color-1}{RGB}{0 200 0}
\definecolor{csf-color-2}{RGB}{255 99 71}

\tikzset{fixed-node/.style={circle,very thick, draw,  text width = {width("10")}, text height = {height("10")}, outer sep = 0, inner sep = 1pt,align=center}}

\tikzset{empty-node/.style={circle,very thick,  text width = {width("10")}, text height = {height("10")}, outer sep = 0, inner sep = 1pt,align=center}}

\definecolor{doublecolor}{RGB}{200 0 0}
\definecolor{singlecolor}{RGB}{0 200 0}

\tikzset{lattline/.style={-,thick}}
\tikzset{fermi/.style={rounded corners}}
\tikzset{line/.style={very thick}}
\tikzset{arrow/.style={->,very thick}}
\tikzset{box/.style={fill=white,rounded corners=1pt, opacity = 0.95, text opacity = 1.0}}

\tikzset{gridline/.style={dashed,opacity = 0.2}, very thin}

\definecolor{navyblue}{RGB}{0 0 128}
\definecolor{cadetblue}{RGB}{95 158 160}
\definecolor{branchingcolor}{RGB}{210 105 30}

\definecolor{back_fill}{gray}{0.9}
\definecolor{back_fill_dark}{gray}{0.7}

\tikzset{end-node/.style = {tree-out, ,text width = 2pt,text height = 2pt}}

\tikzset{l-long/.style= {<-,thick,out = 315, in = 45}}
\tikzset{r-long/.style= {->,thick,out = 315, in = 45}}

\tikzset{l-short/.style= {<-,thick,out = 315, in = 45}}
\tikzset{r-short/.style= {->,thick,out = 315, in = 45}}
\tikzset{bubble/.style= {->,thick,out = 340, in = 20}}
\tikzset{bubble-left/.style= {->,thick,out = 225, in = 110}}

\tikzset{r-long-left/.style= {->,thick,out = 225, in = 135}}
\tikzset{l-long-left/.style= {<-,thick,out = 225, in = 135}}
\tikzset{r-short-left/.style= {->,thick,out = 225, in = 135}}
\tikzset{l-short-left/.style= {<-,thick,out = 225, in = 135}}

\pgfdeclarelayer{background}
\pgfdeclarelayer{foreground}
\pgfsetlayers{background,main,foreground}

\tikzset{flow-box/.style={rectangle,very thick,draw,rounded corners=2pt}}

\tikzset{flow-box-end/.style={rectangle,very thick,draw,rounded corners=2pt}, draw = red}
\tikzset{flow-arrow/.style={->,very thick,rounded corners}}

\begin{document}

\author{Werner Dobrautz}
\email{w.dobrautz@fkf.mpg.de}
\affiliation{%
Max Planck Institute for Solid State Research, Heisenbergstr. 1, 70569 Stuttgart, Germany
}%

\author{Simon D. Smart}
\email{simondsmart@gmail.com}
\affiliation{European Centre for Medium-Range Weather Forecasts, Shinfield Rd, Reading RG2 9AX, United Kingdom}

\author{Ali Alavi}

\affiliation{%
Max Planck Institute for Solid State Research, Heisenbergstr. 1, 70569 Stuttgart, Germany
}%
\affiliation{
 Dept of Chemistry, University of Cambridge, Lensfield Road, Cambridge CB2 1EW, United Kingdom
}%
\email{a.alavi@fkf.mpg.de}

\title{Efficient Formulation of Full Configuration Interaction Quantum Monte Carlo in a Spin Eigenbasis via the Graphical Unitary Group Approach}

\date{\today}
      

\begin{abstract}

We provide a spin-adapted formulation of the Full Configuration Interaction Quantum Monte Carlo (FCIQMC) algorithm, based on the Graphical Unitary Group Approach (GUGA), which enables the exploitation of SU(2) symmetry within this stochastic framework. Random excitation generation and matrix element calculation on the Shavitt graph of GUGA can be efficiently implemented via a biasing procedure on the branching diagram.
The use of a spin-pure basis explicitly resolves the different spin-sectors
and ensures that the stochastically sampled wavefunction is an eigenfunction of the total spin operator $\hat{\mbf S}^2$. The method allows for the calculation of states with low or intermediate spin in systems dominated by Hund's first rule, which are otherwise generally inaccessible.
Furthermore, in systems with small spin gaps, the new methodology enables much more rapid convergence with respect to walker number and simulation time.    
Some illustrative applications of the GUGA-FCIQMC method are provided: computation of the $^2F-^4F$ spin gap of the cobalt atom in large basis sets, achieving chemical accuracy to experiment, 
and the $^1\Sigma_g^+$, $^3\Sigma_g^+$, $^5\Sigma_g^+$,
$^7\Sigma_g^+$ spin-gaps of the stretched N$_2$ molecule, an archetypal strongly correlated system.   

\end{abstract}

\maketitle


\section{\label{sec:introduction}Introduction}



The concept of symmetry is of paramount importance in physics and chemistry. 
The exploitation of the inherent symmetries and corresponding conservation laws in electronic structure calculations not only 
reduces the degrees of freedom by block-diagonalization of the Hamiltonian into different symmetry sectors, but also ensures the conservation of ``good'' quantum numbers and thus the physical correctness of calculated quantities. It also allows to target a specific many-body subspace of the problem at hand. 
Commonly utilized symmetries in electronic structure calculations are discrete translational and point group symmetries, $L_z$ angular momentum and $S_z$ projected spin conservation. 

Due to a non-straight-forward implementation and accompanying increased computational cost, one often ignored symmetry is the global $SU(2)$ spin-rotation symmetry of spin-preserving, nonrelativistic Hamiltonians, 
common to many molecular systems studied. This symmetry arises from the vanishing commutator  
\begin{equation}
\label{eq:hamilton-spin-symm-comm}
\commutator{\hat H}{\hat{\mbf{S}}^2} = 0,
\end{equation}
and leads to a conservation of the total spin quantum number $S$. 
 

In addition to the above-mentioned Hilbert space size reduction and conservation of the total spin $S$, solving for the eigenstates of $\hat H$ in a simultaneous spin-eigenbasis of $\hat{\mbf S}^2$ allows targeting distinct\textemdash even (near-)degenerate\textemdash spin eigenstates, which allows the calculation of spin gaps between states inaccessible otherwise, and facilitates a correct physical interpretation of calculations and description of chemical processes governed by the intricate interplay between them.  
Moreover, by working in a specific spin sector, convergence of projective techniques which rely on the 
repeated application of a propagator to an evolving wavefunction is greatly improved, especially where there are 
near spin-degeneracies in the exact spectrum. 

The Full Configuration Interaction Quantum Monte Carlo (FCIQMC) approach \cite{original-fciqmc,initiator-fciqmc}  
is one such methodology which can be expected to benefit from working in a spin-pure many-body basis. 
Formulated in Slater determinant (SD) Hilbert spaces, at the heart of the FCIQMC algorithm is excitation generation, in which from a given Slater determinant, another Slater determinant (a single or double excitation thereof) is randomly selected to be spawned on, with probability and sign determined by the corresponding Hamiltonian matrix element. Such individual determinant-to-determinant moves cannot, in general, preserve the total spin, which instead would require a {\em collective} move involving several SDs. Therefore, although the FCI wavefunction is a spin eigenvector, this {\em global} property of the wavefunction needs to emerge from the random sampling of the wavefunction, and is not guaranteed from step to step. Especially in systems in which the wavefunctions consist of determinants with many open-shell orbitals, this poses a very difficult challenge. 
If, instead, excitation generation between spin-pure entities could be ensured, this would immensely help in achieving convergence, especially in the aforementioned problems.

To benefit from the above mentioned advantages of a spin-eigenbasis, we present in this work 
the theoretical framework to efficiently formulate FCIQMC in a spin-adapted basis, via the mathematically elegant unitary group approach (UGA) and its graphical (GUGA) extension, and discuss the actual computational implementation in depth.

There are several other schemes to construct a basis of $\hat{\mathbf{S}}^2$ eigenfunctions, 
such as the Half-Projected Hartree-Fock (HPHF) functions \cite{hphf, helgaker}, Rumer spin-paired spin eigenfunctions \cite{Rumer1932, Weyl1932,rumer-3,smart-phd, rumer-non-ortho}, Kotani-Yamanouchi (KY) genealogical spin eigenfunctions \cite{genealogical-1, genealogical-2,Pauncz1979}, Serber-type spin eigenfunctions,  \cite{serber-1,Pauncz1979, serber-me}, L\"{o}wdin spin-projected Slater determinants \cite{lowdin-1} and the Symmetric Group Approach \cite{sga-1,sga-2,pauncz1995symmetric}\textemdash closely related to the UGA\textemdash, which are widely used in electronic structure calculations. Some of these have partially been previously implemented in FCIQMC (HPHF, Rumer, KY and Serber)\textemdash but with severe computational limitations. \cite{neci,hphf-fciqmc,linear-scaling-fciqmc}. The GUGA approach turns out to be quite well suited to the FCIQMC algorithm, and is able to alleviate many of the problems previously encountered.

Concerning other computational approaches in electronic structure theory, there is a spin-adapted version of the Density Matrix Renormalization Group algorithm \cite{spin-adapted-dmrg, spin-adapted-dmrg-2,spin-adapted-dmrg-3, block-dmrg-2, dmrg-proj}, a symmetry-adapted cluster (SAC) approach in the coupled cluster (CC) theory \cite{spin-adapt-cc-1,spin-adapt-cc-2, spin-adapt-cc-3}, where $S$ is conserved due to fully spin- and symmetry-adapted cluster operators and the projected CC method \cite{proj-cc-1,proj-cc-2,proj-cc-3,proj-cc-4}, where the spin-symmetry of a broken symmetry reference state is restored by a projection, similar to the L\"{o}wdin spin-projected Slater determinants \cite{lowdin-1}.

The use of spin-eigenfunctions in the {\tt Columbus} \cite{columbus-1,columbus-2,columbus-3}, \texttt{Molcas} \cite{molcas-general} and {\tt GAMESS} software package \cite{gamess-1,gamess-2} packages rely on the 
graphical unitary group approach (GUGA), where the CI method in \texttt{GAMESS} is based on the loop-driven GUGA implementation of Brooks and Schaefer \cite{brooks-schaefer-1, brooks-schaefer-2}.

Based on the GUGA introduced by Shavitt \cite{Shavitt1977,Shavitt1978}, Shepard et al.\ \cite{Shepard1980,Shepard1981} made extensive use of the graphical representation of spin eigenfunctions in form of Shavitt's distinct row table (DRT). In the multifacet graphically contracted method \cite{Shepard2005, Shepard2006, Shepard2009, Shepard2010, Shepard2014a, Shepard2014b, Shepard2014c} the ground state and excited states wavefunctions are formulated nonlinearly based on the DRT, conserving the total spin $S$. 

In this paper, we begin by reviewing the GUGA approach, concentrating on those aspects of the formalism that are especially relevant to the FCIQMC method, including the concept of branching diagrams in excitation 
generation. We then present a brief overview of the FCIQMC algorithm in the context of the GUGA method, including a discussion of optimal excitation generation and control of the time step. Next we provide
application of this methodology to spin-gaps of the N atom, the N$_2$ molecule and the cobalt atom, which illustrate 
several aspects of the GUGA formulation. 
In Sec.~\ref{chap:guga:conclustion} we conclude our findings and give an outlook to future applications and possible extensions or our implementation.

\section{The Unitary Group Approach \label{sec:uga}}

In this section we discuss the use of the \emph{Unitary Group Approach} (UGA) \cite{Paldus1974} to formulate the FCIQMC method in spin eigenfunctions. 
The UGA is used to construct a spin-adapted basis\textemdash also known as configuration state functions (CSFs)\textemdash, which allows to preserve the total spin quantum number $S$ in FCIQMC calculations. 
With the help of the \emph{Graphical Unitary Group Approach} (GUGA), introduced by Shavitt \cite{Shavitt1977}, an efficient calculation of matrix elements entirely in the space of CSFs is possible, without the necessity to transform to a Slater determinant (SD) basis. 
The GUGA additionally allows effective \emph{excitation generation}, the cornerstone of the FCIQMC method, without reference to a non spin-pure basis and the need of storage of auxiliary information.

In this work we concern ourselves exclusively with \emph{spin-preserving, nonrelativistic} Hamiltonians $\hat H$ in the Born-Oppenheimer approximation \cite{bornoppenheimer1927} in a finite basis set.
The basis of the unitary group approach (UGA), which goes back to
Moshinsky \cite{Moshinsky1968}, is the spin-free formulation of the 
spin-independent, non-relativistic, electronic Hamiltonian in the 
Born-Oppenheimer approximation, given as 
\begin{equation} \label{eq:hamil_1}
 \hat H = \sum_{ij}^n t_{ij} \sum_{\sigma=\u,\d}  a_{i\sigma}^\dagger a_{j\sigma} + \frac{1}{2}
 \sum_{ijkl}^n  \langle ik|r_{12}^{-1}| jl\rangle \sum_{\sigma,\tau=\u,\d}  a_{i\sigma}^\dagger a_{k\tau}^\dagger a_{l\tau} a_{j\sigma},
\end{equation}
where $t_{ij}=\langle i|\hat{h}|j\rangle$.
With the reformulation
\begin{equation*}
 a_{i\sigma}^\dagger a_{k\tau}^\dagger a_{l\tau} a_{j\sigma} = 
 a_{i\sigma}^\dagger a_{j\sigma} a_{k\tau}^\dagger a_{l\tau} - 
 \delta_{jk}\delta_{\sigma\tau} a_{i\sigma}^\dagger a_{l\sigma},
\end{equation*}
we can define
\begin{equation}
\label{eq:singlet-excit-op}
 \sum_{\sigma} a_{i\sigma}^\dagger a_{j\sigma} = \hat E_{ij}
\end{equation}
and
\begin{equation}
\label{eq:singlet-2body}
 \sum_{\sigma\tau} a_{i\sigma}^\dagger a_{k\tau}^\dagger a_{l\tau} a_{j\sigma} = 
 \hat E_{ij} \hat E_{kl} - \delta_{jk}\hat E_{il} = \hat e_{ij,kl}.
\end{equation}
as the \emph{singlet} one- and two-body excitation operators \cite{helgaker}, which 
do not change $S$ and $m_s$ upon acting on a state, $\ket{S,m_s}$, with definite total and z-projection value of the spin. 
With Eqs.~(\ref{eq:singlet-excit-op}) and (\ref{eq:singlet-2body}) the Hamiltonian~(\ref{eq:hamil_1}) can be expressed in terms of these spin-free  excitation operators as \cite{Matsen1964}
\begin{equation}
\label{eq:spin-free}
\hat H = \sum_{ij} t_{ij}\, \hat E_{ij} + \frac{1}{2}\sum_{ij,kl}V_{ij,kl}\,\hat e_{ij,kl}.
\end{equation}
where $V_{ij,kl}= \langle ik|r_{12}^{-1}| jl\rangle$.
An elegant and efficient method to create a spin-adapted basis and calculate the Hamiltonian matrix elements in this basis is based on the important observation that the spin-free excitation operators (\ref{eq:singlet-excit-op}) and (\ref{eq:singlet-2body}) in the non-relativistic Hamiltonian (\ref{eq:spin-free}) obey the same commutation relations as the generators of the \emph{Unitary Group} $U(n)$ \cite{Paldus1974,Paldus1975,Paldus1976}, $n$ being the number of spatial orbitals. 
The commutator of the spin-preserving excitation operators $\hat E_{ij}$ can be calculated as
\begin{align}
 \commutator{\hat E_{ij}}{\hat E_{kl}} =& \sum_{\sigma\tau} a_{i\sigma}^\dagger a_{j\sigma}a_{k\tau}^\dagger a_{l\tau}
 - a_{k\tau}^\dagger a_{l\tau}  a_{i\sigma}^\dagger a_{j\sigma} \nonumber\\
 =& \sum_{\sigma\tau} a_{i\sigma}^\dagger a_{j\sigma}a_{k\tau}^\dagger a_{l\tau} - a_{i\sigma}^\dagger a_{k\tau}^\dagger a_{l\tau} a_{j\sigma} 
 - \delta_{il} a_{k\tau}^\dagger a_{j\sigma} \nonumber\\
 =& \sum_{\sigma\tau} a_{i\sigma}^\dagger a_{j\sigma}a_{k\tau}^\dagger a_{l\tau} - a_{i\sigma}^\dagger a_{j\sigma}a_{k\tau}^\dagger a_{l\tau}\nonumber + \delta_{jk} a_{i\sigma}^\dagger a_{l\tau} - \delta_{il} a_{k\tau}^\dagger a_{j\sigma} \nonumber\\
 \commutator{\hat E_{ij}}{\hat E_{kl}} =& \,\delta_{jk}\,\hat E_{il} - \delta_{il}\,\hat E_{kj},\label{eq:generator-commutator}
\end{align}
which is the same as for the basic matrix units and the generators of the unitary group $U(n)$.

The \emph{Unitary Group Approach} (UGA) was pioneered by Moshinsky \cite{Moshinsky1968}, Paldus \cite{Paldus1974} and Shavitt \cite{Shavitt1977,Shavitt1978}, who introduced the graphical-UGA (GUGA) for practical calculation of matrix elements. 
With the observation that the spin-free, nonrelativistic Hamiltonian (\ref{eq:spin-free}) is expressed in terms of the generators of the unitary group, the use of a basis that is \emph{invariant} and \emph{irreducible} under the action of these generators is desirable. This approach to use \emph{dynamic} symmetry to block-diagonalize the Hamiltonian is different to the case where the Hamiltonian commutes with a symmetry operator. In the UGA $\hat H$ does not commute with the generators of $U(n)$, but rather is \emph{expressed} in terms of them. Block diagonalization occurs, due to the use of an invariant and irreducible basis under the action of these generators. Hence, the UGA is an example of a \emph{spectrum generating algebra} with dynamic symmetry \cite{Iachello1993,Sonnad2016}.

We only want to recap the most important concepts of the UGA here and refer the interested reader to the pioneering work of Paldus~\citen{Paldus1974} and Shavitt~\citen{Shavitt1977,Shavitt1978,Shavitt1981}.

\subsection{\label{sec:gt}The Gel'fand-Tsetlin Basis}
The \emph{Gel'fand-Tsetlin} (GT) \cite{gelfand-1, gelfand-2, gelfand-3} basis is invariant and irreducible under the action of the generators of $U(n)$. 
The group $U(n)$ has $n^2$ generators, $E_{ij}$, and a total of $n$ Casimir operators, commuting with all generators of the group, and the GT basis is based on the group chain
\begin{equation}
\label{eq:group-chain}
U(n) \supset U(n-1) \supset \dots \supset U(2) \supset U(1),
\end{equation}
where $U(1)$ is Abelian and has one-dimensional irreducible representations (irreps). Each subgroup\\
$U(n-1), U(n-2),\dots,U(1)$ has $n-1,n-2,\dots,1$ Casimir operators, resulting in a total of $n(n+1)/2$ commuting operators, named \emph{Gel'fand invariants} \cite{gelfand-3}. 
The simultaneous eigenfunctions of these invariants form the GT basis and are uniquely labeled by a set of $n(n+1)/2$ integers related to the eigenvalues of the invariants. 
Thus, based on the branching law of Weyl \cite{Weyl1931, Weyl1946}, a general $N$-electron CSF can be represented by a \emph{Gel'fand pattern} \cite{gelfand-1}
\begin{equation}
\label{eq:gelfand-pattern}
[\mbf m] = {\small \left[\begin{smallmatrix}
 m_{1,n} & & m_{2,n} & \cdots & m_{n-1,n} & & m_{n,n} \\[1pt]
 & m_{1,n-1} & & \cdots & & m_{n-1,n-1} & \\[-2pt]
 & \phantom{0000}\ddots & & \cdots & &\iddots\phantom{000000} & \\
 & & m_{1,2} & & m_{2,2} & & \\
 & & & m_{1,1} & & & \\
 \end{smallmatrix}
 \right]}.
\end{equation}
The integers $m_{ij}$ in the \emph{top row} (and all subsequent rows) of (\ref{eq:gelfand-pattern}) are nonincreasing, $m_{1n} \geq m_{2n} \geq \dots \geq m_{nn}$, and the integers in the subsequent rows fulfill the condition 
\begin{equation}
\label{eq:inbetween}
m_{i,j+1} \geq m_{ij} \geq m_{i+1,j+1},
\end{equation}
called the ``in-between'' condition \cite{Louck1970}.

The $n$ non-increasing integers of the top row of Eq.~(\ref{eq:gelfand-pattern}), $\mbf m_n = \left(m_{1n},m_{2n},\dots,m_{nn}\right)$, are called the \emph{highest weight} or \emph{weight vector } of the representation and specify the chosen irrep of $U(n)$; the following $n-1$ rows uniquely label the states belonging to the chosen irrep. 



In CI calculations one usually employs a one-particle basis of $2n$ spin-orbitals with creation $\hat a_{i\s}^\dagger$ and annihilation $\hat a_{j\tau}$ operators of electrons in spatial orbital $i,j$ with spin $\s,\tau$. The $(2n)^2$ operators
\begin{equation}
\hat A_{i\s,j\tau} = \hat a_{i\s}^\dagger \hat a_{j\tau}; \quad i,j = 1,\dots, n; \quad \s,\tau = \u,\d
\end{equation}
can be associated with the generators of $U(2n)$ with the commutation relation 
\begin{equation}
\commutator{\hat A_{i\s,j\tau}}{\hat A_{i'\s',j'\tau'}} = \delta_{ji'}\delta_{\tau\s'}\hat A_{i\s,j'\tau'} - \delta_{ij'}\delta_{\s\tau'} \hat A_{i'\s',j\tau}.
\end{equation}
The partial sums over spin or orbital indices of these operators
\begin{equation}
\hat E_{ij} = \sum_{\s = \u,\d} \hat A_{i\s,j\s} \quad \text{and} \quad \hat{\mathcal E}_{\s\tau} = \sum_{i = 1}^n \hat A_{i\s,i\tau}
\end{equation}
are related to the orbital $U(n)$ and spin $U(2)$ generators. 
Since we deal with fermions we have to restrict ourselves to the totally antisymmetric representations of $U(2n)$, denoted as $\Gamma\{1^{2n}\}$.
Since the molecular Hamiltonian~(\ref{eq:spin-free}) is spin independent, we can consider the proper subgroup of the direct product of the spin-free orbital space $U(n)$, with $n^2$ generators $E_{ij}$, and the pure spin space $U(2)$ with the four generators ${\mathcal{E}}_{\s\tau}$ \cite{Paldus1974}, given as 
\begin{equation}
\label{eq:direct-product}
U(2n) \supset U(n) \otimes U(2), 
\end{equation}
where the representations of $U(n)$ and $U(2)$ are 
\emph{mutually conjugate} \cite{Paldus2006, Matsen1974, Matsen1964, Moshinsky1968}. 

Moreover, since the Hamiltonian~(\ref{eq:spin-free}) is spin-independent, $U(2)$ does not contribute to the matrix element evaluation, so we only have to concern ourselves with the irreps of the orbital $U(n)$ subgroup, following Matsen's spin-free approach \cite{Matsen1974,Matsen1964}. 

\subsection{\label{sec:paldus-table}The Paldus tableau}

The consequence of the mutually conjugate relationship between $U(n)$ and $U(2)$ irreps for electronic structure calculations is that the integers $m_{ij}$ in a Gel'fand pattern~(\ref{eq:gelfand-pattern}) for $U(n)$ are related to occupation numbers of spatial orbitals.
This means they are restricted to $0 \leq m_{ij} \leq 2$, due to the Pauli exclusion principle. The highest weight, $\mbf m_n$, indicates the chosen electronic state with the conditions
\begin{equation}
\label{eq:sum-electrons}
\sum_{i = 1}^n m_{in} = N \quad \text {and} \quad \frac{1}{2}\sum_{i = 1}^n \delta_{1,m_{in}} = S,
\end{equation}
with $N$ being the total number of electrons and the number of singly occupied orbitals, $\delta_{1,m_{ij}}$ is equal to twice the total spin value $S$.

This insight led Paldus \cite{Paldus1974} to the more compact formulation of a GT state by a table of $3n$ integers. 
It is sufficient to count the appearances $2's$, $1's$ and $0's$ in each row $i$ of a Gel'fand pattern and store this information, denoted by $a_i, b_i$ and $c_i$ in a table, named a \emph{Paldus tableau}. 

The first column, $a_i$, contains the number of doubly occupied orbitals, the second column, $b_i$, the number of singly occupied and the last one, $c_i$, the number of empty orbitals, as shown by the example of an $n = 8, N = 6, S = 1$ state:
\begin{equation}
\renewcommand\arraystretch{0.7}
\label{eq:paldus-table}
{\footnotesize
\begin{bmatrix}
\phantom{0} \quad \phantom{0} \quad \phantom{0} \quad \phantom{0} \quad \phantom{0} \quad \phantom{0} \quad \phantom{0} \quad \phantom{0} \\
2 \quad 2 \quad 1 \quad 1 \quad 0 \quad 0 \quad 0 \quad 0 \\
2 \quad 2 \quad 1 \quad 1 \quad 0 \quad 0 \quad 0 \\
2 \quad 1 \quad 1 \quad 0 \quad 0 \quad 0 \\
2 \quad 1 \quad 1 \quad 0 \quad 0 \\
2 \quad 1 \quad 0 \quad 0 \\
1 \quad 1 \quad 0 \\
1 \quad 0 \\
1
\end{bmatrix}
\equiv
\left[
\begin{array}{ccc}
a_i & b_i & c_i \\
\hline
2 & 2 & 4 \\
2 & 2 & 3 \\
1 & 2 & 3 \\
1 & 2 & 2 \\
1 & 1 & 2 \\
0 & 2 & 1 \\
0 & 1 & 1 \\
0 & 1 & 0
\end{array}
\right]
\ra
\left[
\begin{array}{ccc}
\Delta a_i & \Delta b_i & \Delta c_i \\
\hline
0 & \phantom{-}0 & 1 \\
1 & \phantom{-}0 & 0 \\
0 & \phantom{-}0 & 1 \\
0 & \phantom{-}1 & 0 \\ 
1 & -1 & 1 \\
0 & \phantom{-}1 & 0 \\
0 & \phantom{-}0 & 1 \\
0 & \phantom{-}1 & 0
\end{array}
\right]}
\end{equation}
where the differences $\Delta x_i = x_i - x_{i-1}$, with $x = a,b,c$, of subsequent rows are also indicated.  
For each row the condition
\begin{equation}
\label{eq:row-equations}
a_i + b_i + c_i = i, \quad (i = 1,\dots,n)
\end{equation}
holds, thus any two columns are sufficient to uniquely determine the state. The top row satisfies the following properties
\begin{equation}
\label{eq:top-a} 
a = a_{n} = {{1}\over{2}}N -S, \quad  b = b_{n} = 2S, \quad c = c_{n} = n - a - b = n - {{1}\over{2}}N -S,
\end{equation}
completely specifying the chosen electronic state as an irrep of $U(n)$. 

The total number of CSFs for a given number of orbitals $n$, electrons $N$ and total spin $S$ is given by the Weyl-Paldus \cite{Paldus1974,Weyl1946} dimension formula 
\begin{equation}
\label{eq:dimension}
N_{CSF} = \frac{b+1}{n+1} {n+1 \choose a} {n+1 \choose c}
 =  \frac{2S+1}{n+1} {n+1 \choose {\frac{N}{2}-S}} {n+1 \choose {n - \frac{N}{2} -S}}.
\end{equation}
As it can be seen from Eq.~(\ref{eq:dimension}), the number of possible CSFs\textemdash of course\textemdash still scales combinatorially with the number of electrons and orbitals, as seen in Fig.~\ref{fig:num-csfs} with a comparison to the total number of possible SDs (without any symmetry restriction). 
The ratio of the total number of SDs  and CSFs for $N = n$ can be estimated by Stirling's formula (for sufficiently large $n$ and $N$) as 
\begin{equation}
\label{eq:sd-csf-ratio}
\frac{N_{SD}}{N_{CSF}} \approx \frac{\sqrt{\pi\, n} n}{2(2S + 1)}, 
\end{equation}
which shows orbital dependent, $\sim n^{3/2}$, decrease of the efficient Hilbert space size for a spin-adapted basis. 
The Paldus tableau also emphasizes the cumulative aspects of the coupling between electrons, with the i-th row providing information on number of electrons, $N_{i}$ (up to i-th level) and the spin, $S_{i}$, by
\begin{equation}
N_{i} = 2 a_{i}+b_{i}, \qquad 
S_{i} = {{1}\over{2}}b_{i}.\label{eq:si}
\end{equation}
As can be seen in Eq.~\ref{eq:paldus-table}, there are four permissible difference vectors $[\Delta a_i, \Delta b_i, \Delta c_i]$ ($\Delta x_i = x_i - x_{i-1}$, with $x = a,b,c$)  between consecutive rows of a Paldus tableau, which 
corresponds to the possible ways of coupling a spatial orbital based on the group chain (\ref{eq:group-chain}). 
This information can be condensed in the four-valued \emph{step value}, shown in Table~\ref{tab:step-vector}. 

All possible CSFs of a chosen irrep can then be encoded by the collection of the step values in a \emph{step-vector}, where starting from the ``vacuum'' $0$'th row $i = 0$, 
an empty spatial orbital is indicated by $d_i = 0$, a ``positively spin-coupled'' orbital, $\Delta S_i = 1/2$, by $d_i = 1$, a ``negatively spin-coupled'', $\Delta S_i = -1/2$, by $d_i = 2$ and a doubly occupied spatial orbital by $d_i = 3$. To retain physically allowed states the condition $S_i \geq 0,\, \forall i$ applies.
(As a side note: Another common notation \textemdash e.g. in \texttt{Molcas}\textemdash is to indicate positive spin-coupling as $d_i = u$, negative spin-coupling by $d_i = d$ and a doubly occupied orbital by $d_i = 2$.)

\begin{table}
\centering
\caption{\label{tab:step-vector}Four possible ways of coupling an orbital $i$.}
\begin{tabular}{cccccc}
\toprule
$d_i$ & $\Delta a_i$ & $\Delta b_i $ & $\Delta c_i$ & $\Delta N_i$ & $\Delta S_i$\\
\hline
0 & 0 & \phantom{-}0 & 1 & 0 & 0 \\
1 & 0 & \phantom{-}1 & 0 & 1 & \phantom{-}1/2 \\
2 & 1 &-1 & 1 & 1 &-1/2 \\
3 & 1 & \phantom{-}0 & 0 & 2 & 0 \\
\botrule
\end{tabular}
\end{table}
The \emph{step-value} $d_i$ in Tab.~\ref{tab:step-vector} is given by $d_{i} = 3\Delta a_{i} + \Delta b_i $ and the collection of all $d_i$ into the 
\textbf{step-vector} $\mathbf d$ representation is the \emph{most compact} form of representing a CSF, with the same storage cost as a Slater determinant, with 2 bits per spatial orbital. One can create all basis function of a chosen irrep of $U(n)$ by constructing all possible distinct step-vectors $\ket{\mathbf d}$ which lead to the same top-row of the Paldus tableau (\ref{eq:top-a}), specifying the chosen irrep with definite spin and number of electrons, with the restriction $S_i \geq 0,\, \forall i$. 

\section{\label{sec:guga}The Graphical Unitary Group Approach (GUGA)}

\begin{figure}
\centering 
\includegraphics{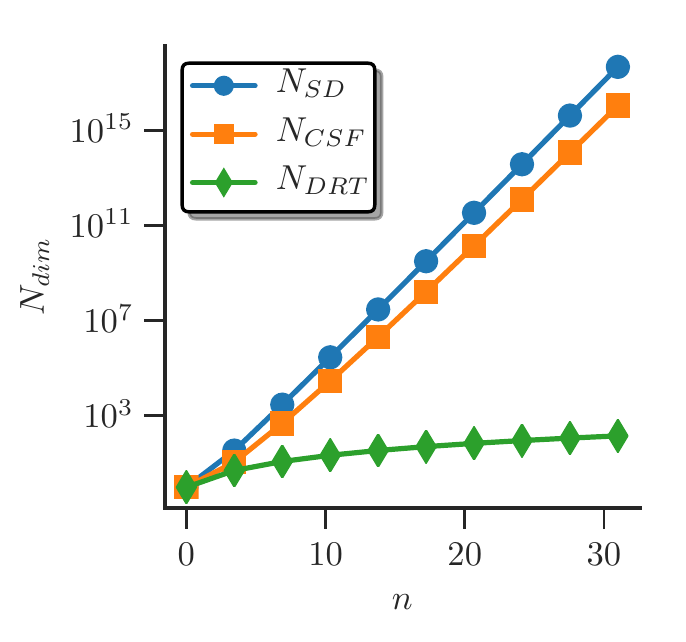}
\caption{\label{fig:num-csfs}(Color online) Number of total SDs (without any symmetry restrictions), CSFs and entries of the distinct row table (DRT) for $S=0$ and $N=n$ as a function of $n$.}
\end{figure}

The graphical unitary group approach (GUGA) of Shavitt \cite{Shavitt1977,Shavitt1981} is based on this step-vector representation and the observation that there is a lot of repetition of possible rows in the Paldus tableaux specifying the CSFs of a chosen irrep of $U(n)$.
Instead of all possible Paldus tableaux, Shavitt suggested to just list the possible sets of distinct rows in a table, called the \emph{distinct row table} (DRT). The number of possible elements of this table is given by \cite{Shavitt1977}
\begin{align}
\label{eq:num-drt}
N_{DRT} =& \left(a + 1\right)\left(c + 1\right)\left(b + 1 + \frac{d}{2}\right) - \frac{d(d+1)(d+2)}{6}  \nonumber \\ =& \left(\frac{N}{2} - S + 1\right)\left(n - \frac{N}{2} - S+1\right)\left(2S + 1 + \frac{d}{2}\right) - \frac{d(d+1)(d+2)}{6},
\end{align}
with $d = \min(a,c) = \min(N/2-S,n-N/2-S)$, which is drastically smaller than the total number of possible CSFs~(\ref{eq:dimension}) or Slater determinants (without any symmetry restrictions) as seen in Fig.~\ref{fig:num-csfs}.
Each row in the DRT is identified by a pair of indices $(i,j)$, with $i = a_{ij} + b_{ij} + c_{ij}$ being the {\itshape level index}, related to the orbital index and $j$ being the {\itshape lexical row index} such that $j < j'$ if $a_{ij} > a'_{ij}$ or if $a_{ij} = a'_{ij}$ and $b_{ij} > b'_{ij}$. 

A simple example of the DRT of a system with $n=3$, $N=4$ and $S=0$ is shown in Table~\ref{tab:DRTTablen3N4}.
\begin{table}
\centering
\caption{\label{tab:DRTTablen3N4}Distinct row table for $n=3$, $N=4$ and $S=0$.
  }
\begin{tabular}{cccccccccccccccc}
\toprule
 a & b & c && i & j && $k_0$ & $k_1$ & $k_2$ & $k_3$ && $l_0$ & $l_1$ & $l_2$ & $l_3$  \\
\cline{1-3} \cline{5-6} \cline{8-11} \cline{13-16}
 2 & 0 & 1 && 3 & 1 && 2 & 0 & 3 & 4 &&  - & - & - & - \\
\cline{1-3} \cline{5-6} \cline{8-11} \cline{13-16}
 2 & 0 & 0 && 2 & 2 && 0 & 0 & 0 & 5 &&  1 & 0 & 0 & 0 \\
 1 & 1 & 0 && 2 & 3 && 0 & 5 & 0 & 6 &&  0 & 0 & 1 & 0 \\
 1 & 0 & 1 && 2 & 4 && 5 & 0 & 6 & 7 &&  0 & 0 & 0 & 1 \\
\cline{1-3} \cline{5-6} \cline{8-11} \cline{13-16}
 1 & 0 & 0 && 1 & 5 && 0 & 0 & 0 & 8 &&  4 & 3 & 0 & 2  \\
 0 & 1 & 0 && 1 & 6 && 0 & 8 & 0 & 0 &&  0 & 0 & 4 & 3  \\
 0 & 0 & 1 && 1 & 7 && 8 & 0 & 0 & 0 &&  0 & 0 & 0 & 4  \\
\cline{1-3} \cline{5-6} \cline{8-11} \cline{13-16}
 0 & 0 & 0 && 0 & 8 &&- & - & - & - &&  7 &  6 & 0 &  5  \\
\botrule
\end{tabular}
\end{table}
Relations between elements of the DRT belonging to two neighboring levels $k$ and $k-1$ are indicated by the so called \emph{downward}, $k_{d_k}$, and \emph{upward}, $l_{d_k}$, \emph{chaining indices}, with $d_k = 0,1,2,3$. These indices indicate the connection to a lexical row index in a neighboring level by a step-value $d_k$, where a zero entry indicates an invalid connection associated with this step-value. Given a DRT table any of the possible CSFs can be generated by connecting distinct rows linked by the chaining indices.

\begin{figure}
\centering
\includegraphics{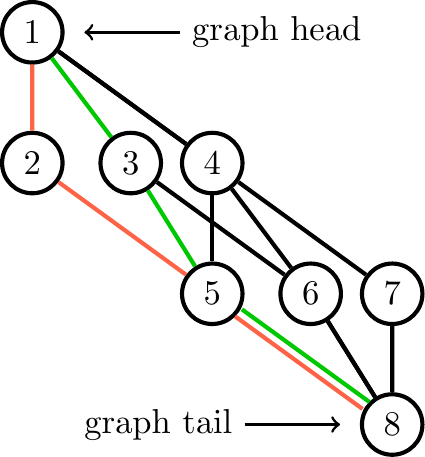}
\caption{\label{Fig:graphn3N4}(Color online) Graph representing the DRT of Table~\ref{tab:DRTTablen3N4}. The orange line corresponds to the CSF {\color{csf-color-2} $\ket d_1 = \ket{3,3,0}$} and the green line to {\color{csf-color-1}$\ket d_2 = \ket{3,1,2}$} in the step-vector representation.}
\end{figure}

This DRT table can be represented as a graph, see Fig.~\ref{Fig:graphn3N4}, where each distinct row is represented by a \emph{vertex} (node) and nonzero chaining indices are indicated by an \emph{arc} (directed edge). The vertices are labeled according to the lexical row index $j$, starting at the unique \emph{head} node at the top, which corresponds to the highest row $(a,b,c)$. It ends at the second unique null row $(0,0,0)$, which is called the \emph{tail} of the graph. Vertices with the same $i$-value of Table~\ref{tab:DRTTablen3N4} are at the same level on this grid. The highest $i$-value is on top and the lowest at the bottom. 
Vertices also have left-right order with respect to their $a_i$ value and vertices that share the same $a_i$ value are further ordered\textemdash still horizontally\textemdash with respect to their $b_i$ value.
With the above mentioned ordering of the vertices according to their $a_i$ and $b_i$ values, the slope of each arc is in direct correspondence to the step-value $d_i$, connecting two vertices. $d = 0$ corresponds to vertical lines, and the tilt of the other arcs increases with the step-value $d_i$.

Each CSFs in the chosen irrep of $U(n)$, is represented by a {\it directed walk} through the graph starting from the tail and ending at the head, e.g.\ the green and orange lines in Fig.~\ref{Fig:graphn3N4} (color online), representing the states $\ket{\mbf d_1} = \ket{3,3,0}$ and $\ket{\mbf d_2} = \ket{3,1,2}$ in step-vector representation. 
Such a walk spans $n$ arcs (number of orbitals) and visits one node at each level $i$. 
There is a direct correspondence between the Paldus tableau, Gel'fand patterns and directed walks on \emph{Shavitt graphs} for representing all possible CSFs in a chosen irrep of $U(n)$.

\subsection{\label{sec:guga:me}Evaluation of Nonvanishing Hamiltonian Matrix Elements}

Given the expression of the nonrelativistic spin-free Hamiltonian in \eqref{eq:spin-free} a matrix element between two CSFs, $\ket{m'}$ and $\ket m$, is given by:
\begin{equation}
\label{eq:hamil-matel}
\braopket{m'}{\hat H}{m} = \sum_{ij} t_{ij} \braopket{m'}{\hat E_{ij}}{m}  + \frac{1}{2}\sum_{ij,kl}V_{ij,kl} \braopket{m'}{\hat e_{ij,kl}}{m}.
\end{equation}
The matrix elements, $\braopket{m'}{\hat E_{ij}}{m}$ and $\braopket{m'}{\hat e_{ij,kl}}{m}$, provide the \emph{coupling coefficients} between two given CSFs and $t_{ij}$ and $V_{ij,kl}$ are the \emph{integral contributions}. The coupling coefficients are independent of the orbital shape and only depend on the involved CSFs, $\ket{m'}$ and $\ket m$- Therefore, for a given set of integrals the problem of computing Hamiltonian matrix elements in the GT basis is reduced to the evaluation of these coupling coefficients. The graphical representation of CSFs has been proven a powerful tool to evaluate these coupling coefficients thanks to the formidable contribution of Paldus, Boyle, Shavitt and others \cite{Paldus1980,Shavitt1978,DownwardRobb1977}. 

The great strength of the graphical approach is the identification and evaluation of nonvanishing matrix elements of the excitation operators (generators) $\hat E_{ij}$, between two GT states (CSFs), $\braopket{m'}{\hat E_{ij}}{m}$. 
The generators are classified according to their indices, with $\hat E_{ii}$ being diagonal \emph{weight} (W) and $\hat E_{ij}$ with $i < j$ being \emph{raising} (R) and $i > j$ \emph{lowering} (L) operators (or generators).
In contrast to Slater determinants, $\hat E_{ij}$ applied to $\ket m$ yields a \emph{linear combination} of CSFs $\ket{m'}$,
\begin{equation}
\label{eq:gen-action}
\hat E_{ij} \ket {m} = \sum_{m'} \ket{m'} \braopket{m'}{\hat E_{ij}}{m},
\end{equation}
with an electron moved from spatial orbital $j$ to orbital $i$ without changing the spin of the resulting states $\ket{m'}$. They are called raising (lowering) operators since the resulting $\ket{m'}$ will have a higher (lower) lexical order than the starting CSF $\ket m$. 

\begin{figure}
\centering
\includegraphics[width = 0.48\textwidth]{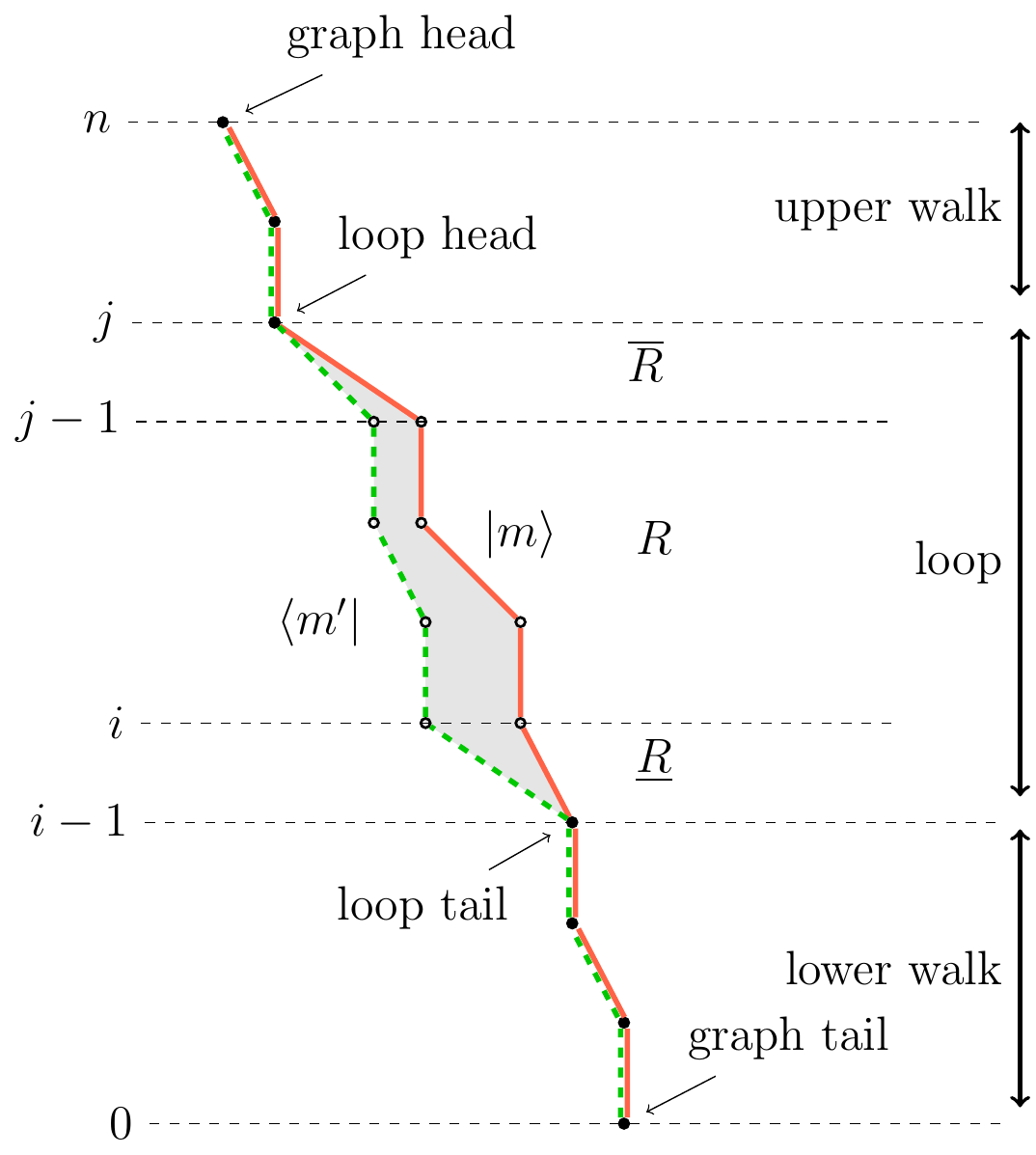}
\caption{\label{fig:loop}(Color online) Graphical representation of a matrix element $\braopket{m'}{\hat E_{ij}}{m}$ as a loop shape created by two CSFs $\ket{m'}$ and $\ket m$ on a Shavitt graph.}
\end{figure}

The distance, $S_0$, from $\min(i,j)-1$ to $\max(i,j)$, is an important quantity and is called the \emph{range of the generator} $\hat E_{ij}$. For the one-body term in (\ref{eq:spin-free}) Shavitt \cite{Shavitt1977} was able to show that the walks on the graph, representing the CSFs $\ket m$ and $\ket{m'}$,  must coincide outside of this range $S_0$ to yield a non-zero matrix element. The two vertices in the DRT graph, related to orbital $i-1$ and $j$ (with $i < j$) represent the points of separation of the walks and they are named {\it loop head} and {\it loop tail}. And the matrix element $\braopket{m'}{\hat E_{ij}}{m}$ only depends on the \emph{shape of the loop} formed by the two graphs in the range $S_0$, shown in Fig.~\ref{fig:loop}. 
Shavitt \cite{Shavitt1978} showed that the relations
\begin{equation}
\label{eq:single-cond}
N'_k = N_k \pm 1 \quad \text {and} \quad b'_k = b_k \pm 1 \; \Leftrightarrow \;  S'_{k} = S_{k} \pm {{1}\over{2}} \quad \text{for}\quad  k \in S_0,
\end{equation}
between $\ket m$ and $\ket{m'}$ must be fulfilled to yield a nonzero matrix element ($N'_k = N_k + 1$ for a raising and $N'_k = N_k - 1$ for a lowering generator). 
 
This allows \emph{two} possible relations between the vertices at each level in terms of Paldus array quantities depending on the type of generator (R,L). 
For \emph{raising generators} R:
\begin{align}
a_k' = a_k,\quad b_k' = b_k + 1, \quad c_k' = c_k - 1, \quad &\rightarrow \underline{\Delta b_k = -1},\label{eq:R-dbk-1} \\
a_k' = a_k + 1, \quad b_k' = b_k - 1, \quad c_k' = c_k \quad &\rightarrow \underline{\Delta b_k = +1},\label{eq:R-dbk+1}
\end{align}
where $\Delta b_k = b_k - b_k'$ and for \emph{lowering generators} L:
\begin{align}
a_k' = a_k - 1, \quad b_k' = b_k + 1, \quad c_k' = c_k \quad &\rightarrow \underline{\Delta b_k = -1}, \label{eq:L-dbk-1}\\
a_k' = a_k, \quad b_k' = b_k - 1, \quad c_k' = c_k + 1 \quad &\rightarrow \underline{\Delta b_k = + 1}. \label{eq:L-dbk+1}
\end{align}
At each vertex of the loop in range $k$ one of the relations~(\ref{eq:R-dbk-1}-\ref{eq:L-dbk+1}) must be fulfilled for the one-body matrix element to be non-zero.

\begin{figure*}
\centering
\includegraphics[width = \textwidth]{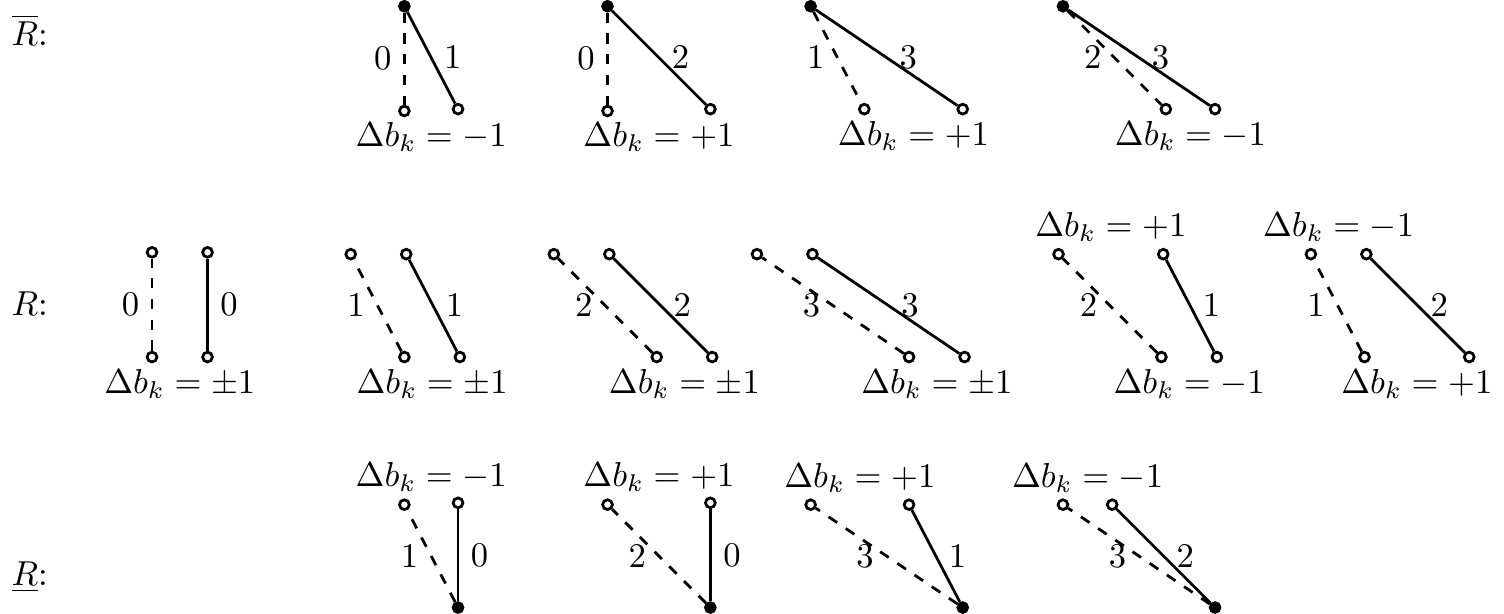}
\caption{\label{fig:single-loops}Nonzero segment shapes of a raising generator $\hat E_{ij}$. The numbers next to the lines indicate the step-values $d'$ and $d$. $\underline{R} (\overline{R})$ correspond to the loop tail (head) segments and $R$ to shapes inside the generator range $S_0$. $\Delta b_k$ indicates the possible difference of $b'_k$ and $b_k$ leading to nonzero matrix elements.}
\end{figure*}

Based on the graphical approach, Shavitt \cite{Shavitt1978} showed that the matrix elements of the generators $\hat E_{ij}$ can be factorized in a product, where each term corresponds to a segment of the loop in the range $S_0$ and is given by
\begin{equation}
\label{eq:single-prod}
\braopket{m'}{\hat E_{ij}}{m} = \prod_{k = i}^j W(Q_k;d'_k,d_k,\Delta b_k ,b_k),
\end{equation}
where $b_k$ is the $b$ value of state $\ket m$ at level $k$. $W(Q_k;d'_k,d_k,\Delta b_k ,b_k)$ additionally depends on the \emph{segment shape} of the loop at level $k$, determined by the type of the generator $Q_k = W, R, L$, the step values $d'_k$ and $d_k$ and $\Delta b_k = b_k - b'_k$. The nonzero segment shapes for a raising (R) generator are shown in Fig.~\ref{fig:single-loops}. In Table~\ref{tab:single-mateles} the nonzero matrix elements of the one-electron operator $\hat E_{ij}$\textemdash an over/under-bar indicates the loop head/tail\textemdash depending on the segment shape symbol, the step-values and the $b$-value are given in terms of the auxiliary functions 
\begin{equation}
\label{eq:a-c-functions}
A(b,x,y) = \sqrt{\frac{b + x}{b + y}} , \quad 
C(b,x) = \frac{\sqrt{(b+x-1)(b+x+1)}}{b + x}.
\end{equation}

\begin{table*}
\centering
\caption{\label{tab:single-mateles}Nonzero matrix elements of the one-body operator $\hat E_{ij}$ in terms of the auxiliary functions (\ref{eq:a-c-functions}.}

{\small \renewcommand{\arraystretch}{1.2}
\begin{tabular}{cccccccccc}
\cline{1-2} \cline{4-6} \cline{8-10}
 $d' d$ & W & & $d' d$ & $\overline R$ & $\underline L$ &  & $d' d$ & $\underline R$ & $\overline L$ \\ 
\cline{1-2} \cline{4-6} \cline{8-10}
 00 & 0  & & 01 & 1 & 1 &  & 10 & 1 & 1 \\ 
 11 & 1 & & 02 & 1 & 1 &  & 20 & 1 & 1 \\ 
 22 & 1 & & 13 & $A(b,0,1)$ & $A(b,2,1)$ &  & 31 & $A(b,1,0)$ & $A(b,0,1)$ \\ 
 33 & 2 & & 23 & $A(b,2,1)$ & $A(b,0,1)$ &  & 32 & $A(b,1,2)$ & $A(b,2,1)$ \\ 
\cline{1-2} \cline{4-6} \cline{8-10}
 & & & & & & & \\
  \cline{4-10}
  & & & &\multicolumn{2}{c}{R} & &  & \multicolumn{2}{c}{L}   \\ 
   \cline{4-10}
 & & &$d' d$ & $\Delta b = -1$ & $\Delta b = +1$ & & & $\Delta b = -1$ & $\Delta b = +1$  \\ 
  \cline{4-10} 
 & & & 00 & 1 & 1 & & & 1 & 1 \\ 
 & & & 11 & -1 & $C(b,0)$ & & & $C(b,1)$ & -1 \\ 
 & & & 12 & -$1/(b+2)$ & - & & &  $1/(b+1)$ & - \\ 
 & & & 21 & - & $1/b$ & & &  - & -$1/(b+1)$ \\ 
 & & & 22 & $C(b,2)$ & -1 & & & -1 & $C(b,1)$ \\ 
 & & & 33 & -1 & -1 & &  & -1 & -1 \\ 
  \cline{4-10}
\end{tabular} 
}

\end{table*}

\subsection{\label{sec:guga:two-body}Two-Body Matrix Elements}

\begin{figure*}
\centering
\includegraphics[width=0.85\textwidth]{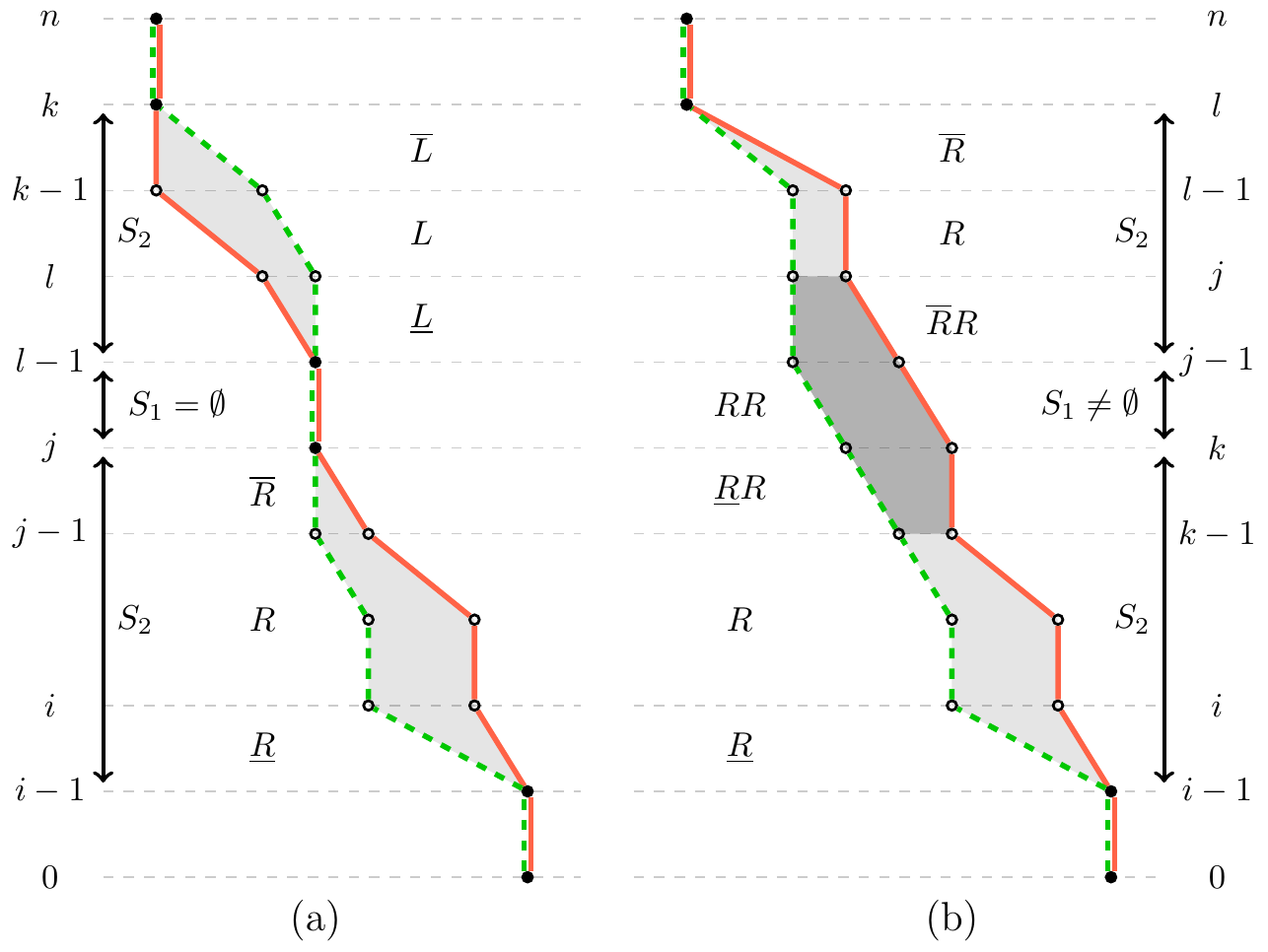}
\caption{\label{fig:two-body-loop}(Color online) Two examples of nonzero two-body matrix elements. (a) shows a non-overlapping ($i < j < l < k$) raising-lowering (RL) example and (b) shows a overlapping ($i < k < j < l$) loop with two raising generators (RR).}
\end{figure*}

The matrix elements of the two-body operators $\hat e_{ij,kl}$ are more involved than the one-body operators, especially the product of singlet excitation generators, $\hat E_{ij}\hat E_{kl}$. Similar to the one-electron operators, the GT states $\ket m$ and $\ket{m'}$ must coincide outside the total range $\min(i,j,k,l)$ to $\max(i,j,k,l)$ for $\braopket{m'}{\hat e_{ij,kl}}{m}$ to be nonzero. The form of the matrix element depends on the \emph{overlap range} of the two ranges 
\begin{equation}
\label{eq:overlap}
S_1 = (i,j)\cap (k,l).
\end{equation}
One possibility to calculate the matrix element would be to sum over all possible intermediate states, $\ket{m''}$, 
\begin{equation}
\label{eq:two-matel-sum}
\braopket{m'}{\hat E_{ij}\hat E_{kl}}{m} = \sum_{m''} \braopket{m'}{\hat E_{ij}}{m''}\braopket{m''}{\hat E_{kl}}{m},
\end{equation}
but in practice this is very inefficient. For non-overlapping ranges $S_1 =\emptyset$ the matrix element just reduces to the product 
\begin{equation}
\label{eq:two-matel-1}
\braopket{m'}{\hat e_{ij,kl}}{m} = \braopket{m'}{\hat E_{ij} \hat E_{kl}}{m} = \bra{m'} \hat E_{ij} \ket{m''}\bra {m''} \hat E_{kl} \ket m, 
\end{equation}
where $\ket{m''}$ must coincide with $\ket m$ in the range $(i,j)$ and with $\ket {m'}$ in range $(k,l)$. The same rules and matrix elements as for one-body operators apply in this case. An example of this is shown in the left panel of Fig.~\ref{fig:two-body-loop}.\\
For $S_1 \neq \emptyset$, we define the \emph{non-overlap range}
\begin{equation}
\label{eq:non-overlap}
S_2 = (i,j) \cup (k,l) - S_1,
\end{equation}
where the same restrictions and matrix elements as for one-body operators apply. In the overlap range, $S_1$, different restrictions for the visited Paldus tableau vertices $p$ apply for the matrix element to be nonzero. This depends on the type of the two generators involved and were worked out by Shavitt \cite{Shavitt1981}. For two raising generators (RR) the following conditions apply
\begin{align}
a_p' = a_p, \quad b_p' = b_p + 2, \quad c_p' = c_p - 2 \quad &\ra \underline{\Delta b_p = -2} \label{eq:r-db-2}\\ 
a_p' = a_p + 2, \quad b_p' = b_p + 2, \quad c_p' = c_p \quad &\ra \underline{\Delta b_p = +2} \label{eq:r-db+2}\\ 
a_p' = a_p + 1, \quad b_p' = b_p, \quad c_p' = c_p - 1 \quad &\ra \underline{\Delta b_p = 0}. \label{eq:r-db0}
\end{align}
For two lowering generators (LL):
\begin{align}
a_p' = a_p + 2, \quad b_p' = b_p + 2, \quad c_p' = c_p, \quad &\ra  \underline{\Delta b_p = -2} \label{eq:l-db-2}\\ 
a_p' = a_p, \quad b_p' = b_p - 2, \quad c_p' = c_p + 2 \quad &\ra  \underline{\Delta b_p = +2} \label{eq:l-db+2}\\ 
a_p' = a_p - 1, \quad b_p' = b_p, \quad c_p' = c_p + 2 \quad &\ra  \underline{\Delta b_p = 0}. \label{eq:l-db0}
\end{align}
And for a mixed combination of raising and lowering generators (RL)
\begin{align}
a_p' = a_p - 1, \quad b_p' = b_p + 2, \quad c_p' = c_p - 1,\quad &\ra  \underline{\Delta b_p = -2} \label{eq:m-db-2}\\ 
a_p' = a_p + 1, \quad b_p' = b_p - 2, \quad c_p' = c_p + 1\quad &\ra  \underline{\Delta b_p = +2} \label{eq:m-db+2}\\ 
a_p' = a_p, \quad b_p' = b_p, \quad c_p' = c_p\quad &\ra  \underline{\Delta b_p = 0}. \label{eq:m-db0}
\end{align}
Drake and Schlesinger \cite{Drake1977}, Paldus and Boyle \cite{Paldus1980}, Payne \cite{Payne1982} and Shavitt and Paldus \cite{Shavitt1981} were able to derive a scheme, where the two-body matrix elements can be computed as a product of \emph{segment values} similar to the one-body case (\ref{eq:single-prod}) 
\begin{equation}
\label{eq:double_matele}
\braopket{m'}{\hat e_{ij,kl}}{m} = \prod_{p\in S_2} W(Q_p;d'_p,d_p,\Delta b_p,b_p) 
\times \sum_{x=0,1} \prod_{p \in S_1} W_x(Q_p;d'_p,d_p,\Delta b_p,b_p),
\end{equation}
where $S_1$ and $S_2$ are the overlap (\ref{eq:overlap}) and non-overlap (\ref{eq:non-overlap}) ranges defined above.\\
$W(Q_p;d'_p, d_p, \Delta b_p, b_p)$ are the already defined single operator segment values, listed in Table~\ref{tab:single-mateles}, and  $W_x(Q_p;d'_p,d_p,\Delta b_p,b_p)$ are new segment values of the overlap range (their listing is omitted for brevity here, but can be found in Refs.~[\citen{Shavitt1981,dobrautz-phd}].
The sum over two products in $S_1$ corresponds to the \emph{singlet coupled} intermediate states ($x=0$), with a nonzero contribution if $\Delta b_p = 0, \forall p \in S_1$ and the \emph{triplet intermediate coupling} ($x = 1$). 

This product formulation of the two-body matrix elements in a spin-adapted basis is the great strength of the graphical unitary group approach, which allows an efficient implementation of the GT basis in the FCIQMC algorithm. 
The details of the matrix element calculation in this basis are, however, tedious and will be omitted here for brevity and clarity. More details on the matrix element calculation, especially the contributions of the two-body term to diagonal and one-body matrix elements can be found in Appendix~\ref{app:mat-eles} or in Refs.~[\citen{Shavitt1981, dobrautz-phd}].


\section{\label{sec:guga+fciqmc}Spin-Adapted Full Configuration Interaction Quantum Monte Carlo}

The Full Configuration Interaction Quantum Monte Carlo (FCIQMC) method \cite{original-fciqmc,initiator-fciqmc} attempts to obtain the exact solution of a quantum mechanical problem in a given single-particle basis set by an efficient sampling of a stochastic representation of the wavefunction\textemdash originally expanded in a discrete antisymmetrised basis of Slater determinants (SDs)\textemdash through the random walk of walkers, governed by imaginary-time the Schr\"odinger equation. 
For brevity of this manuscript we refer the interested reader to Refs.~[\citen{original-fciqmc,initiator-fciqmc}] and [\citen{linear-scaling-fciqmc}] for an in-depth explanation of the FCIQMC method.

Having introduced the theoretical basis of the unitary group approach (UGA) and its graphical extension (GUGA) to permit a mathematically elegant and computationally efficient incorporation of the total spin symmetry in form of the Gel'fand-Tsetlin basis, here we will present the actual implementation of these ideas in the FCIQMC framework, termed GUGA-FCIQMC.

Fundamentally, the three necessary ingredients for an efficient spin-adapted formulation of FCIQMC are:
\begin{enumerate}[label = \bf(\roman*)]
\item Efficient storage of the spin-adapted basis
\item Efficient excitation identification and matrix element computation
\item Symmetry adapted excitation generation with manageable computational 
cost 
\end{enumerate}
The first point is guaranteed with the UGA, since storing
the information content of a CSF and a SD amounts to the same memory requirement, with CSFs represented in the step-vector representation. 
Efficient identification of valid excitations is rather technical and explained in Appendix~\ref{app:excit_info} and in Ref.~[\citen{dobrautz-phd}]. 
For the present discussion we simply need to know, although it is more involved to determine if two CSFs are connected by a single application of $\hat H$ than for SDs, it is possible to do so efficiently. 
Matrix element computation is based on the product structure of the one-~(\ref{eq:single-prod}) and two-body~(\ref{eq:double_matele}) matrix elements derived by Shavitt \cite{Shavitt1978} explained above and presented in more detail in App.~\ref{app:mat-eles} and in Ref.~[\citen{dobrautz-phd}].
Concerning point (iii): symmetry adaptation in FCIQMC is most efficiently implemented at the excitation generation step, by creating only symmetry-allowed excitations. 
For the continuous $SU(2)$ spin symmetry this is based on Shavitt's DRT and the restriction for nonzero matrix elements in the GUGA. 
This, in addition to the formulation in a spin-pure GT basis, ensures that the total spin quantum number $S$ is conserved in a FCIQMC calculation.

\subsection{\label{sec:excit-gen-singles}Excitation Generation: Singles}
The concept of efficient excitation generation in the spin-adapted GT 
basis via the GUGA will be explained in detail by the example of 
single excitations. Although more complex, the same concepts apply for generation of double excitation, which are discussed below. 

In contrast to excitation generation for SDs, there are now two steps involved for a CSF basis. 
The first, being the same as in a formulation of FCIQMC in Slater determinants, is the choice of the two spatial orbitals $i$ and $j$, with probability $p(i)\,p(j|i)$. 
This should be done in a way to ensure the generation probability to be proportional to the Hamiltonian matrix element involved. 
However, here comes the first difference of a CSF-based implementation compared to a SD-based one. 
For Slater determinants, the choice of an electron in spin-orbital $(i,\sigma)$ and an empty spin-orbital $(j,\sigma)$ is sufficient to 
uniquely specify the excitation $\ket{D_j} = a_{j,\sigma}^\dagger a_{i,\sigma} \ket{D_i}$, and to calculate the involved matrix element $\braopket{D_j}{\hat H}{D_i}$. 
However, in a CSF basis, the choice of an occupied \emph{spatial} orbital $j$, and empty or singly occupied \emph{spatial} orbital $i$, only determines the type of excitation generator $\hat E_{ij}$ acting on an CSF basis state $\ket{m}$ as well as the involved \emph{integral contributions} $t_{ij}, V_{ikjk}$ and $V_{ikkj}$ of the matrix element $\braopket{m'}{\hat H}{m}$.
To ensure $p(i)p(j|i) \propto \abs{t_{ij} +\frac{1}{2} \sum_{k\in \text {occ}} (V_{ikjk} - V_{ikkj})}$, the occupied orbital $j$ and (partially) empty $i$ are picked in the same way as for SDs, but with an additional restriction to ensure $\hat E_{ij} \ket{m} \neq 0$.
However, the choice of $(i,j)$ does not uniquely determine the excited CSF as there are multiple possible ones, as explained above. 
%

As a consequence, the choice of spatial orbitals $i$ and $j$ does not determine the \emph{coupling coefficient} $\braopket{m'}{\hat E_{ij}}{m}$ of the matrix element $H_{m'm}$. 
Optimally, for a given $\ket{m}$ and generator $\hat E_{ij}$, the connected CSF $\ket{m'}$ has to be created with a probability $p(m'|m)$ proportional to the coupling coefficient $\braopket{m'}{\hat E_{ij}}{m}$. 
By ensuring $p(i)p(j|i)$ is proportional to the integral contributions and $p(m'|m)$ to the coupling coefficients, the total spawning probability 
\begin{equation}
\label{eq:csf-pgen}
p_s(m'|m) = p(i)\, p(j|i)\, p(m'|m)
\end{equation}
will be proportional to the magnitude of Hamiltonian matrix element $\abs{H_{m'm}}$.
The efficiency of the FCIQMC algorithm depends on the ratio of the Hamiltonian matrix element $\abs{H_{m'm}}$ between two connected states and the probability $p_s(m'|m)$ to choose the excitation $\ket m \rightarrow \ket {m'}$, as the imaginary timestep $\Delta \tau$ of the simulation is adapted to faithfully account for all excitations
\begin{equation}\label{eq:time-step}
\Delta \tau^{-1} \propto \frac{\abs{H_{m'm}}}{p_s(m'|m)}.
\end{equation}
In a primitive implementation, $\Delta \tau$ is determined by the ``worst-case'' $\max{\abs{H_{m'm}/p_s(m'|m)}}$ ratio during a simulation. 
A less strict approach to this problem is discussed below.
By choosing nonzero $\hat E_{ij} \ket{m} \neq 0$ and ensuring 
$p(m'|m)$ is achieved by a \emph{branching tree} approach, 
we obtain one of the different possible walks on the Shavitt graph with nonzero loop contributions with the starting CSF $\ket m$.

\subsection{\label{sec:branching-tree}The Branching Tree}
 
\begin{figure}
\centering
\includegraphics[width=0.48\textwidth]{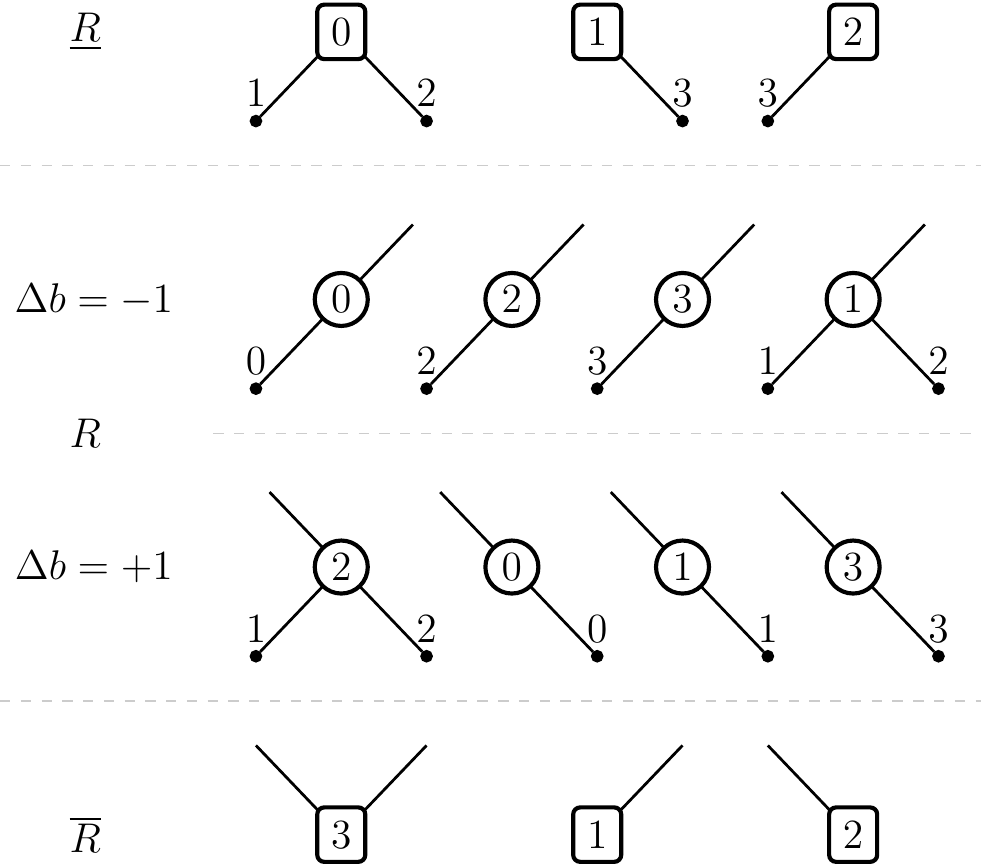}
\caption{\label{fig:single-tree-elements}Branching tree elements of a one-body operator $\hat E_{ij}$.}
\end{figure}

In the spin-adapted excitation generation, after a certain generator $\hat E_{ij}$ is picked with a probability $p(i)p(j|i)$ based on the integral contributions of the Hamiltonian matrix element, the type of generator is determined, raising (R) if $i < j$ and lowering (L) if $i > j$. 
One connecting single excitation is then chosen by looping from starting orbital $\min(i,j)$ to $\max(i,j)$ and stochastically choosing a valid nonzero Shavitt graph, based on the restrictions~(\ref{eq:R-dbk-1}-\ref{eq:L-dbk+1}), mentioned in the GUGA section above.
As an example, let us have a closer look at a chosen \emph{raising} generator. 
As can be seen in the single segment value Table~\ref{tab:single-mateles} there are 4 possible nonzero starting $\underline{R}$ matrix elements. 
These starting segments are associated with a relative difference of the total spin $\Delta S_i = S_i(m') - S_i(m)$ and $\Delta b_i = b_i(m') - b_i(m)$ between the two CSFs $\ket{m}$ and $\ket{m'}$ at level $i$, as shown in Table~\ref{table:R-start}. 
For certain step-values ($d_i = 0$ for raising and 
$d_i = 3$ for lowering generators) two possible excited CSFs with different $\Delta b_i$ are possible. 
This can be represented pictorially as elements 
of a branching tree, as seen in Fig.~\ref{fig:single-tree-elements} for raising generator, where the number in the boxes represent the step-value $d_i$ of $\ket{m}$ and the direction of the outgoing lines the $\Delta b_i$ value (left going lines correspond to $\Delta b_i = -1$ and right going ones $\Delta b_i = +1$).
The number above the small dots represent the associated $d'_i$ value of the excited $\ket{m'}$.

\begin{table*}
\centering
\caption{\label{table:R-start}Nonzero starting segments for $\underline{R}$ with the number of electrons $N_i' = N_i +1$ in all cases.}
\small
\begin{threeparttable}
\renewcommand{\arraystretch}{1.3}
\begin{tabular}{ccccccccccccccccc}
\toprule
\multicolumn{4}{c}{$\underline{R}$} & & \multicolumn{3}{c}{$\overline{R}$} & & \multicolumn{4}{c}{$\underline{L}$} & & \multicolumn{3}{c}{$\overline{L}$}\\
\cline{1-4} \cline{6-8} \cline{10-13} \cline{15-17}
$d_i$ & $d_i'$  & $\Delta S_i$ & $\Delta b_i$ & & $d_j$ & $d'_j$ & $\Delta b_{j-1}$\tnote{a} & & $d_i$ & $d_i'$  & $\Delta S_i$ & $\Delta b_i$ & & $d_j$ & $d'_j$ & $\Delta b_{j-1}$\tnote{a} \\
\cline{1-4} \cline{6-8} \cline{10-13} \cline{15-17}
0 & 1  & $+1/2$  & $-1$ & & 1 & 0 & $-1$ & & 1 & 0 & $-1/2$ & $+1$\tnote{b} & & 0 & 1 & $+1$ \\ 
0 & 2  & $-1/2$  & $+1$\tnote{c} & & 2 & 0 & $+1$ & & 2 & 0  & $+1/2$  & $-1$ & & 0 & 2 & $-1$ \\ 
1 & 3  & $-1/2$  & $+1$\tnote{b} & & 3 & 1 & $+1$ & & 3 & 1  & $+1/2$  & $-1$ & & 1 & 3 & $-1$ \\ 
2 & 3  & $+1/2$  & $-1$ & & 3 & 2 & $-1$ & & 3 & 2 & $-1/2$ &  $-1$\tnote{c} & & 2 & 3 & $+1$\\ 
\botrule 
\end{tabular}
\begin{tablenotes}
\footnotesize
\item [a] Necessary $\Delta b$ value for a valid CSF.
\item [b] Here $b_i > 0$ is ensured, due to $d_i = 1$.
\item [c] Only for $b_i > 0$ otherwise $S_i' < 0$ would be a non-valid CSF.
\end{tablenotes}
\end{threeparttable}
\end{table*}

The intermediate contributions to the coupling coefficients $R/L$, see 
Table~\ref{tab:single-mateles}, have similar properties. 
Depending on the current $\Delta b_k$ value of the excitation $\ket{m'}$ relative to $\ket{m}$ there are \emph{branching} possibilities for singly occupied spatial orbitals in $\ket{m}$, corresponding to possible spin-recouplings in the \emph{excitation range} of $\hat E_{ij}$. 
An excitation with $\Delta b_{k-1} = -1$ can branch at $d_k = 1$ values, into $d'_k = 1$ with $\Delta b_{k-1} = \Delta b_k = -1$ or change the spin-coupling to $d'_k = 2$ accompanied by a change to $\Delta b_k = +1$. 
At empty or doubly occupied orbitals only $d'_k = d_k$ and $\Delta b_{k} = \Delta b_{k-1}$ leads to nonzero excitations.
These relations are tabulated in Table~\ref{tab:non-zero-inter} and pictorially represented in Fig.~\ref{fig:single-tree-elements}.

The possible single excitations of a given CSF can be represented by a branching diagram, where each node is a successive element of $d_k$ and a left going branch represents a $\Delta b_k = -1$ value and a right going branch $\Delta b_k = +1$. 
The end value $d_j = 1$ requires an incoming $\Delta b_{j-1} = -1$ value, whereas $d_j = 2$ requires $\Delta b_{j-1} = +1$ to ensure $\Delta b_j = 0$ at the end of the excitation, indicated by the directions of the ingoing lines  of the elements at the bottom of Fig.~\ref{fig:single-tree-elements}.  
For a raising generator both $\Delta b_{j-1}$ values are possible for $d_j = 3$.
The restrictions on the end segments $\oR/\oL$ are listed in Table~\ref{table:R-start} and pictorially represented in Fig.~\ref{fig:single-tree-elements}. 
These restrictions are a direct consequence of the conservation of the total spin quantum number $S$ in the GUGA.

\begin{figure*}
\centering
\begin{minipage}[c]{0.4\textwidth}
\centering
\includegraphics[width=0.6\columnwidth]{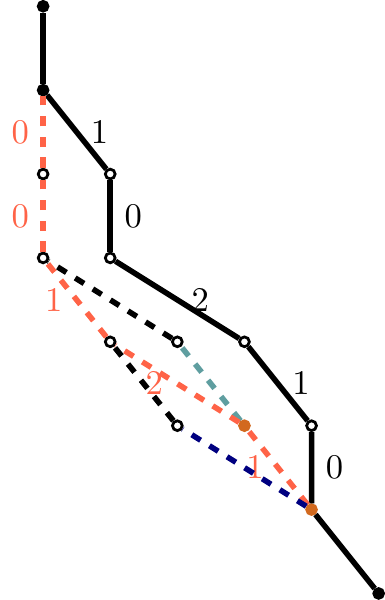}
\end{minipage}
\begin{minipage}[c]{0.55\textwidth}
\centering
\includegraphics[width=0.7\columnwidth]{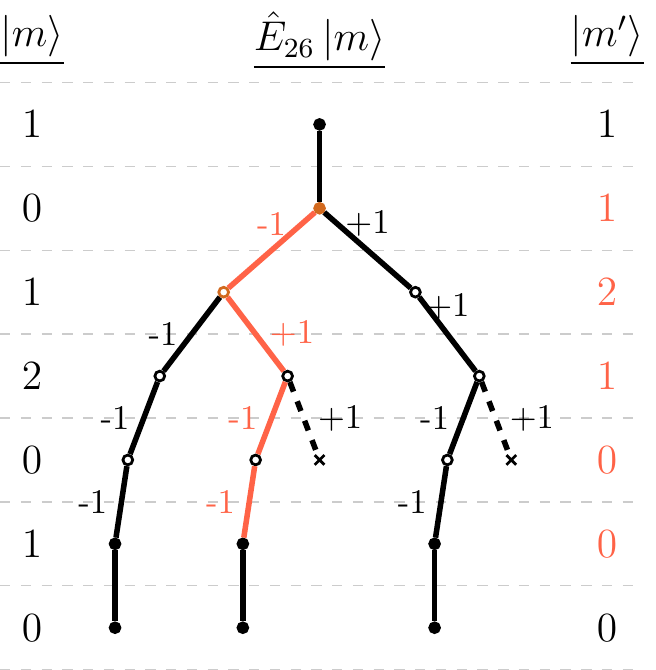}
\end{minipage}
\caption{\label{fig:branching-tree-combined}(Color online) Example of different branching possibilities for a raising single excitation $\hat E_{26}\ket{1,0,1,2,0,1,0}$ on the left. Branching tree form of the possible single excitation of $\hat E_{2,6}\ket{1,0,1,2,0,1,0}$ on the right.}
\end{figure*}

\begin{table}
\centering
\caption{\label{tab:non-zero-inter}Nonzero intermediate $R$ and $L$ segments.}
{\small
\begin{threeparttable}
\renewcommand{\arraystretch}{1.1}
\begin{tabular}{cccc}
\toprule
\multirow{2}{*}{$d_k$} & \multirow{2}{*}{$d'_k$}& $\Delta b_{k-1} = -1$ & $\Delta b_{k-1} = +1$ \\
\cline{3-4}
 & & $\Delta b_k$ & $\Delta b_{k}$ \\
 \hline
 0 & 0 & $-1$ & $+1$ \\
 1 & 1 & $-1$ & $+1$ \\
 1 & 2 & $-1$\tnote{a} & -\tnote{c} \\
 2 & 1 & -\tnote{c} & $-1$ \\
 2 & 2 & $-1$ & $-1$\tnote{b}\\
 3 & 3 & $-1$ & $+1$\\
 \botrule
\end{tabular}
\begin{tablenotes}
\footnotesize
\item [a] $b_k > 0$ is ensured due to $d_k = 1$.
\item [b] Only possible if $b_k > 1$.
\item [c] Not possible otherwise $\abs{\Delta b_k} > 1$.
\end{tablenotes}
\end{threeparttable}
}
\end{table}
A very simple implementation to create a single excitation $\ket{m'}$ would be to loop from orbital $i$ to $j$ and depending on the step-value $d_k$ of $\ket m$, at each orbital $k\in(i,j)$ choose one possible $\Delta b_k$ path at random if there are multiple possible ones. 
However, this would totally neglect that there are certain branching choices which would lead to a dead end, due to incompatible end-segments $\overline{R},\overline{L}$ and would not relate the probability to create a certain CSF $\ket {m'}$ to the magnitude of coupling coefficient.
An example of the excitation generation based on the \emph{branching tree} is given in Fig.~\ref{fig:branching-tree-combined}, for the raising generator $\hat E_{26}$ acting on the CSF $\ket{m} = \ket{1,0,1,2,0,1,0}$, moving an electron from spatial orbital $6$ to $2$. 
The left panel of Fig.~\ref{fig:branching-tree-combined} shows this excitation in the Shavitt graph form based on the DRT and the right panel shows the branching tree representation (with the orbitals ordered from 
top to bottom now, as this is the usual representation of trees.), with $\pm 1$ indicating the $\Delta b_k$ value associated to the possible branches. 
The orange path (color online) in both the Shavitt graph and branching tree representation show one valid single excitation $\ket{m'}$ of $\hat E_{26}\ket{m}$. 
The above mentioned dead ends are indicated with dashed lines and crossed out vertices in the right panel of Fig.~\ref{fig:branching-tree-combined}. 

As one can see the number of connected CSFs $n_{m'}$ to $\ket m$ via a single application of $\hat E_{ij}$ depends on the number of singly occupied orbitals $n_s$ within the excitation range $(i,j)$ and grows approximately as $n_{m'} \approx 1.6^{n_s+2}$.
The highest number of possible connected CSFs is given for a starting segment $\uR/\uL$ with two possible branches, exclusively alternating singly occupied orbitals $d_k = \{1,2\}$ in the excitation range with $b_k > 0$ and an end-segment $\oR/\oL$ with nonzero contributions for both $\Delta b_k = \pm 1$. 
In this case the number of connected CSFs is related to the Fibonacci series and given by the Fibonacci number 
\begin{equation}
N_S^{max} = F_{n+2} = \sum_{k=0}^{\floor{\frac{n-1}{2}}} {{n-k-1}\choose{k}}. 
\end{equation}
Calculating all possible excitations would lead to an exponential wall for highly open-shell CSFs $\ket m$, but since we only need to obtain \textbf{one} connected CSF in the excitation generation of FCIQMC, this exponential scaling is not an immediate problem. 
However, since the overall generation probability $\sum_{m'} p(m'|m)$ is normalized to unity, a specific $p(m'|m)$ will be negligibly small for numerous possible excitations. 
As the timestep is directly related to this probability (\ref{eq:time-step}), a small $p(m'|m)$ directly causes a lowering in the usable $\Delta \tau$ in a FCIQMC calculation. 

Since this is a consequence of the inherent \emph{high connectivity} of a spin-adapted basis, systems with many open-shell orbitals are difficult to treat in such a basis. 
In general this restricts common implementations of spin-eigenfunctions to a maximum of 18 open-shell orbitals. 
However, similar to the avoidance of the exponential wall associated with the FCI solution to a system, the stochastic implementation of the CSF excitation generation in FCIQMC, avoids the exponential bottle-neck caused by the high connectivity of a CSF basis. 

\subsection{\label{sec:rem-switches}Remaining Switches}

To avoid ending up in incompatible dead-end excitations it is convenient, for a given excitation range $(i,j)$, to determine the vector of remaining switch possibilities $s_k({\Delta b_k})$ for the $\Delta b_k = \pm 1$ branches.
$s_k(\Delta b_k)$ is the number of $d_{k'} = 1$ for $\Delta b_k = -1$ and $d_{k'} = 2$ for $\Delta b_k = +1$ to come in $k' = k + 1, \dots j-1$ (with the already mentioned restriction of $b_{k'} > 1$ for $d_{k'} = 2$ to be a valid $\Delta b_k = +1$ switch)
\begin{equation}
\label{eq:remaining-switch}
s_k(\Delta b_k) = \begin{cases}
\sum_{l = k+1}^{j-1} \delta_{d_l,1} \quad \text {for} \quad \Delta b_k = -1 \\
\sum_{l = k+1}^{j-1} \delta_{d_l,2} \quad \text {for} \quad \Delta b_k = +1.
\end{cases}
\end{equation} 
The quantity $s_k(\Delta b_k)$ can be used to decide if a possible $\Delta b_k$ branch is taken or not, depending on if it will end up in a dead-end of the branching tree. 

\subsection{\label{sec:me-on-fly}On-The-Fly Matrix Element Calculation}

To pick the connecting CSF $\ket {m'}$ with a probability $p(m'|m)$ relative to the magnitude of the generator matrix element $\abs{\bra {m'} \hat E_{ij} \ket m}$
we have to investigate the matrix element $\bra{m'} \hat E_{ij} \ket m$ between a given CSF $\ket m$ and an excitation $\ket{m'}$. 
As the coupling coefficient is calculable as a product of terms, which depend on the type of excitation (lowering, raising) and 
is determined by the step-vector values $d_k$, $d'_k$, the $b_k$ and the $\Delta b_k$ associated to each level of the excitation, see Eq.~(\ref{eq:single-prod}).
One of the major advantages of the GUGA in FCIQMC is that this matrix 
element can be calculated \textbf{on-the-fly} during the creation 
of the excitation. As one can see in Table~\ref{tab:single-mateles} there is a relation between the matrix element amplitude 
and the number of direction switches of $\Delta b_k$ in the excitation range.
Most product contributions are of order $\bigO 1$, except the elements related to a switch of $\Delta b_{k+1} \leftarrow -\Delta b_{k}$, which are of order $\bigO{1/b_k}$
\begin{equation}
W(Q_k;d'_k,d_k,\Delta b_k,b_k)= 
\begin{cases}
\bigO{1} & \text{for } d_k = d'_k \\
\bigO{b_k^{-1}} & \text{for } d_k \neq d'_k.
\end{cases}
\end{equation}
So for a higher intermediate value of $b_k$, which in the end also means more possibly pathways in the branching tree, it should be less favorable to change the current $\Delta b_k$ value. 
In order to create an excitation $\ket{m'}$ with a probability proportional 
to the coupling coefficient $\abs{\bra {m'} \hat E_{ij} \ket m}$ this fact is included in the decision of the chosen branch and is achieved by the use of \emph{branch weights.}

\subsection{\label{se:branch-weights}Branch Weights}
\begin{figure*}
\centering
\includegraphics[width=0.9\textwidth]{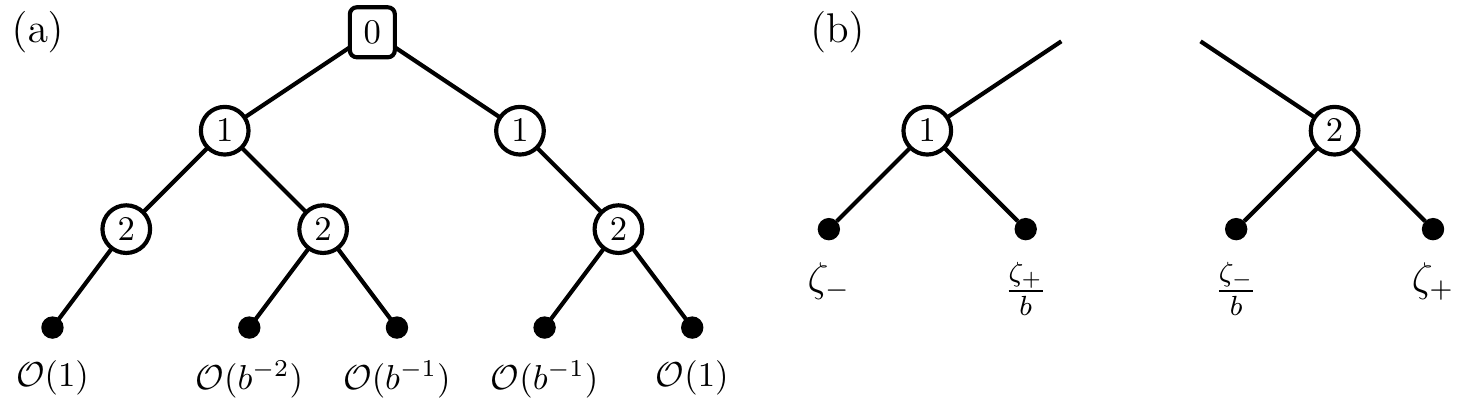}
\caption{\label{fig:branch-comb}\small (a) Scaling of one-body coupling coefficients with changes in $\Delta b$ along the branching tree. (b) Future branch weights at branching possibility.}
\end{figure*}

It is possible to take into account the ``probabilistic weight'' of each tree 
branches at a possible branching decision. 
As one can see in the left panel of Fig.~\ref{fig:branch-comb}, the starting $\Delta b_i = \pm 1$ branches each have one contribution of order $\bigO 1$. 
For each branching possibility there is a resulting branch with opposite $\Delta b$ and weight of order $\bigO{b^{-1}}$. 
However, it also depends on the end-segment determined by $d_j$, if a given 
branch can be chosen. The following branch weights
\begin{equation}
\label{eq:singleWeights}
\zeta_- = f(d_j) + \frac{s_k(-1)}{b} g(d_j) + \bigO{\frac{1}{b^2}}, \qquad
\zeta_+ = g(d_j) + \frac{s_k(+1)}{b} f(d_j) + \bigO{\frac{1}{b^2}}
\end{equation}
with
\begin{equation}
\label{eq:end-cond}
 f(d_j) = \begin{cases}
           0 & \text{if } d_j = 2 \\
           1 & \text{else}
          \end{cases}
 ,\qquad
g(d_j) = \begin{cases}
          0 & \text{if } d_j = 1 \\
          1 & \text{else}
         \end{cases}
\end{equation}
where $s_k(\pm 1)$ is the number of remaining switches (\ref{eq:remaining-switch}), can be used to determine the probability of each $\Delta b$ branch to be chosen. 
The right panel of Fig.~\ref{fig:branch-comb} shows the influence of the 
matrix element on the branching probabilities in the excitation range. 
By choosing the $\Delta b = -1$ with a probability $p_- = \frac{\zeta_-}{\zeta_- + \zeta_+}$ at the start of an excitation
and in the excitation region choose to \emph{stay} on the current $\Delta b$ branch according to
\begin{equation}
 \label{eq:stay}
 p_s^{\pm} = \frac{b \zeta_\pm}{b \zeta_\pm + \zeta_\mp},
\end{equation}
the overall probability to choose the specific excitation $\ket{m'}$ is given by 
\begin{equation}
\label{eq:overall-pgen}
p(m'|m) = p_-(i) \prod_{k = i + 1}^{j - 1} p_s^{\pm}(k). 
\end{equation}
With this choice of branching probabilities it is possible to retain an \emph{almost linear} ratio between $p(m'|m)$ and coupling coefficient amplitudes $\abs{\braopket{m'}{\hat E_{ij}}{m}}$. 
Additionally, because of the $f(d_j)$ and $g(d_j)$ functions and inclusion of the remaining switches (\ref{eq:remaining-switch}) in Eq.~(\ref{eq:singleWeights}), this approach avoids dead-ends and thus choosing invalid excitations. 

An important note on the matrix element calculation of single excitations: 
there are of course contractions of the two-body operator in Eq.~(\ref{eq:hamil-matel}), which contribute to the matrix element of a single excitation $\braopket{m'}{\hat H}{m}$. These contractions have to be taken into account in the ``on-the-fly matrix element computation'' and are explained in more detail in Appendix~\ref{app:mat-eles} or can be found in Ref.~[\citen{dobrautz-phd}]. 

\section{\label{sec:excit-gen-doubles}Excitation Generation: Doubles}

The generation of double excitation in the GUGA formalism is much more 
involved than single excitations and the detailed background on matrix 
element computation and weighted orbital choice can be found in Appendix~\ref{app:pick-orb} or Ref.~[\citen{dobrautz-phd}] for conciseness of this manuscript. 
Here we will only present the general ideas
involved in doubly excitation generation.

\begin{figure*}
\centering
\includegraphics[width=\textwidth]{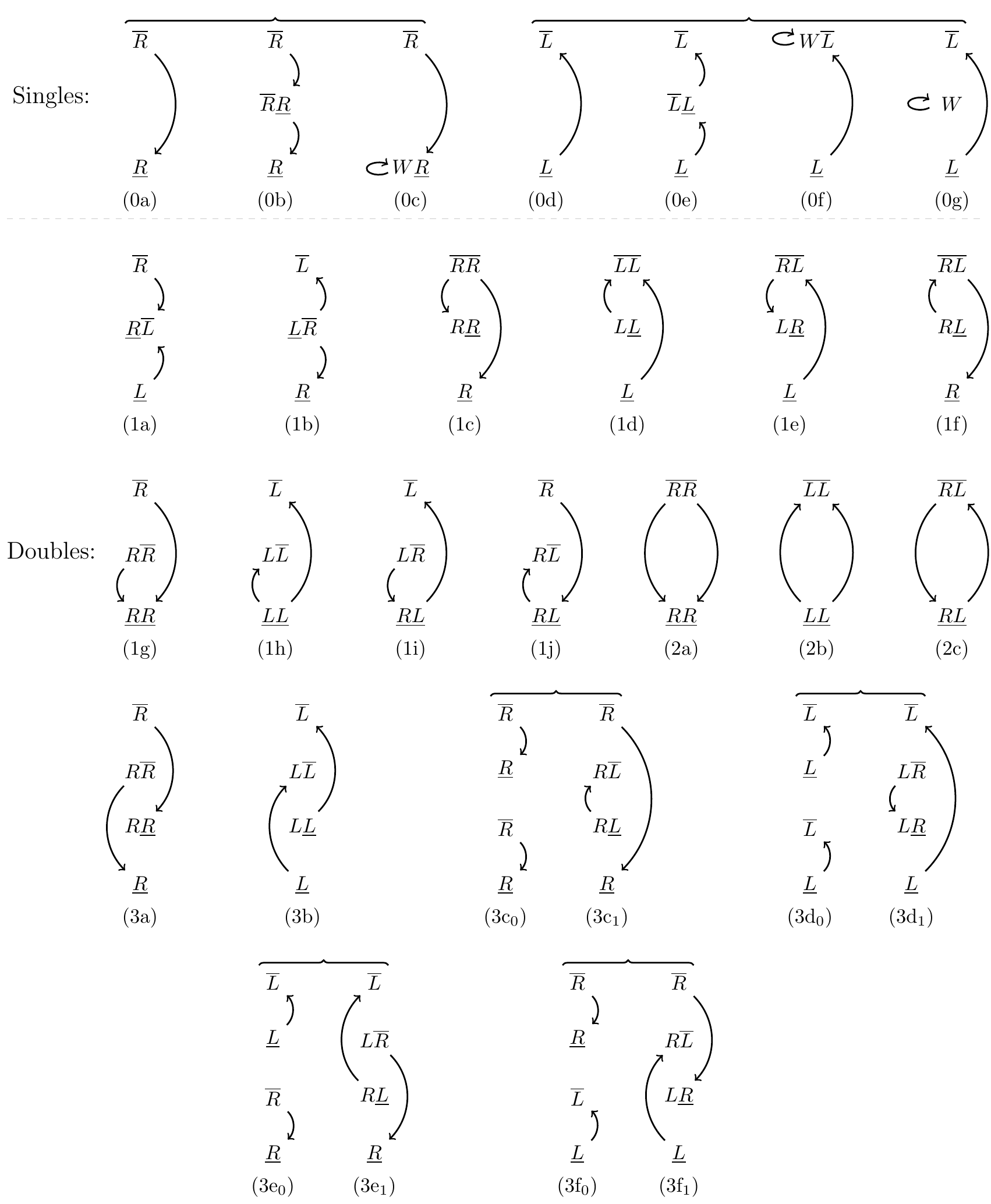}
\caption{\label{fig:excit-types}The 30 different types of single and double excitations, where the equivalent excitations are grouped together and reduce the number of distinct types to 21. The indices correspond to the entries in Table~\ref{tab:types-of-doubles}.}
\end{figure*}

Depending on the ordering of the involved spatial orbitals of the one- and
two-body generators, $\hat E_{ij}$ and $\hat e_{ij,kl}$, thirty different 
excitation types, involving different combinations of lowering (L) and raising (R) generators, can be identified and are listed in Table~\ref{tab:types-of-doubles}.
\begin{table}
\centering
\small
\caption{\label{tab:types-of-doubles}Distinct types of double excitations. $i < j < k < l$ in all cases and $e_{ij,kl} = e_{kl,ij}$ in mind.}
\resizebox{0.48\textwidth}{!}{
{\renewcommand{\arraystretch}{1.2}
\begin{tabular}{lcc}
\toprule
Label & Generator order & Operator \\
\hline
0a & $\uR(i) \ra \oR(j)$ & $\hat E_{ij}$\\
0b & $\uR(i) \ra \oR\uR(j) \ra \oR(k)$ & $\hat e_{ij,jk}$\\
0c & $W\uR(i) \ra \oR(j)$ & $\hat e_{ii,ij}$\\

0d & $\uL(i) \ra \oL(j)$ & $\hat E_{ji}$\\
0e & $\uL(i) \ra \oL\uL(j) \ra \oL(k)$ & $\hat e_{ji,kj}$\\
0f & $\uL(i) \ra W\oL(j)$ & $\hat e_{ji,jj}$\\
0g & $\uL(i) \ra W(j) \ra \oL(k)$ & $\hat e_{ji,kk}$\\
\hline
1a & $\ul L(i) \ra \ol L \ul R(j) \ra \ol R(k) $ &$\hat e_{ji,jk}$  \\
1b & $\ul{R}(i) \ra \ol{R}\ul L(j) \ra \ol L(k)$ & $\hat e_{ij,kj}$ \\
1c & $\ul R(i) \ra \ul R R(j) \ra \ol{RR}(k) $ &$\hat e_{jk,ik} $ \\
1d & $\ul L(i) \ra L \ul L(j) \ra \ol{LL}(k) $ &$\hat e_{ki,kj} $  \\ 
1e & $\ul L(i) \ra \ul R L(j) \ra \ol{RL}(k) $ & $\hat e_{jk,ki} $ \\ 
1f & $\ul R(i) \ra \ul L R(j) \ra \ol{RL}(k) $ & $\hat e_{kj,ik}$ \\

1g & $\ul{RR}(i) \ra R \ol R(j) \ra \ol R(k) $ & $\hat e_{ik,ij} $\\
1h & $\ul{LL}(i) \ra \ol L L(j) \ra \ol L(k) $ & $\hat e_{ji,ki} $ \\
1i & $\ul{RL}(i) \ra \ol R L(j) \ra \ol L(k) $ & $\hat e_{ij,ki} $ \\ 
1j & $\ul{RL}(i) \ra \ol L R(j) \ra \ol R(k) $ & $\hat e_{ji,ik}$  \\

2a & $\ul {RR}(i) \ra \ol {RR}(j)$ & $\hat e_{ij,ij} $ \\
2b & $\ul {LL}(i) \ra \ol {LL}(j)$ & $\hat e_{ji,ji} $ \\
2c & $\ul{RL}(i) \ra \ol {RL}(j) $ & $\hat e_{ij,ji} $\\

3a & $\ul R(i) \ra \ul R R(j) \ra R \ol R(k) \ra \ol R(l) $ & $\hat e_{jl,ik}/\hat e_{jk,il}$\\
3b & $\ul L(i) \ra L\ul L(j) \ra \ol L L(k) \ra \ol L(l) $ & $\hat e_{ji,lk}/\hat e_{li,kj}$\\
3c$_0$ & $\uR(i) \ra \oR(j) \ra \uR(k) \ra \oR(l)$ & $\hat e_{ij,kl}$\\
3c$_1$ & $\ul R(i) \ra \ul L R(j) \ra \ol L R(k) \ra \ol R(l) $ & $\hat e_{kj,il}$ \\ 
3d$_0$& $\uL(i) \ra \oL(j) \ra \uL(k) \ra \oL(l)$ & $\hat e_{ji,lk}$\\
3d$_1$& $\ul L(i) \ra \ul R L(j) \ra \ol R L(k) \ra \ol L(l) $ & $\hat e_{jk,il}$ \\
3e$_0$ & $\uR(i) \ra \oR(j) \ra \uL(k) \ra \oL(l)$ & $\hat e_{ij,lk}$\\
3e$_1$ & $\ul R(i) \ra \ul L R(j) \ra \ol R L(k) \ra \ol L(l) $ & $\hat e_{lj,ik}  $ \\
3f$_0$ & $\uL(i) \ra \oL(j) \ra \uR(k) \ra \oR(l)$ & $\hat e_{ji,kl}$\\
3f$_1$ & $\ul L(i) \ra \ul R L(j) \ra \ol L R(k) \ra \ol R(l) $ & $\hat e_{jl,ki}$ \\

\botrule
\end{tabular}
}
}
\end{table} 
Some of them are equivalent, in the sense that they lead to the same excitations, such as the 7 single excitation (0a-0g) in Table~\ref{tab:types-of-doubles}, which reduce to the two distinct raising $\uR \ra \oR$ and lowering $\uL \ra \oL$ generators. 
The pictorial representation of these generators are shown 
in Fig.~\ref{fig:excit-types}, where the ordering of orbitals is from 
bottom to top and arrows indicate the replacement of electrons. 
The two-body operators $\hat e_{ij,kl}$, which contribute to single 
excitations, (0c-0g) in Table~\ref{tab:types-of-doubles}, are already accounted for in the single excitation matrix element calculation, see Sec.~\ref{sec:excit-gen-singles} and Appendix \ref{app:off-diag}. 
These also include the single overlap excitation (0b) and (0e) with two alike generator types. 

Double excitation with a single overlapping index $j$ but two different 
generators (1a) and (1b) can be treated  in a similar way to single excitations, with the same weighting functions (\ref{eq:singleWeights}) and classification of remaining switches (\ref{eq:remaining-switch}), but with a change of generator type at the overlap site, $L \leftrightarrow R$. 
Double excitations with an empty overlap range $S_1$ (\ref{eq:overlap}) (3c$_0$, 3d$_0$, 3e$_0$ and 3f$_0$) can be calculated as the product of two 
single excitations (\ref{eq:two-matel-1}). 
However, e.g.\ for excitation (3c$_0$), the two-body operators $\hat e_{ij,kl}$ and $\hat e_{kj,il}$ contribute to the same Hamiltonian matrix element. 
We made the decision to treat these \emph{non-overlap} excitations by using the corresponding two-body generators with a nonzero overlap range $S_1$, see App.~\ref{app:pick-orb} for more details. 

For ``proper'' double excitations, we separate the excitation range 
$\min(i,j,k,l)\ra \max(i,j,k,l)$ into the lower non-overlap range $S_2$ below the overlap range $S_1$ and the upper non-overlap range $S'_2$ above $S_1$, 
as depicted in Fig.~\ref{fig:two-body-loop}. 
We introduce the terminology of a \emph{full-start} of mixed generators 
$\uR\uL$ and alike generators $\uRR/\uLL$, a \emph{semi-start} corresponds 
to the segment types like $R\uR$ or $R\uL$, a \emph{semi-stop} indicates 
generator combination like $L\oL$ or $L\oR$ and a \emph{full-stop} is where
both generators end on the same orbital, e.g.\ $\oLL$ or $\oR\oL$.

\begin{figure}
\centering
\includegraphics[width=0.48\textwidth]{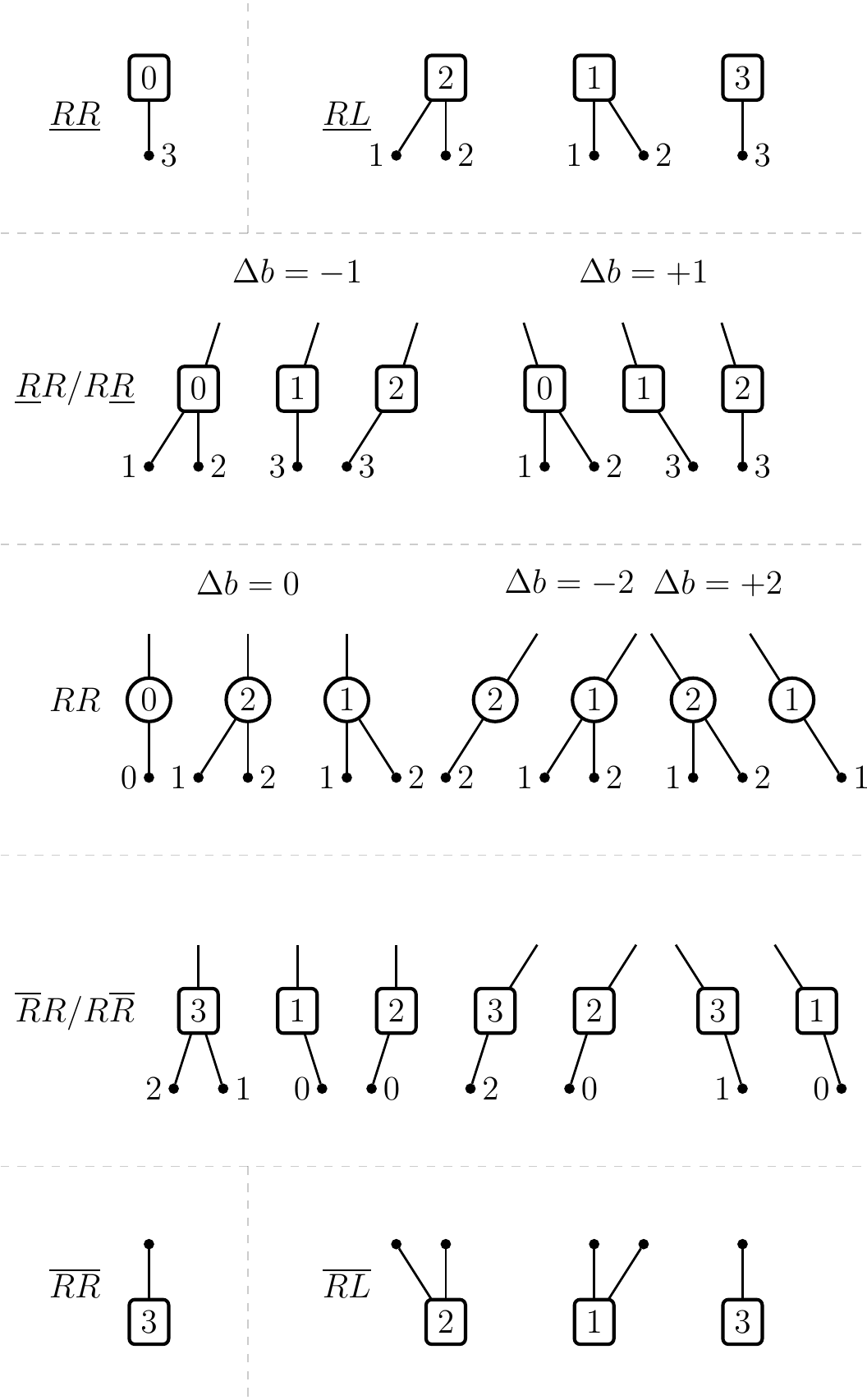}
\caption{\label{fig:double-branch-elements}An example of possible new 
branching tree elements for double excitations. Vertical lines now indicate the $\Delta b = 0$ branch, while left going lines in the overlap range correspond to $\Delta b = -2$ and right going ones to $\Delta b = +2$.}
\end{figure}

The excitation generation for doubles is again performed by choosing a 
valid path in a branching tree with modified rules in the 
overlap range $S_1$ of the double excitation. 
As can be seen by the restrictions for nonzero two-body matrix elements, Eq.~(\ref{eq:r-db-2}-\ref{eq:m-db0}), the 
allowed $\Delta b$ values in $S_1$ are now $\pm 2$ and $0$. 
This leads to new elements of the branching tree in $S_1$, which are shown 
by the example of alike raising and mixed generators in Fig.~\ref{fig:double-branch-elements}, where vertical lines indicate the new $\Delta b = 0$ branch, and left (right) going lines in $S_1$ correspond to 
$\Delta b = \pm 2$. The rules for the intermediate elements $RR, LL$ and $RL$ are the same for all combinations of generators. 

The calculation of the remaining switch possibilities (\ref{eq:remaining-switch}) essentially is the same as for single excitations, except they are calculated for each segment, $S_2, S_1$ and $S'_2$ of the excitation separately. In $S_1$ a $\Delta b_k = -2$ branch 
can switch at $d_k = 1$, a $\Delta b_k = +2$ at $d_k = 2$ and the 
$\Delta b_k = 0$ branch at both open-shell step-values 
\begin{equation}
\label{eq:remaining-switch-double}
s_k(\Delta b_k) = \begin{cases}
\sum_{l > k \in S_1} \delta_{d_l,1} \quad \text {for} \quad \Delta b_k = -2 \\
\sum_{l > k \in S_1} \delta_{d_l,2} \quad \text {for} \quad \Delta b_k = +2 \\
\sum_{l > k \in S_1} \delta_{d_l,1} + \delta_{d_l,2}  = s_k(-2) + s_k(+2) \quad \text {for} \quad \Delta b_k = 0. \\
\end{cases}
\end{equation}
The remaining switches in $S_2$ are calculated up until the index of the 
start of $S_1$, as, similar to single end segments, e.g. $\oR$, there are the restrictions for nonzero matrix elements for semi-start segments, e.g. $R\uL$, to guarantee the total spin is conserved. 
Similarly, for the end of the overlap range, depending on the step-value at e.g. $L\oL$, the mentioned restrictions apply so the remaining switches (\ref{eq:remaining-switch-double}) are calculated 
until the start of $S'_2$. 


To relate $p(m'|m)$ to the generator matrix element $\braopket{m'}{\hat e_{ij,kl}}{m}$ we again use branching weights to determine which paths 
of the tree are chosen. 
For a full-start $\ul{RL}(i)$ into full-stop $\ol{RL}(j)$ excitation the weights of the different $\Delta b$ branches in terms of the intermediate $b$-values and remaining switch possibilities are
\begin{align}\label{eq:fullstart-fullstop-weights}
\Sigma_{-2}^{(k)} &= f(d_j) + \frac{s_k(-2)}{b_k} + \bigO{b_k^{-2}}, \\
\Sigma_{+2}^{(k)} &= g(d_j) + \frac{s_k(+2)}{b_k} + \bigO{b_k^{-2}},\\
\Sigma_0 &= 1 + \frac{1}{b_k}\left(s_k(-2)g(d_j) + s_k(+2)f(d_j) \right) + \bigO{b_k^{-2}}, 
\end{align}
with $f(d_j)$ and  $g(d_j)$ given by Eq.~(\ref{eq:end-cond}).
We bias towards the $\Delta b_k = 0$ branch at the start of the excitation range with
\begin{equation}\label{eq:RL-RL-start-weight}
p_0 = \frac{\Sigma_0}{\Sigma_0 + \Sigma_{\pm 2}},
\end{equation}
depending if $d_i = \{1,2\}$ and weight to stay on the current $\Delta b_k$ excitation branch in $S_1$ with
\begin{equation}\label{eq:RL-RL-stay-weight}
p_{\Delta b} = \frac{b_k \Sigma_{\Delta b}}{b_k \Sigma_{\Delta b} + \Sigma_{\Delta b}}.
\end{equation}
For a full-start into semi-stop excitation, e.g.\ $\ul{RL} \ra \ol L R \ra \ol R$, the weights of the branches in the overlap region $S_1$ are given by 
{\small
\begin{align}\label{eq:fullstart-semistop-weight}
\Sigma_{-2}^{(k)} &= f(d_j) \zeta_{-1}(j) + \frac{s_k(-2)}{b_k}\left[g(d_j) \zeta_{+1}(j) + f(d_j)\zeta_{+1}(j)\right] + \bigO{b_k^{-2}}, \\
\Sigma_{+2}^{(k)} &= g(d_j) \zeta_{+1}(j) + \frac{s_k(+2)}{b_k}\left[g(d_j)\zeta_{-1}(j) + f(d_j)\zeta_{+1}(j) \right] + \bigO{b_k^{-2}}, \\
\Sigma_0^{(k)} &= f(d_j)\zeta_{+1}(j) + g(d_j)\zeta_{-1}(j) + \frac{1}{b_k}\left[s_k(-2)g(d_j)\zeta_{+1}(j) + s_k(+2) f(d_j)\zeta_{-1}(j) \right] + \bigO{b_k^{-2}},
\end{align}
}
where $\zeta_{\pm1}(j)$ are the single weights (\ref{eq:singleWeights}) for the non-overlap region $S'_2$ at the end of the excitation, evaluated with the $b_j$ and $s_j(\pm 2)$ values at the semi-stop. 
The biasing function towards a certain branch at the beginning of an excitation and to stay at a chosen $\Delta b$ branch are the same as (\ref{eq:RL-RL-start-weight}) and (\ref{eq:RL-RL-stay-weight}) and in the non-overlap region $S'_2$ the single excitation weights and biasing factors (\ref{eq:singleWeights}, \ref{eq:stay}) apply.

The weighting functions in the non-overlap region $S_2$ for a semi-start into full-stop excitation, e.g.\ $\ul R(i) \ra \ul LR(j) \ra \oR\oL(k)$, are given by 
\begin{align}\label{eq:semistart-fullstop-weights}
\sigma_{-1}^{(k)} =& f(d_j) \Sigma_0(j) + g(d_j)\Sigma_{-2}(j) + \frac{s_k(-2)}{b_k} \left[f(d_j)\Sigma_{+2}(j) + g(d_j)\Sigma_0(j) \right] + \bigO{b_k^{-2}}, \\ 
\sigma_{+1}^{(k)} =& g(d_j)\Sigma_0(j) + f(d_j) \Sigma_{+2}(j) + \frac{s_k(+2)}{b_k} \left[g(d_j)\Sigma_{+2}(j) + f(d_j)\Sigma_0(j) \right] + \bigO{b_k^{-2}}, 
\end{align}
with $\Sigma_x$ being the weights of the full-stop excitation (\ref{eq:fullstart-fullstop-weights}) evaluated with the $b_j$ and $s_j(\pm 2)$ values at the start of the overlap region $j$. The biasing function for the start and staying probabilities are the same as in the single excitation case (\ref{eq:stay}) evaluated with $\sigma_{\pm1}$ instead of $\zeta_{\pm1}$.

For a ``full'' double excitation, e.g.\ $\ul R(i) \ra \ul R R(j) \ra \ol R R(k) \ra \ol R(l)$, the weights and biasing functions for the first non-overlap $S_2$ region $i\ra j-1$ are the same as for the semi-start into full-stop excitation $(\ref{eq:semistart-fullstop-weights})$, but evaluated with the full-start into semi-stop weights $\Sigma_x$(\ref{eq:fullstart-semistop-weight}). In the overlap region $S_1$, $j\ra k-1$, the weights and biasing functions are the same as for full-start into semi-stop excitations (\ref{eq:fullstart-semistop-weight}), where the $f(d_k), g(d_k)$ and $\zeta_{\pm1}$ functions are evaluated at the semi-stop index $k$ now. And finally for the final non-overlap region $S'_2$, $k\ra l-1$, the weights and biasing functions for single excitations (\ref{eq:singleWeights}, \ref{eq:stay}) apply. 

By using this biasing we ensure to create a valid \emph{spin conserving} excitation, avoid ending up in a dead-end of the branching tree and create excitations with a probability $p(m'|m)$ proportional to the coupling coefficient magnitude $\abs{\braopket{m'}{\hat e_{ij,kl}}{m}}$. 
The used weight functions are set up before an excitation in terms of the $b_k$ and remaining switch possibilities with, if necessary, the precomputed switch possibilities for the remaining overlap and non-overlap contributions and $\Delta b$ conditions. It is not necessary to recompute the whole setup at each step of the excitation. 
The computational effort to set up this weight objects, as it needs the information of the remaining switches, is $\bigO n$, in the worst case of an excitation spanning the whole orbital range. 
An analysis of the increase in computational effort of the GUGA-FCIQMC method compared to the SD based implementation can be found below.

\begin{figure*}
\centering
\includegraphics[width=\textwidth]{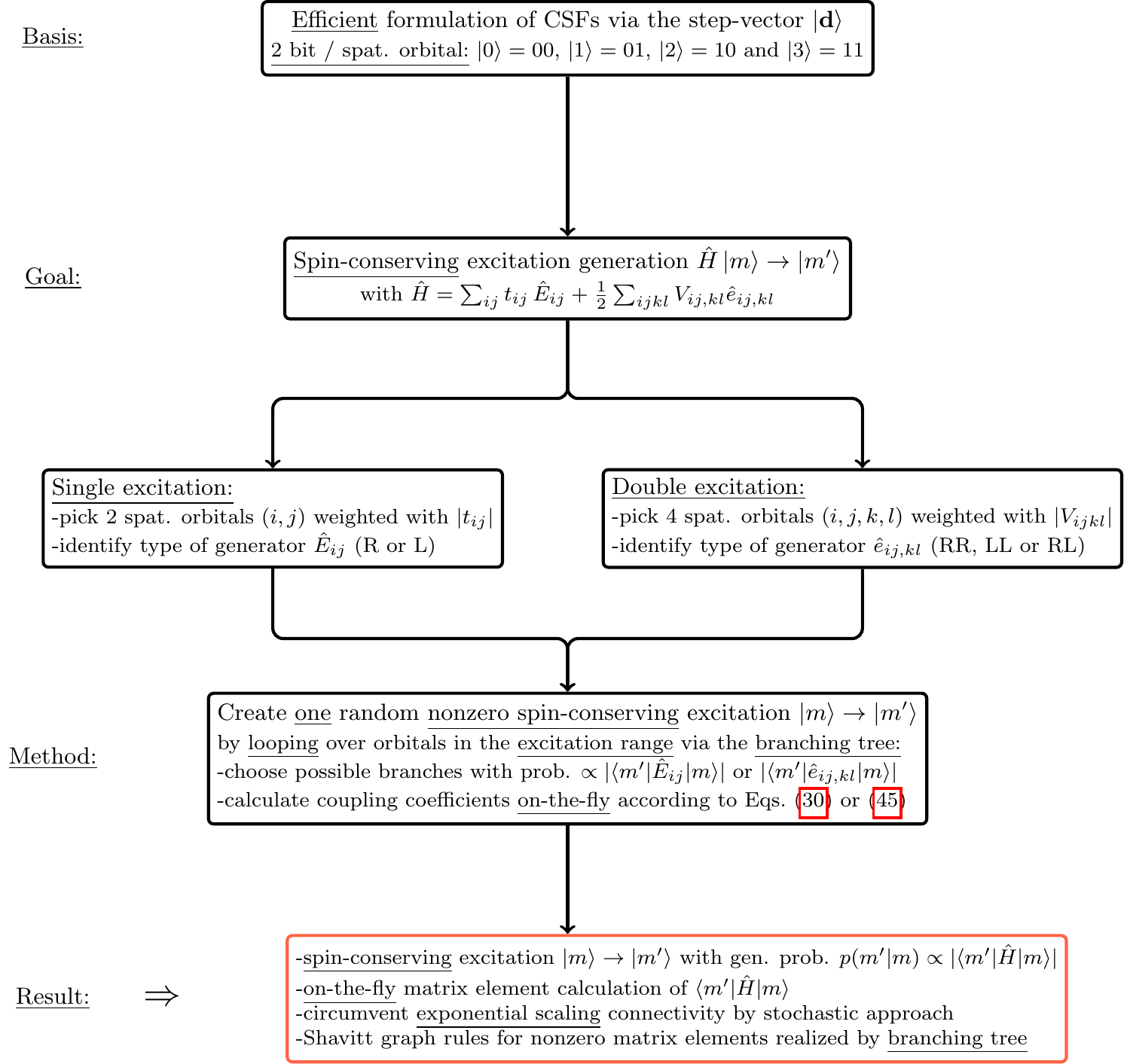}
\caption{\label{fig:guga-flow}(Color online) Flow chart of the GUGA-FCIQMC implementation.}
\end{figure*}

\section{\label{sec:histogram}Histogram based Timestep Optimization}

Due to the increased connectivity of CSFs compared to SDs, the generation probability, $p(m'|m)$, to spawn a new walker on state $\ket{m'}$ from an occupied CSF $\ket{m}$, is in general much lower than between SDs. An efficient sampling of the off-diagonal Hamiltonian matrix elements and stable dynamics of a simulation, demand the quantity $\Delta \tau \abs{H_{m'm}}/p(m'|m)$ to be close to unity. In the original determinant-based FCIQMC algorithm this is ensured by a dynamically adapted timestep $\Delta \tau(t)$, taking on the value of the ``worst-case'' $p(m'|m)/\abs{H_{m'm}}$ ratio encountered during a simulation. 

\begin{figure}
\centering
\includegraphics{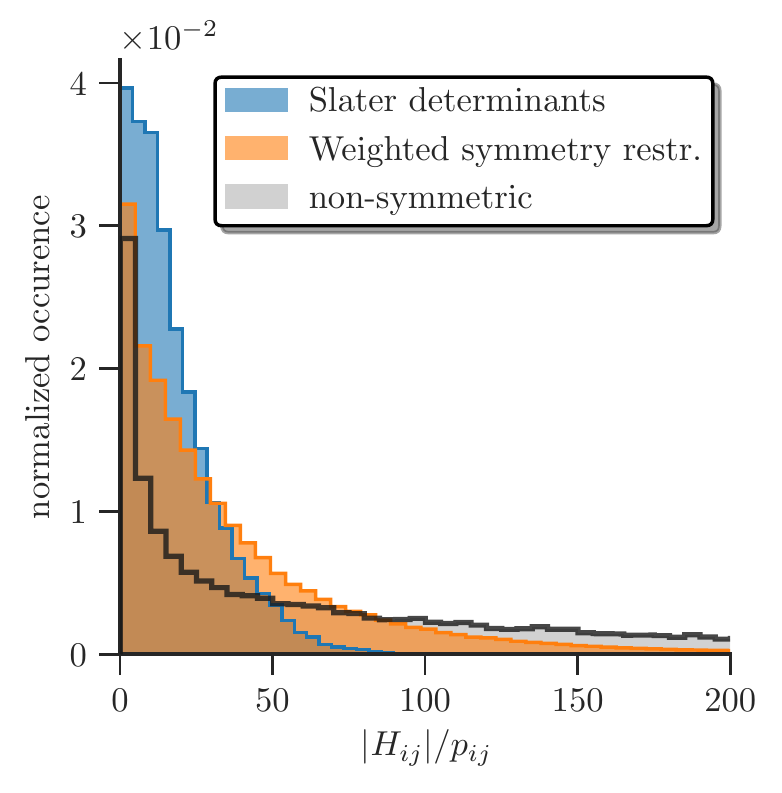}
\caption{\label{fig:histogram-guga}(Color online) Histogram of the ratio of matrix element magnitude and generation probability for the determinant- and CSF-based FCIQMC method for N$_2$ at the equilibrium geometry in a cc-pVDZ basis set. Both the optimized (orange) and unoptimized GUGA (black) results are shown.}
\end{figure}

However, due to the large number of possible connections between CSFs, this causes the timestep to drop dramatically.
At the same time a tiny spawning probability $p(m'|m)$ means that these problematic excitation only happen a minuscule fraction of times compared to more ``well-behaved'' excitations. Through the timestep, the global dynamics of all the walkers are affected by possibly only one ill-sampled excitation with a large $\abs{H_{m'm}}/p(m'|m)$ ratio. 
The optimized excitation generation mentioned in the sections above, ameliorates this issue, but
still cannot avoid the inherent ``connectivity problem'' of a CSF based implementation. 
If we store all $\abs{H_{m'm}}/p(m'|m)$ of all successful excitation attempts in a histogram of certain bin width, we can see that the majority of excitations are well represented by the optimized generation probability, see Fig.~\ref{fig:histogram-guga}.

The SD based method has a fast exponential decaying tail. This is the reason the ``worst-case'' timestep adaptation does not cause any problems for the original FCIQMC implementation. 
The GUGA implementation on the other hand, especially in the unoptimized version (uniform choice of branching possibilities and now weighting according to the molecular integrals), has a very slow decay and much larger maximum $\abs{H_{m'm}}/p(m'|m)$ ratios, over 10000 in the N$_2$ example shown in Fig.~\ref{fig:histogram-guga} (not displayed for clarity). 
The optimized CSF excitation scheme, explained above, greatly improves the $p(m'|m)$ to  $\abs{H_{m'm}}$ relation, but expectedly behaves worse than the SD based method. 
The timestep obtained with the ``worst-case'' optimization are given in Table~\ref{tab:guga-timestep}. 

To avoid this hampering of the global dynamics by a few ill-behaved excitations, we implemented a new automated timestep adaptation by storing the $\abs{H_{m'm}}/p(m'|m)$ ratios off all successful excitation attempts in a histogram, and setting the timestep $\Delta \tau$ to ensure $\Delta \tau \abs{H_{m'm}}/p(m'|m) \leq 1$ for a certain percentage of all excitations. 
The results of this ``histogram-tau-search'' are listed in Table~\ref{tab:guga-timestep} for a SD based and unoptimized (vanilla) and optimized GUGA-FCIQMC implementation for simulations of the nitrogen dimer at equilibrium geometry in a cc-pVDZ basis. 
For an SD-based implementation there is not much difference between the two approaches. Similar, for the vanilla GUGA implementation, due to the slow decaying tail in the histograms, see Fig.~\ref{fig:histogram-guga}. There is a two order of magnitude difference between the SD-based and the unoptimized GUGA-based timestep, which in practice would make the GUGA-FCIQMC implementation useless. 
However, with the optimized CSF excitation generation, the histogram-based $\Delta \tau_h$-adaptation yields a timestep two orders of magnitude larger than the ``worst-case'' $\Delta \tau_w$-optimization. 
The obtained $\Delta \tau_h$ is still half that of the SD-based FCIQMC, but due to a smaller Hilbert space size, and possibly faster convergence for spin-degenerate systems, this makes the GUGA-FCIQMC applicable for real systems. 

\begin{table}
\centering
\caption{\label{tab:guga-timestep}Automatically obtained timesteps for an SD- and GUGA-based (optimized and vanilla) simulation of N$_2$ at equilibrium distance in a cc-pVDZ basis. Results for the ``worst-case'' optimization $\Delta \tau_w$ and for the integrated histogram based optimization $\Delta \tau_h$ covering 99.99\% of all excitations.}
\resizebox{0.48\textwidth}{!}{
\renewcommand{\arraystretch}{1.0}
\begin{tabular}{cccc}
\toprule
& $\Delta \tau_w$ & $\Delta \tau_h$ & $\Delta \tau_h / \Delta \tau_w $ \\
\hline
SD & $5.59\cdot 10^{-3}$  & $6.20\cdot 10^{-3}$   & 1.11  \\
GUGA van. & $4.78\cdot 10^{-5}$ & $8.62\cdot 10^{-5}$ & 1.80 \\
GUGA opt. & $5.20\cdot 10^{-5}$ & $1.12 \cdot 10^{-3}$ & 21.50 \\
\hline 
GUGA van. / opt. & 0.92 & 0.08  & \\
SD / GUGA opt. & 107.51 & 5.55 & \\
\botrule
\end{tabular}
}
\end{table}

\section{Results and Discussion\label{sec:results}}

\subsection{\label{sec:n-atom-results}Nitrogen Atom}

To benchmark the GUGA-FCIQMC implementation we first investigated the nitrogen atom. 
The ground state configuration of N is 1s$^2$2s$^2$2p$^3$ with the 3 electrons in the p-shell forming a $S = 3/2$ quartet $^4S^o$ state. 
The first excited state is the $S = 1/2$ $^2D^o$ doublet, $2.384$ eV above the ground state \cite{nist-db,n_spin_gap_exp}, with  spin-orbit effects neglected.
This setup of a half-integer high-spin ground state with low-spin excited state is the prime playground of the GUGA-FCIQMC method. 
Previous spin-adapted implementations in FCIQMC, using half-projected and projected Hartree-Fock (HPHF) states \cite{hphf}, are only applicable to an even number of electrons.
At the same time, restricting the total $m_s$ quantum number to target an excited state only works if the low-spin state is the ground state with excited states being high-spin, since the high-spin ground state also contains contributions of energetically lower $m_s$ states, causing the projective FCIQMC to converge to the latter one. 

\begin{figure}
\centering
\includegraphics[width=\jctcsingle]{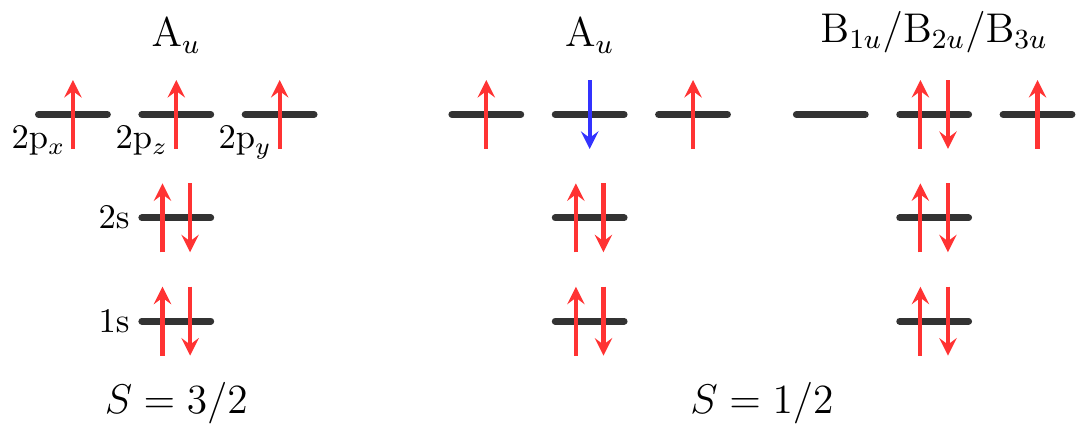}
\caption{\label{fig:nitrogen-orbitals}Schematic orbital diagram of nitrogen 1s$^2$2s$^2$2p$^3$ for the doublet ($S = 1/2$) and quartet ($S = 3/2$) states. A$_u$ and B$_{iu}$, $i = 1,2,3$, represent the irrep of D$_{2h}$ from the reduction of SO(3) symmetry.}
\end{figure}

We prepared all-electron ab-initio Hamiltonians with \texttt{MOLPRO} \cite{molpro-general-1, molpro-general-2} for N in a cc-pV$n$Z basis set, with $n$ = D, T, Q, 5 and 6. 
The maximal symmetry point group in \texttt{MOLPRO} is D$_{2h}$ and thus the 
much larger SO(3) symmetry of N gets reduced to the one-dimensional irreps of D$_{2h}$. 
The $S = 3/2$ quartet with singly occupied 2p orbitals belongs to the 
irrep A$_u$. While the $S = 1/2$ doublet splits into one A$_u$ state with three open-shell 2p orbitals and three states belonging to B$_{1u}$, B$_{2u}$ and B$_{3u}$ with one doubly occupied and one open-shell 2p orbital, see Fig.~\ref{fig:nitrogen-orbitals}.

We calculated the quartet-doublet, $^4S^o - {^2D^o}$, spin-gap with the spin-adapted i-FCIQMC method (GUGA-FCIQMC) for basis sets up to cc-pV6Z and compared our results to unrestricted coupled cluster singles and doubles with perturbative triples ((U)CCSD(T)) and FCI calculations up to cc-pVTZ obtained with \texttt{MOLPRO} \cite{molpro-fci-1, molpro-fci-2, molpro-uccsdt-1,molpro-uccsdt-2} and experimental results \cite{nist-db,n_spin_gap_exp}. 
The CCSD(T) calculations are based on restricted open-shell Hartree-Fock \cite{rohf} (ROHF) orbitals, which for the $S = 1/2$ state are only possible to be done for the B$_{iu}$, $i = 1, 2, 3$, states. Although GUGA-FCIQMC calculations for the B$_{iu}$ irreps yield the same results as for the $S = 1/2$ A$_u$ state, the CCSD(T) results are far off the FCI results and the experimental gap, due to the multi-reference character of these states.

The results are given in Table~\ref{tab:n-spin-gap-IP} with a complete basis set (CBS) extrapolation given by a two-parameter inverse cube fit \cite{Helgaker1997}
\begin{equation}
\label{eq:inverse-cube}
E(n) = E_{CBS} + \frac{A}{n^3}
\end{equation}
using $n$ = T, Q and 5. The GUGA-FCIQMC CBS results shows excellent agreement with the experimental value within chemical accuracy, 
while the CCSD(T) calculations are not able to obtain the correct result, due to the multiconfigurational character of the $^2D^o$ excited state. 

We also calculated the ionization potential (IP) of the nitrogen atom in the CBS limit with GUGA-FCIQMC and compared our results to CCSD(T) calculations and experimental data. 
The ground state of the N$^+$ cation is the $S = 1$ triplet $^3P_0$ state. 
The results from GUGA-FCIQMC and CCSD(T) calculations up to cc-pV6Z basis set are shown in Table~\ref{tab:n-spin-gap-IP}. Since CCSD(T) \cite{molpro-fci-1, molpro-fci-2, molpro-uccsdt-1, molpro-uccsdt-2} can treat both the $^4S^o$ and $^3P_0$ well, coupled cluster results and GUGA-FCIQMC CBS limit values, using $n$ = Q, 5 and 6, agree within \emph{chemical accuracy} with experimental values. 

\begin{table*}
\centering
\caption{\label{tab:n-spin-gap-IP}Spin gap $^2D^o - {^4S^o}$ and ionization potential (IP) $^3P_0 - {^4S^o}$ of the nitrogen atom obtained with GUGA-FCIQMC and CCSD(T) \cite{molpro-uccsdt-1, molpro-uccsdt-2} for different basis set sizes cc-pV$n$Z, $n = 2, 3, 4, 5, 6$ (2 = D, 3 = T, 4 = Q) and CBS limit extrapolations, with Eq.~(\ref{eq:inverse-cube}) using the $n= 3, 4, 5$ and $6$ results, compared with experimental results \cite{nist-db,n_spin_gap_exp,nitrogen-ip}. Energies are given in atomic units.} 
\small
\renewcommand{\arraystretch}{1.2}
\small
\begin{tabular}{cccccc}
\toprule
& \multicolumn{2}{c}{$^2D^o - {^4S^o}$ spin gap} & & \multicolumn{2}{c}{N$^+$ {$^3P_0$} - N ${^4S^o}$ IP} \\
\cline{2-3}\cline{5-6}
$n$  & CCSD(T) & GUGA-FCIQMC & &  CCSD(T) & GUGA-FCIQMC \\
\hline
2 &  0.1061699 & 0.099951(11)\phantom{0}& & 0.5216483 & 0.5215711(23) \\
3 &  0.1013301 & 0.0923719(90)& & 0.5310023 & 0.5310210(83) \\ 
4 &  0.0994312 & 0.0896661(73)&  & 0.5333069 & 0.5333855(25) \\
5 &  0.0986498 & 0.0885878(67)&  & 0.5341573 & 0.534204(12)\phantom{0} \\
6 &  0.0983705 & 0.0881762(69)&  & 0.5344735 & 0.5345070(62) \\
\hline
CBS& 		 \phantom{()0}0.097950(35) & 0.0875830(80)& &\phantom{(1)}0.534987(43) & 0.534971(13)\phantom{0} \\
Experiment & \multicolumn{2}{c}{0.08746(37)\phantom{00}} && \multicolumn{2}{c}{0.5341192(15)}\\
$\Delta E$ & \phantom{()}0.01034(40) & 0.00003(38)\phantom{00} && \phantom{(1)}0.000868(46) & 0.000852(16)\phantom{0} \\
\botrule
\end{tabular}

\end{table*}

\subsection{\label{ssec:nitrogen-dimer}Nitrogen Dimer}

The breaking of the strong triple bond of N$_2$ is accompanied by a change of single-reference to multiconfigurational character of the electronic structure and the concomitant strong electron correlation effects pose a difficult problem for quantum chemical methods. 
The ground state of the nitrogen molecule at equilibrium bond distance, $r_0 \approx 2.1 a_0$, is a singlet $^1\Sigma_g^+$, where all bonding molecular orbitals (MOs) formed from the 2p atomic orbitals (AOs) of the constituent N atoms are doubly occupied

At large bond distances the ground states of the $S = 0, 1, 2$ and $3$ states are degenerate, since the coupling of the independent nitrogen atoms $A$ and $B$, $^4S^o_A \otimes ^4S^o_B$, are all degenerate. 
We calculated the dissociation energy of N$_2$ as the difference of the $^1\Sigma_g^+$ N$_2$ ground state at equilibrium geometry $r_0 = 2.074 a_0$ and the $^7\Sigma_u^+$ state at $r = 30 a_0$ in the cc-pV$n$Z basis set, up to $n = 5$, with four core electrons frozen and performed a CBS limit extrapolation using Eq.~(\ref{eq:inverse-cube}) with the $n$ = T, Q and 5 results. 
The results are shown in Table~\ref{tab:n2-dissociation} with CCSD(T) results obtained with \texttt{MOLPRO} \cite{molpro-general-1, molpro-general-2,molpro-uccsdt-1,molpro-uccsdt-2} and compared with experimental results \cite{n2-diss-exp-1}, which are corrected to remove scalar relativistic, spin-orbit and core correlation effects according to Refs.~\citen{Bytautas2005,Feller2000}.
We also checked the convergence of the $r = 30 a_0$ results with the independent atom calculations with a frozen core and found excellent agreement. Both the GUGA-FCIQMC and CCSD(T) results agree with experimental values within chemical accuracy. 

To investigate the correct accounting of core and core-valence correlation effects, we performed all-electron GUGA-FCIQMC and CCSD(T) calculations in the cc-pCV$n$Z basis set and calculated the dissociation energy of N$_2$ as the difference of the independent nitrogen atom $^4$S$^o$ ground state results in the same basis set and the  $^1\Sigma_g^+$ N$_2$ ground state at equilibrium geometry $r_0 = 2.074 a_0$, as $E_{diss} = 2 E_{atom} - E_{dimer}$. 
The results are shown in Table~\ref{tab:n2-dissociation} and the CBS limit extrapolations of the N$_2$ dissociation energy agree within chemical accuracy with experiment\cite{n2-diss-exp-1} for both the GUGA-FCIQMC and CCSD(T) calculations. 
We also performed a counterpoise correction \cite{Boys1970}, but found the basis set superposition error to be negligibly small. 

\begingroup
\squeezetable
\begin{table*}
\centering
\caption{\label{tab:n2-dissociation}N$_2$ dissociation energy obtained with GUGA-FCIQMC in the frozen-core approximation in a cc-pV$n$Z basis set and all-electron calculations in a cc-pCV$n$Z basis set for increasing cardinal number $n$ compared to CCSD(T) \protect\cite{molpro-general-1, molpro-general-2,molpro-uccsdt-1,molpro-uccsdt-2} and experimental results \protect\cite{huberherzberg1979}. The frozen-core experimental results are corrected for scalar relativistic, spin-orbit and frozen-core effects according to \protect\citen{Bytautas2005,Feller2000}. The CBS limit is obtained with Eq.~(\ref{eq:inverse-cube}) for the $n$ = 3, 4 and 5 data points where available and with a Helgaker two-point extrapolation \protect\cite{Helgaker1997} for the T and Q GUGA-FCIQMC all-electron results. All energies are given in $E_h$.}
\begin{threeparttable}
\renewcommand{\arraystretch}{1.2}
\begin{tabular}{cccccc}
\hline
 & \multicolumn{2}{c}{Frozen-core cc-pV$n$Z} & &  \multicolumn{2}{c}{All-electron cc-pCV$n$Z} \\
 \cline{2-3} \cline{5-6} 
$n$ & CCSD(T) & GUGA-FCIQMC & & CCSD(T) & GUGA-FCIQMC \\
 \hline
2 & 0.3184752\phantom{()} &   0.3198257(42) & & 			0.3203232\phantom{()} 		& 0.3215439(80)  \\
3 & 0.3448342\phantom{()} &   0.345412(26)\phantom{0} & & 0.3472842\phantom{()} 		& 0.347445(35)\phantom{0} \\
4 & 0.3551440\phantom{()} &   0.355565(55)\phantom{0} & & 0.3566507\phantom{()} 		& 0.356759(46)\phantom{0} \\
5 & 0.3587423\phantom{()} & 0.358984(49)\phantom{0} & & 0.3601695\phantom{()} 		&  \\
\hline
CBS & \phantom{0}0.362603(47)\tnote{a} & 0.362797(53)\tnote{a}\phantom{0} & & 0.3634857\tnote{a}\phantom{()} 	& 0.363555(84)\tnote{b}\phantom{0} \\
Experiment & \multicolumn{2}{c}{0.362700(10)\tnote{c}} & & \multicolumn{2}{c}{0.364002(10)\tnote{d}} \\
$\Delta E$ & \phantom{0}0.000097(57) & -0.000097(63)\phantom{0-} & & \phantom{0}0.000516(10) 	& 0.000447(94)\phantom{0} \\
\hline
\end{tabular}
\begin{tablenotes}
\footnotesize
\item [a] Using the $n = 3, 4$ and $5$ data points with Eq.~\ref{eq:inverse-cube}
\item [b] Using the Helgaker two-point extrapolation \protect\cite{Helgaker1997} based on Eq.~\ref{eq:inverse-cube}
\item [c] Valence-only dissociation energy from \protect\citen{Bytautas2005}
\item [d] From Huber and Herzberg \protect\citen{huberherzberg1979}
\end{tablenotes}
\end{threeparttable}
\end{table*}
\endgroup

\begin{figure*}
\centering
\includegraphics{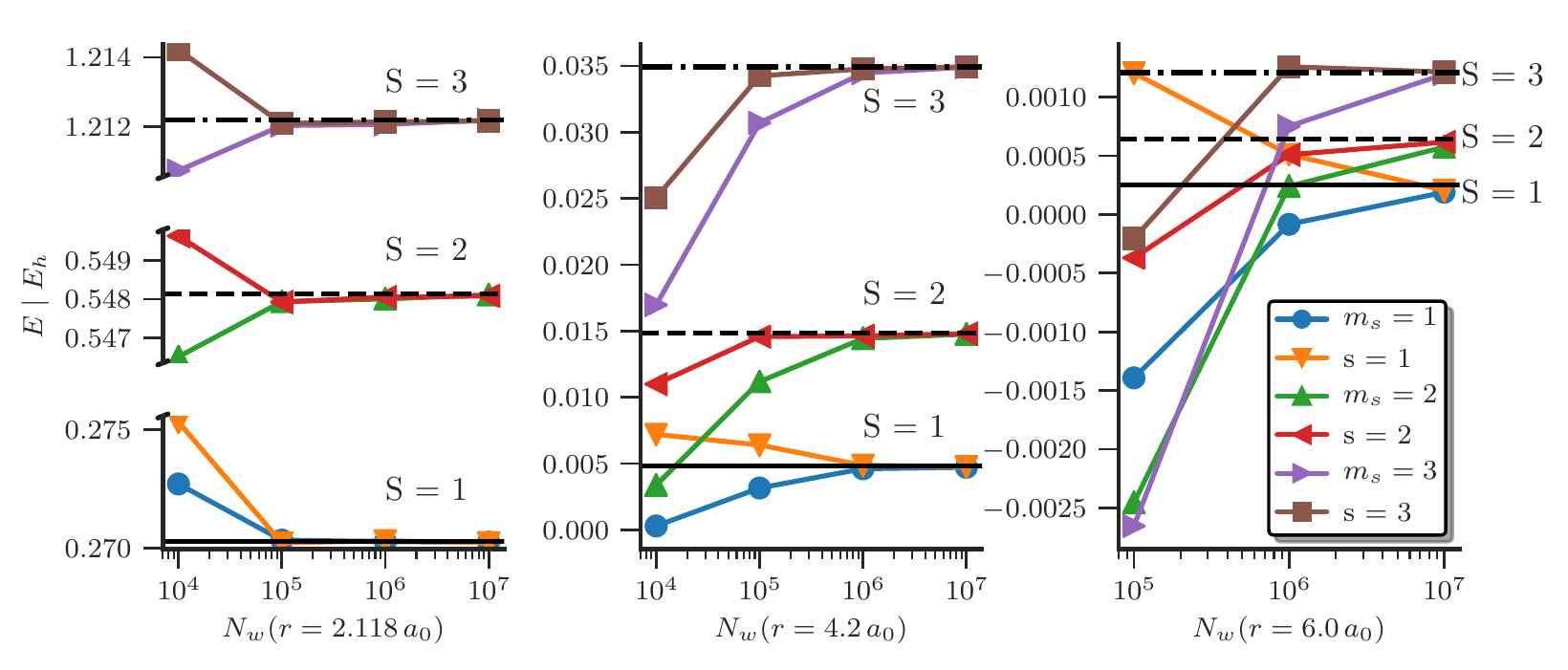}
\caption{\label{fig:spin_gaps}The spin gaps between the singlet $^1\Sigma_g^+$ ground state of N$_2$ to the triplet $^3\Sigma_u^+$, quintet $^5\Sigma_g^+$ and septet $^7\Sigma_u^+$ state obtained with the determinant based (indicated with $m_s = x$) and spin-adapted (indicated with $s = x$) FCIQMC method as a function of total walker number $N_w$ compared with DMRG \cite{block-dmrg-1,block-dmrg-2} reference results at bond distance $r = 2.118, 4.2$ and $6.0\, a_0$ in a cc-pVDZ basis set.}
\end{figure*}

To show the improved convergence behavior of the spin-adapted FCIQMC method for systems with near-degenerate spin-eigenstates, we calculated the gap of the singlet $^1\Sigma_g^+$ ground-state to the triplet $^3\Sigma_u^+$, quintet $^5\Sigma_g^+$ and septet $^7\Sigma_u^+$ excited states of N$_2$\textemdash which are all degenerate at dissociation\textemdash for the equilibrium bond distance $r = 2.118\, a_0$\footnote{In the cc-pVDZ basis set.} and two stretched geometries $r = 4.2\, a_0$ and $r = 6.0\, a_0$ in a cc-pVDZ basis set with the original SD-based and GUGA-FCIQMC method. 
Figure~\ref{fig:spin_gaps} shows the gaps between the ground state and three excited states as a function of the total walker number compared with DMRG reference results \cite{block-dmrg-1,block-dmrg-2}. 
Since the energy of the spin states are ordered according to their total spin quantum number, it is possible to obtain the spin-gaps in the original determinant based FCIQMC method by restricting the $m_s$ quantum number alone. 
At cc-pVDZ equilibrium bond distance $r = 2.118\, a_0$ the determinant based and spin-adapted FCIQMC implementation are equivalent in their convergence behavior of the spin-gaps w.r.t. the walker number. 
However, as the bond distance increases, and thus the spin-gaps decrease, the spin-adapted FCIQMC implementation shows a dramatically improved convergence, especially for the singlet-quintet and singlet-septet gaps, where around an order of magnitude fewer walkers are necessary to obtain the same accuracy as the SD-based FCIQMC method.

\begin{figure}
\centering
\includegraphics{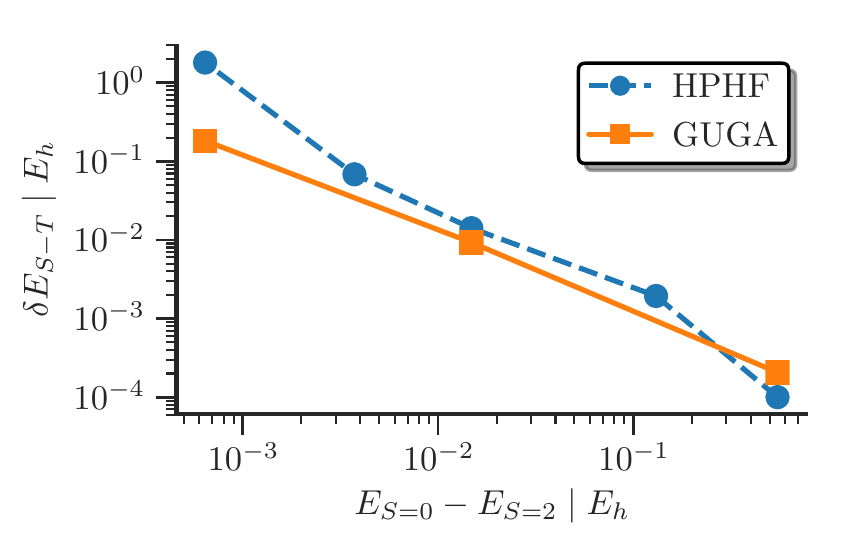}
\caption{\label{fig:hphf_gap}Relative error of the singlet-triplet gap of N$_2$ in a cc-pVDZ basis versus the singlet-quintet gap obtained with the HPHF and GUGA FCIQMC implementation with $N_w^{tot} = 10^7$ compared to DMRG reference results \protect\cite{block-dmrg-1,block-dmrg-2}.}
\end{figure}

The second common option to obtain spin-gaps with the FCIQMC method is based on HPHF functions, which allow targeting spin-states with an even (singlet, quintet, ...) or odd (triplet, septet, ...) total spin $S$, allowing to obtain the singlet-triplet gap in the case of N$_2$. 
Figure~\ref{fig:hphf_gap} shows the relative error of the singlet triplet gap, obtained with the HPHF based and spin-adapted FCIQMC implementation with $N_w^{tot} = 10^7$ as a function of the singlet-quintet gap, compared to DMRG reference results \cite{block-dmrg-1,block-dmrg-2} on a double logarithmic scale. 
As both even-spin singlet and quintet states belong to the same spatial point group irrep A$_g$, the HPHF solution is spin-contaminated by an increasing amount for a decreasing singlet-quintet gap. 
This fact prohibits the HPHF-based FCIQMC implementation to obtain the correct singlet-triplet gaps for increasing bond distance for the nitrogen dimer.

\subsection{\label{sec:cost}Computational Effort and Scaling of GUGA-FCIQMC}

To analyze the additional computational cost associated with the GUGA-based CSF implementation in FCIQMC, we compare the time per iteration, $t_{iter}$, and timestep, $\Delta \tau$, with the original SD-based FCIQMC method for the nitrogen atom and dimer, mentioned above. 
Since FCIQMC is formally linear-scaling with the walker number $N_w$\cite{linear-scaling-fciqmc} we removed the bias of walker number differences by comparing the time per iteration and per walker. 

\begin{figure*}
\centering
\includegraphics{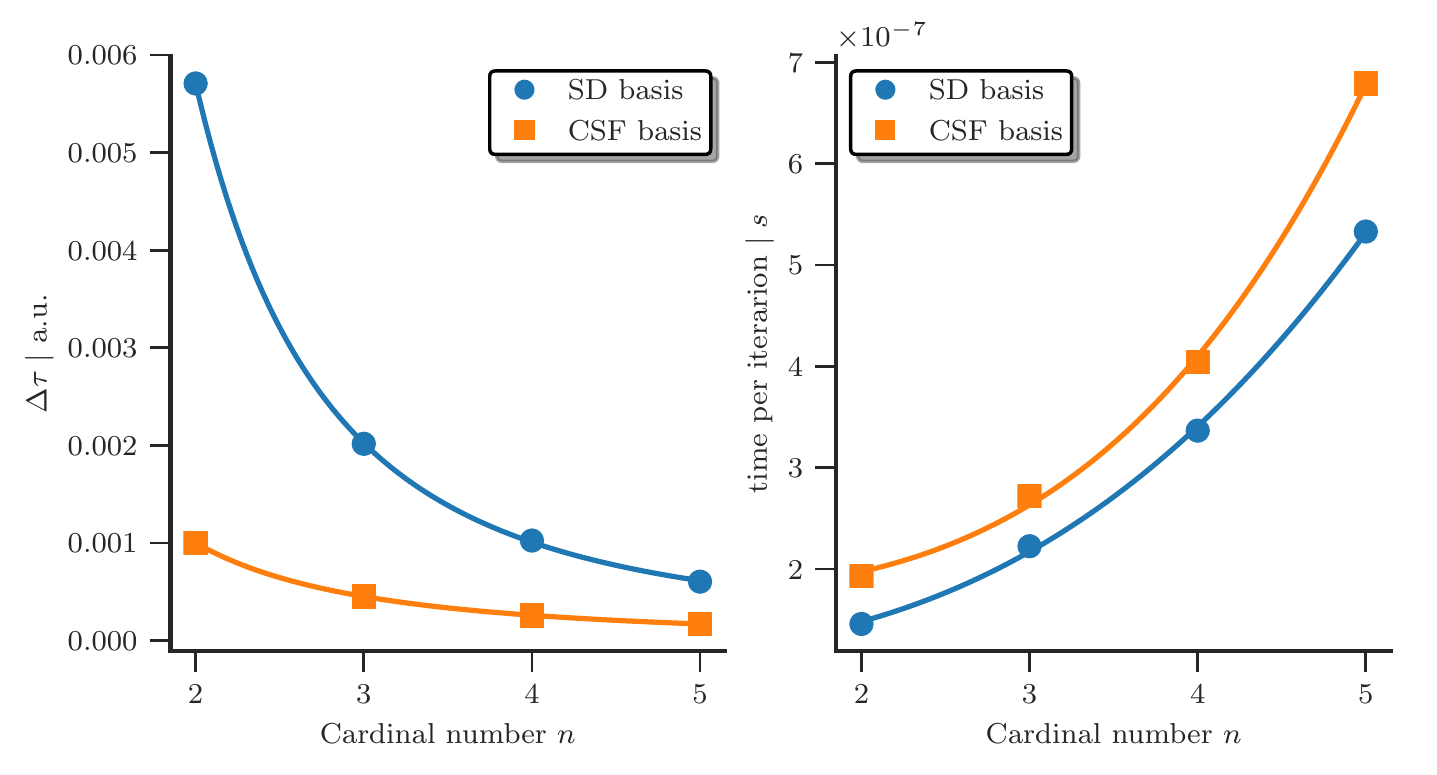}
\caption{\label{fig:guga-time-and-cost}SD- and CSF-based results for N$_2$ at $r = 4.2 a_0$ for cc-pV$n$Z basis sets, $n$ = D, T, Q, 5. (left) Time-step $\Delta \tau$ adapted with the histogram-based optimization with an integration threshold of 0.9999 and (right) time per iteration vs. the cardinal number of the basis set. The results were obtained on identical 20 core Intel Xeon E5-2680 nodes with 2.8GHz clock rate and with $N_w^{tot} = 100k$. The lines are fits to the data explained in the main text.}
\end{figure*}

The left panel of Fig.~\ref{fig:guga-time-and-cost} shows the timestep $\Delta \tau$ obtained with the histogram based optimization, see Sec.~\ref{sec:histogram}, for N$_2$ at $r = 4.2\,a_0$ vs. the cardinal number $n$ of the cc-pV$n$Z basis set. 
As expected the usable timestep in the SD-based simulation is higher compared to the CSF-based calculation, with roughly twice the possible $\Delta \tau$. However, rather surprisingly the difference between the two decreases with increasing basis set size. 
The right panel of Fig.~\ref{fig:guga-time-and-cost} shows the time per iteration and walker for the same simulations. 
The additional computational cost of the GUGA implementation roughly doubles the time per iteration compared to the original FCIQMC method. 
While there seems to be a steeper increase with increasing basis set size for the CSF-based implementation, it is nowhere near the formally $\bigO{n}$ cost, with $n$ being the number of orbitals, mentioned above. 
In total, with twice the timestep and twice the time per iteration, the spin-pure GUGA implementation amounts to a fourfold increase in computational cost compared to the original SD-based FCIQMC method. \footnote{All simulations for this comparison were performed on identical 20 core Intel Xeon E5-2680 nodes with 2.8GHz clock rate, 20MB cache and 128GB memory.} 

To examine the scaling in more detail, a least-squares fit to the polynomial $f(n) = a + b \cdot n^{c}$, with 3 parameters $a$, $b$ and $c$, was performed on the available data points with $n$ being the cardinal number of the basis set. 
The lines in Fig.~\ref{fig:guga-time-and-cost} represent this fit for
the timestep $\Delta \tau(n)$ and time per iteration $t_{iter}(n)$, as a function of the cardinal number $n$ of the basis set. The results of the least-squares fit for the determinant- and CSF-based calculations are shown in Table~\ref{tab:fit}.
\begin{table}
\caption{\label{tab:fit}Results of the least-squares polynomial fit $f(n) = a + b \cdot n^{c}$ of the timestep $\Delta \tau(n)$ and time per iteration $t_{iter}(n)$ as a function of the cardinal number $n$ of the basis set for the Slater determinant (SD) based and spin-adapted (CSF) FCIQMC implementation.}
{\small 
\begin{tabular}{ccccc}
\toprule
& & $a$ & $b$ & $c$ \\ 
\hline
\multirow{2}{*}{$\Delta \tau$}  & SD: & $1.25\cdot 10^{-4}$ & $3.53\cdot 10^{-2}$ & -2.66 \\ 
& CSF: & $1.05\cdot 10^{-5}$ & $3.95\cdot 10^{-3}$ & -2.00 \\ 
\hline
\multirow{2}{*}{$t_{iter}$} & SD: & $1.14\cdot 10^{-7}$ & $5.04\cdot 10^{-9}$ & \phantom{-}2.74 \\ 
& CSF: & $1.75\cdot 10^{-7}$ & $2.08\cdot 10^{-9}$ & \phantom{-}3.41 \\ 
\botrule
\end{tabular}}
\end{table}
The scaling of the decrease in the 
possible timestep $\Delta \tau$ is almost less than a factor of $n$ smaller in the CSF based implementation 
and the increase of the time per iteration $t_{iter}$ less than $n$ larger compared to the 
determinant based implementation. However, the combination of these two effects causes the spin-adapted FCIQMC implementation to scale by an additional factor of $\approx \bigO{n^{1.3}}$ for this specific system, compared to the original SD-based FCIQMC method. 

Table~\ref{tab:guga-averaged-comp} shows the averaged timestep and time per iteration ratios between GUGA- and SD-based simulations for the nitrogen atom and dimer.
Compared to the CSF-based 
FCIQMC the maximum possible timestep in the original determinant based implementation is larger by a factor of $2.55$ to $3.87$ and the time per iteration is smaller by a factor of $0.63$ to $0.90$. 
The combination of these effects result in a slow down by a factor of $2.8$ to $5.0$ of the spin-adapted FCIQMC method. 

\begin{table}
\centering
\caption{\label{tab:guga-averaged-comp}Averaged timestep $\Delta \tau$ and time per iteration $t$ ratios of CSF- and SD-based FCIQMC calculations for the nitrogen atom and dimer with sample sizes $n_{s}$. The standard errors $\delta t$ and $\delta \Delta \tau$ are also given.}
\resizebox{0.49\textwidth}{!}{
\renewcommand{\arraystretch}{1.1}
\small

\begin{tabular}{lccccc}
\toprule
System & $n_{s}$ & $\Delta \tau_{SD} / \Delta \tau_{CSF} $ & $\delta \Delta \tau$ & $t_{SD} / t_{CSF} $ & $\delta t$  \\
\hline
N & 10 & 2.55 & 0.12 & 0.90 & 0.08 \\
N$_2$ & 12 & 3.87 & 0.48 & 0.78 & 0.05 \\
\botrule
\end{tabular}
}
\end{table}

\subsection{\label{sec:transition-metal}The Cobalt Atom}
As with most open-shell transition metals, the cobalt atom has a high-spin ground state, due to Hund's first rule. 
This prohibits the calculation of the spin-gap to low-spin excited states by restriction of the $m_s$ quantum number, as inevitably these excited state calculations will converge to the high-spin ground state in the projective procedure of FCIQMC.

We compare our results to coupled cluster calculations, which are not so easily applicable, due to the multireference character of the excited states of these systems. 

The ground state electronic configuration of the neutral cobalt atom is [Ar]$3s^23p^63d^74s^2$ and is a quartet $^4F$ state. 
We calculated the spin gap to the first doublet excited state $^2F$ with the [Ar]$3s^2 3p^6 3d^8 4s$ configuration with the GUGA-FCIQMC method, correlating 17 electrons in all available orbitals. We employed an ANO basis set \cite{ano-basis} with primitive contractions corresponding to a comparable V$n$ZP basis with $n$ = D, T and Q and the \emph{full} and completely uncontracted \emph{primitive} ANO basis set with 2nd order Douglas-Kroll scalar relativistic corrections \cite{douglas-kroll}.
The ANO molecular integral files were computed with \texttt{MOLCAS} \cite{molcas-general}. 
We also prepared ab-initio integrals with an augmented correlation consistent core-valence basis set with 2nd order Douglas-Kroll scalar relativistic corrections \cite{douglas-kroll},
aug-cc-pwCV$n$Z-DK (denoted as cc-basis in Table~\ref{tab:co-gap-results}), up to $n$ = Q. The cc-basis molecular integrals were computed with \texttt{MOLPRO} \cite{molpro-general-1,molpro-general-2}. 
We also performed 2nd order complete active space perturbation theory \cite{caspt2-1,caspt2-2} (CASPT2) calculation on the ANO basis set with \texttt{MOLCAS} and CCSD(T) calculation in the cc-basis with \texttt{MOLPRO}. 

\begin{figure}
\centering
\includegraphics[width=\jctcsingle]{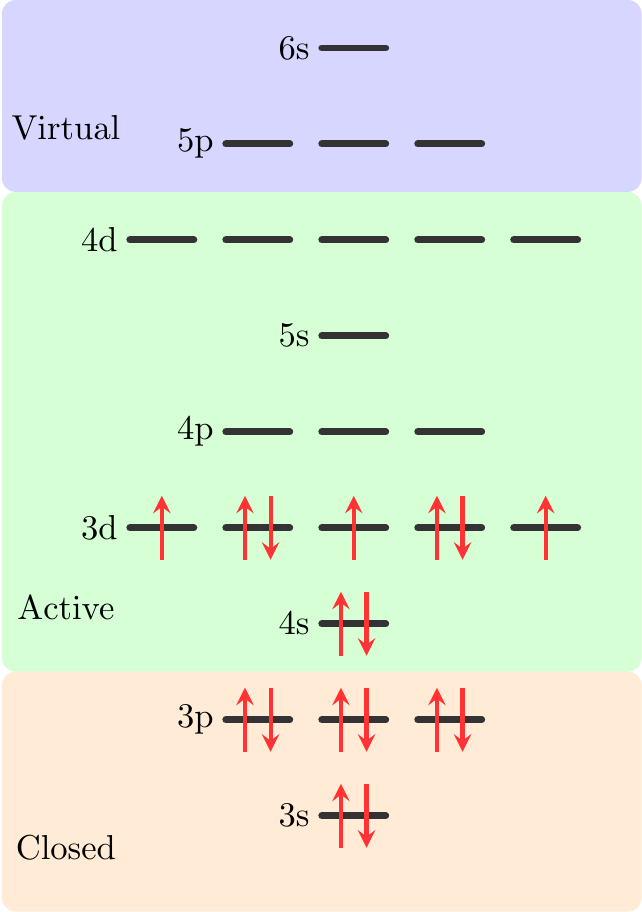}
\caption{\label{fig:sc-co-energy-diagram}Schematic orbital energy diagram of and ground state electron configuration of the $^4F$ state of cobalt. The chosen active spaces for the CASSCF calculation are shown in orange (closed), green (active) and blue (virtual).}
\end{figure}

The starting orbitals for the Co $^2F$ and $^4F$ calculation with FCIQMC were CASSCF \cite{olsen2011,Hegarty1979} orbitals with the $1s^2 2s^2 2p^6 3s^2 3p^6$ orbitals frozen, 9 active electrons in the active space of $4s, 3d, 4p, 5s$ and $4d$, CAS(9,15) and further orbitals being virtuals, see Fig.~\ref{fig:sc-co-energy-diagram}. 
The CASSCF calculations were performed with \texttt{MOLCAS} \cite{molcas-general} and \texttt{MOLPRO} \cite{molpro-casscf-1, molpro-casscf-2}. 
Similar to the nitrogen atom the SO(3) symmetry of Co is reduced to the D$_{2h}$ symmetry implemented in \texttt{MOLPRO} and \texttt{MOLCAS}. We chose the B$_{1g}$ irrep for the $^2F$- and the A$_g$ irrep for the $^4F$-state. 

Similar to the nitrogen atom the odd number of electrons and high-spin ground state to low-spin excited state setup makes previous spin-pure methods implemented in FCIQMC not applicable. 
However, as the results in Table~\ref{tab:co-gap-results} show, the GUGA-FCIQMC implementation is able to provide energies within chemical accuracy close to the experimental result \cite{nist-db, cobalt-spingap, cobalt-spingap-error}. 
For both GUGA-FCIQMC and CASPT2, the CBS limit extrapolation of the spin-gap, using the VTZP and VQZP results for Eq.~(\ref{eq:inverse-cube}), in the ANO basis set agree within 1 kcal/mol (chemical accuracy) with the experimental result.

For the aug-cc-pwCV$n$Z-DK we performed separate CBS limit extrapolations of the $^2$F and $^4$F ground state energy, using the Hartree-Fock energy of a $n = 5$ calculation and a two-point extrapolation of the correlation energy, according to Eq.~(\ref{eq:inverse-cube}), using the $n = $T and Q data points. 
The resulting estimated spin-gap lies approximately $0.0024\,E_h \approx 1.519$ kcal/mol above the experimental result, see Table~\ref{tab:co-gap-results}.

Similar to the spin gap of nitrogen, see Sec.~\ref{sec:n-atom-results}, coupled cluster is not able to provide correct results of the doublet $^2F$ state of cobalt. The CCSD(T) calculations are based on ROHF orbitals and the valence electronic configuration of the $^2F$ state, $3d^8 4s$, enforces the $4s$ orbital to be singly occupied with all the $d$-electrons being in a closed shell conformation. This obviously violates Hund's rule and thus the CCSD(T) results give a too high energy for the $^2F$ state and thus the spin-gap is immensely overestimated, see Table~\ref{tab:co-gap-results}. 
Further investigations
are being conducted on the performance of CCSD(T) by varying the
reference determinant, and will be reported elsewhere.

\begin{table*}
\centering
\caption{\label{tab:co-gap-results}GUGA-FCIQMC, CASPT2 \cite{molcas-general} and CCSD(T) \protect\cite{molpro-general-1,molpro-general-2} results for the $^2F - {^4F}$ spin gap of Co in an ANO \cite{molcas-general} and cc-basis set \protect\cite{molpro-general-1,molpro-general-2} compared with the experimental values \protect\cite{nist-db, cobalt-spingap, cobalt-spingap-error}. CBS limit extrapolations were obtained with Eq.~(\ref{eq:inverse-cube}) with the used data points of the basis sets in parentheses.
The CCSD(T)
results are obtained by running \texttt{MOLPRO} in default mode, without further
specification of the reference configuration. Further investigations
are being conducted on the performance of CCSD(T) by varying the
reference determinant, and will be reported elsewhere.
}
\begin{threeparttable}
\small
\renewcommand{\arraystretch}{1.1}
\begin{tabular}{ccccc}
\toprule
& & \multicolumn{3}{c}{Co $^2F - {^4F}$ spin gap $\mathrel{\vert} E_h$} \\
\cline{3-5}
& Basis set & GUGA-FCIQMC & CASPT2 & CCSD(T) \\
\hline
ANO-basis 	& VDZP & 0.04895(32)\phantom{0} & 0.04667 & \\
			& VTZP & 0.043358(40) 			& 0.04373 & \\
			& VQZP & 0.03655(29)\phantom{0} & 0.03675 & \\
			& Full & 0.03626(21)\phantom{0}	& 0.03565 & \\
			& Primitive & 	   	    		& 0.03565 & \\
\hline 
			& CBS  & 0.03158(50)\tnote{a}\phantom{0}	& 0.03165\tnote{a}  & \\
\hline 
cc-basis   	& n = D & 0.046448(88)& 	& 0.1057278\phantom{(00)} \\
 			& n = T & 0.03855(22)\phantom{0} & 			& 0.1054354\phantom{(00)} \\ 
 			& n = Q & 0.03685(27)\phantom{0} & 			& 0.1052032\phantom{(00)} \\
\hline
			& CBS 	& 0.0347(50)\tnote{b}\phantom{00} &  	& 0.1050555\tnote{b}\phantom{(00)} \\
\hline
			& Experiment &  							& 0.032285 & \\
			& $\Delta E_{ANO}$ & 0.00070(50)\phantom{0} & 0.00063\phantom{0} & \\
			& $\Delta E_{cc}$ & -0.00242(50)\phantom{0-} & & 0.0729115(92) \\
\botrule
\end{tabular}
\begin{tablenotes}
\footnotesize
\item[a] Direct two-point extrapolation of VTZP and VQZP spin-gap results according to Eq.~(\ref{eq:inverse-cube})
\item[b] Separate CBS extrapolation of $^2$F and $^4$F state with HF energy of aug-cc-pwCV5Z-DK basis set and fit of the correlation energy according to Eq.~\ref{eq:inverse-cube} with the $n =$ T and Q results.
\end{tablenotes}
\end{threeparttable}
\end{table*}

\section{\label{sec:discussion}Discussion}

The spin-adapted FCIQMC implementation was tested and benchmarked for the nitrogen atom and dimer, where excellent agreement with exact results\textemdash where available\textemdash and other quantum chemical methods was observed. 
We found that the additional computational cost associated with the more complicated and highly connected Hilbert space of CSFs is manageable and applications of this approach for large basis sets was demonstrated; eradicating the severe limitations of previous spin-adapted approaches in general and in FCIQMC in particular.  

The validity of the approach was proven and the direct targeting of specific spin states is possible;
enabling us to obtain results previously not accessible to the FCIQMC method. 
These are gaps of high-spin ground and low-spin excited state systems with an odd number of electrons and the excitation energies within an explicit spin symmetry sector.
For system with near-degenerate spin-eigenstates we observe an accelerated convergence of spin-gap results with respect to the total walker number in the spin-adapted FCIQMC implementation.
 
However, the additional scaling with the number of spatial orbitals in the GUGA-FCIQMC method, starts to become relevant for large basis set expansions, limiting the applicability, where the SD based implementation remains preferable.
The increased connectivity of a spin-pure basis reduces the generation probabilities in the spawning step of the FCIQMC method and thus limits the possible timestep of a simulation and causing stability issues in the sampling process.
In this regard, the scope of application of this method is to target specific, interesting spin states, which allows a clearer chemical and physical interpretation of results. 
As a consequence, more insight in chemical processes governed by the interplay of different spin states is possible.

\section{\label{chap:guga:conclustion}Conclusion and Outlook} 

The efficient usage of a spin-adapted basis in FCIQMC has been made possible within the (graphical) unitary group approach (GUGA) and the severe limitations of previous implementations have been overcome.
When formulated in such a basis, simulations conserve the total spin quantum number and the Hilbert space size of the problem is reduced.
As another positive consequence, targeting specific many-body subspaces of the Hamiltonian and getting access to their excitation energies is possible, and thus one is able to study phenomena governed by the interplay of different\textemdash even degenerate\textemdash spin sectors.
Additionally, the use of a spin-adapted basis improves the convergence of the projective FCIQMC method, for systems with near-degenerate spin states. 

We benchmarked the spin-adapted FCIQMC method and compare results with other computational approaches, for the nitrogen atom and dimer, where we find excellent agreement with reference results, when available. 
For the nitrogen atom we obtain the spin gap of the $^4$S$^o$ ground- and $^2$D$^o$ excited state and the ionization potential, and the dissociation energy of the nitrogen dimer within chemical accuracy to experiment.
We apply the method to study the 3d-transition metal cobalt, targeting properties, which defy a simple single-reference description. 
For cobalt, the spin-gap of the high-spin ground state (single-reference wavefunction) and low-spin excited state (\emph{multi-reference} wavefunction) was determined within chemical accuracy to experiment. 

This spin-adapted implementation brings FCIQMC \emph{en par} with many other quantum chemical methods, which already utilize the inherent total spin conservation of nonrelativistic, spin-independent molecular Hamiltonians. 

To combine the spin-adapted FCIQMC with the stochastic CASSCF method, 
the final missing piece is the spin-pure implementation of an efficient sampling of reduced density matrices \cite{rdm-fciqmc}, which would enable us to solve active spaces of unprecedented size in a spin-pure fashion, extending even further the applicability of the method. 
Unfortunately, the sampling of RDMs in the spin-adapted formulation based on the GUGA is unfortunately a highly non-trivial task. Although there is no theoretical problem of density matrices in the 
unitary group formalism \cite{GouldPaldusChandler1990, PaldusGould1993, Shepard2006, Robb2018}, from a practical standpoint there is. 
Due to the increased connectivity within a CSF basis and the possibility 
of generators with different spatial indices contributing to the 
same density matrix element, there is a large overhead involved in sampling RDMs in the spin-adapted FCIQMC method. 
However, we are optimistic to solve these problems in due time, which would allow us to use GUGA-FCIQMC as a spin-pure FCI solver in the stochastic CASSCF method \cite{Giovanni2016}. This would enable us to solve active spaces of unprecedented size in a spin-pure fashion, extending even further the applicability of the method.
Furthermore, the unitary group formalism is extendable to spin-dependent operators \cite{Drake1977, KentSchlesingerDrake1981, GouldChandler1984, KentSchlesinger1990, GouldPaldus1990,GouldBattle1993, KentSchlesinger1994,  Li2014, Yabushita2014}, and an extension of the spin-adapted FCIQMC method 
to this approach is currently investigated to enable us to study systems with spin-orbit coupling and explicit spin dependence. 
Along this line, another interesting problem to be investigated is the application of GUGA-FCIQMC to the two-dimensional $t$-$J$ and Heisenberg models. 

\appendix

\section{\label{app:excit_info}CSF Excitation Identification}

Efficiently identifying the difference between two given CSFs and the type of excitation (generator types), listed in Table~\ref{tab:types-of-doubles} of the main text, is crucial for an optimized matrix element calculation. 
For CSFs this operation is more involved compared to Slater determinants. 
This is because not only occupancy differences but also changes in the singly occupied orbitals (different spin-couplings) must be taken into account, as they can also lead to non-zero \emph{coupling coefficients}.
The defining difference for the excitation is the difference in spatial occupation numbers. The step-values, $d_i = \{0,1,2,3\}$, of the spatial orbitals of a CSF are efficiently encoded by two bits per spatial orbital
\[
d_i = 0: 00, \quad d_i = 1: 01, \quad d_i = 2: 10, \quad d_i = 3: 11,
\]
in an integer of length $2n$. This is equivalent to the memory requirement of storing the occupied spin-orbitals of a Slater determinant.
The spatial occupation difference, $\abs{\Delta n}$, can be efficiently obtained by shifting all negatively spin-coupled, $d_i = 2: 10$, to the right, and computing the bit-wise \code{xor}-operation on two given CSFs and counting the number of set bits in $\abs{\Delta n}$, e.g.\ by the \code{Fortran 2008} intrinsic \code{popcnt}:
\begin{center}
{\renewcommand{\arraystretch}{1.2}
\begin{tabular}{rc}
$\ket m = \ket{0,1,2,3}:$ &$00\;01\;10\;11$ \\
$\ket{m'} = \ket{1,2,1,2}:$ & $01\;10\;01\;10$ \\
\cline{2-2}
$n(m):$ &  $00\;01\;01\;11$\\
$n(m'):$ & $01\;01\;01\;01$\\
\cline{2-2}
$\abs{\Delta n}:$ \code{xor}: & $01\;00\;00\;01$\\
\hline
$\Sigma \abs{\Delta n}$: \code{popcnt}($\Delta n$): & 2\\
\hline
$\Delta n = n(m')-n(m):$ & $\ket{+1,0,0,-1}$\\ 
\end{tabular}}
\end{center}
With $\Sigma\abs{\Delta n}$ we can identify the excitation level, which would be a single excitation from orbital $4$ to $1$ in the example above, and $\Delta n$ gives us information, in which spatial orbitals electrons got removed or added. For CSFs the orbital occupation difference alone is not enough to completely identify an excitation between two CSFs, since for excitations of exchange type, involving $\uR\uL$ and $\oR\oL$ generators, there can be a change in the spin-coupling, without an actual change in orbital occupation. So additionally we also need information of the step-vector difference, $\Delta d$, which is just obtained by the \code{xor}-operation on the bit-representation of two given CSFs:
\begin{center}
{\renewcommand{\arraystretch}{1.2}
\begin{tabular}{rcc}
$\ket m = \ket{1,1,0,3}:$ & $01\,01\,00\,11$ & \\
$\ket{m'} = \ket{1,2,1,1}:$ & $01\,10\,01\,01$& \\
\cline{2-2}
$\abs{\Delta n}:$ & $00\,00\,01\,01$ & $\Sigma = 2$ \\
\cline{2-2}
$\Delta d:$ & $00\,11\,01\,10$ 
\end{tabular}}
\end{center}
In this example it can be seen, that the $\abs{\Delta n}$ information alone would lead us to believe a single excitation connects $m$ and $m'$, but this is not compatible with the change in step-vector at orbital $2$. 
So in addition, we need to determine if there are step-vector changes below the first, $\Delta d_b$, or above the last, $\Delta d_a$, occupation change in $\Delta n$. This can be done efficiently with the \code{Fortran 2008} intrinsic bit-operations, \code{leadz(I)} (\code{trailz(I)}), which give the number of leading(trailing) zeros in integer \code{I}. The case that there are only step-vector changes, $\Delta d$, within, the first and last $\Delta \neq 0$ cases, is encoded by $\Delta d_b = \Delta d_a = 0$. 

$\Sigma \abs{\Delta} n > 4$ indicates a higher excitation than double, so the two CSFs are not possibly connected by a single Hamiltonian application and can be disregarded. The non-zero Hamiltonian matrix elements can be identified by following combinations of $\Delta n$ and $\Delta d$:

\underline{$\Sigma \abs{\Delta n} = 0 \; \& \; \Delta d \neq 0$:}\\
This combination indicates, that there is no difference in the occupation number between two CSFs $m$ and $m'$, but a change in the spin-coupling of the singly occupied orbitals. Only a mixed generator $\uR\uL \ra \oR\oL$ generator combination, corresponding to the type (2c) in Table~\ref{tab:types-of-doubles}, can lead to those types of excitations. Details on the matrix element calculation in general are discussed below.

\underline{$\Sigma \abs{\Delta n} = 2\; \& \; \Delta d_b = \Delta d_a = 0$:}\\
This combination indicates a regular single excitation and the order of the removed and added electron determines the type of generator $\hat E_{ij}$, corresponding to type (0a) in Table~\ref{tab:types-of-doubles},
\begin{align*}
\Delta n &= +1 \ra -1: \quad \ul R \ra \ol R \\
\Delta n &= -1 \ra +1: \quad \ul L \ra \ol L.
\end{align*}
   
\underline{$\Sigma \abs{\Delta n} = 2 \;\&\; \Delta d_b \neq 0 \;\text{or}\; d_a \neq 0$:}\\
This indicates step-vector changes above or below the occupation differences, which identifies a mixed start $\uR\uL$ or end $\oR\oL$ segment. Again the order of the orbital occupation and step-vector changes below or above $\Delta n$ identifies the type of excitation 
\begin{align*}
\text {(1j)} \quad \Delta d_b \ra \Delta n_{+1} \ra \Delta n_{-1}&: \quad \ul{RL} \ra \ol L R \ra \ol R  \\
\text {(1i)} \quad \Delta d_b \ra \Delta n_{-1} \ra \Delta n_{+1}&: \quad \ul{RL} \ra \ol R L \ra \ol L \\
\text {(1f)} \quad  \Delta n_{+1} \ra \Delta n_{-1} \ra \Delta d_a&: \quad \ul R \ra \ul L R \ra \ol{RL} \\
\text {(1e)}\quad \Delta n_{-1} \ra \Delta n_{+1} \ra \Delta d_a&: \quad \ul L \ra \ul R L \ra \ol{RL}  ,
\end{align*}
with $\Delta n_{\pm 1} = \Delta n = \pm1$ and the reference to the entries of Table~\ref{tab:types-of-doubles}.

\begin{figure}
\centering
\includegraphics[width=0.47\textwidth]{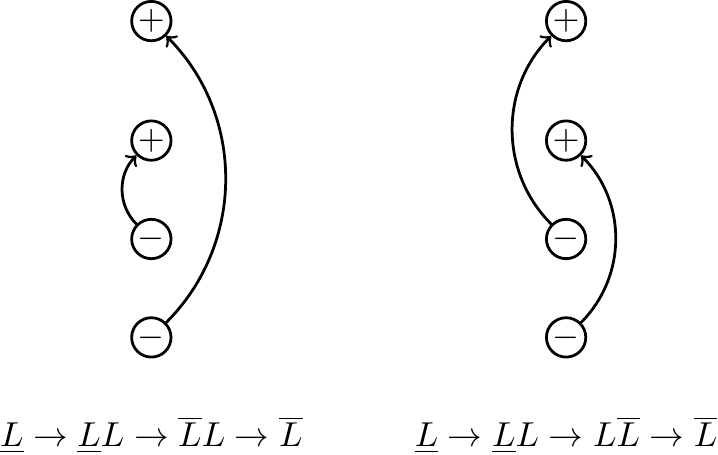}
\caption{\label{fig:two-body-sign}Equivalence of the two possible type (3b) double excitations $\hat e_{il,jk}$ and $ \hat e_{ik,jl}$ with $i < j < k < l$. The minus (plus) indicates the removal (addition) of an electron.}
\end{figure}

\underline{$\Sigma \abs{\Delta n} = 4 \; \& \Delta d_b = \Delta d_a = 0$:}\\
In this case it is necessary to have $\Delta d_b = \Delta d_a = 0$, otherwise this would indicate more than a double excitation, which would lead to a vanishing Hamiltonian matrix element. Again the order of the occupation differences gives information on the type of generators involved. The following combinations are identifiable only with $\Delta n$ (with reference to the entries of Table~\ref{tab:types-of-doubles})
{\small
\begin{align*}
\text {(2b)} \quad   &\Delta n = -2 \ra +2:  &&\ul{LL} \ra \ol{LL}\\
\text {(2a)} \quad   &\D n = +2 \ra -2:  &&\ul{RR} \ra \ol{RR}\\
\text {(1h)} \quad   &\Delta n = -2 \ra +1 \ra +1:  &&\ul{LL} \ra L\ol L / \ol L L \ra \ol L\\
\text {(1g)} \quad   &\D n = +2 \ra -1 \ra -1:  &&\ul{RR} \ra R\ol  / \ol R R \ra \ol R\\
\text {(1d)} \quad   &\D n = -1 \ra -1 \ra +2: &&\ul L \ra L\ul L / \ul L L\ra \ol{LL}\\
\text {(1c)} \quad   &\D n = +1 \ra +1 \ra -2:  &&\ul R \ra R\ul R / \ul R R \ra \ol{RR}\\
\text {(1a)} \quad   &\D n = -1 \ra +2 \ra -1:  &&\ul L \ra \ol L \ul R \ra \ol R\\
\text {(1b)} \quad   &\D n = +1 \ra -2 \ra +1:  &&\ul R \ra \ol R \ul L \ra \ol L\\
\text {(3b)} \quad   &\D n = -1 \ra -1 \ra +1 \ra +1: &&\ul L \ra \ul L L / L \ul L \ra \ol L L / L \ol L \ra \ol L\\
\text {(3a)} \quad   &\D n = +1 \ra +1 \ra +1 \ra +1: &&\ul R \ra \ul R R / R \ul R \ra \ol R R / R \ol R \ra \ol R,
\end{align*}} 
where, e.g.\ $\uL L/L\uL$, indicates that the order of indices of equivalent generators, $\hat e_{il,jk}/\hat e_{ik,jl}$ see Fig.~\ref{fig:two-body-sign}, can not be determined by $\Delta n$ alone. In the case of alike generators $RR(LL)$ this order does have influence on the sign of the matrix element \cite{Shavitt1981}, see below.

\begin{figure*}
\centering
\includegraphics[width=\textwidth]{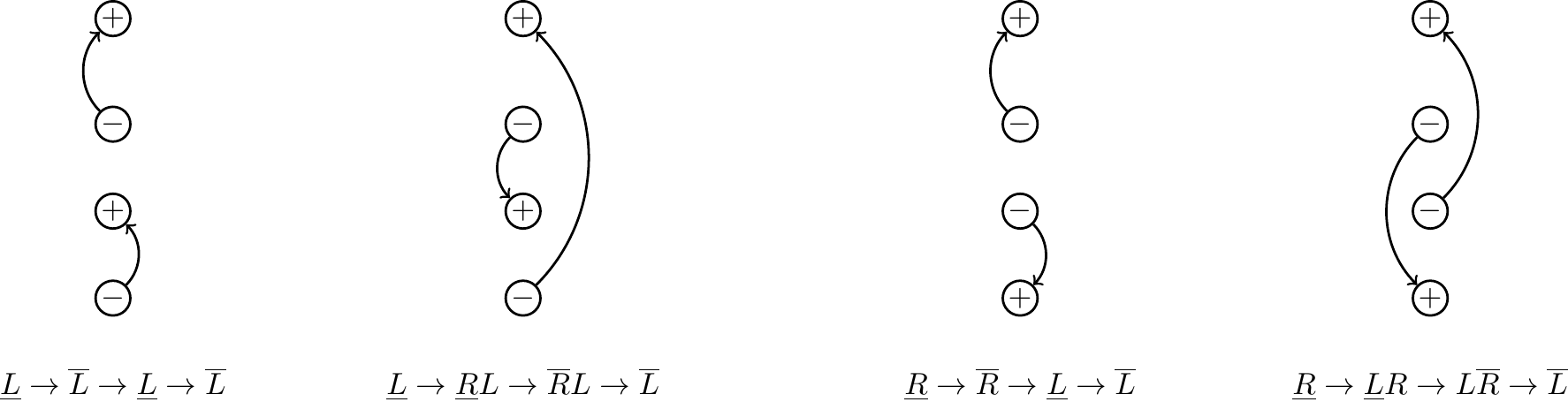}
\caption{\label{fig:two-body-order}Equivalence of the 3d$_0$ and 3d$_1$ (left) and 3e$_0$ and 3e$_1$ (right) generator combinations leading to the same orbital occupation difference between two CSFs.}
\end{figure*}

There are combinations of occupation differences where additionally the step vector differences have to be checked, since there are multiple two-body operators $\hat e_{ij,kl}$ possible, which can lead to the same excitation:
{\small
\begin{align*}
\D n &= -1 \ra +1 \ra \Delta d \ra -1 \ra +1:
\begin{cases}
\ul L \ra \ul R L \ra \ol R L \ra \ol L  \qquad\quad\; \text {(3d$_1$)} \\
\ul L \ra \ol L \ra \ul L \ra \ol L, \, \Delta d \mustbe 0 \quad \text {(3d$_0$)}
\end{cases} \\
\D n &= -1 \ra +1 \ra \Delta d \ra +1 \ra -1:
\begin{cases}
\ul L \ra \ul R L \ra \ol L R \ra \ol R \qquad\quad\; \text {(3f$_1$)}\\
\ul L \ra \ol L \ra \ul R \ra \ol R, \, \Delta d \mustbe 0\quad \text {(3f$_0$)}
\end{cases}\\
\D n &= +1 \ra -1 \ra \Delta d \ra +1 \ra -1:
\begin{cases}
\ul R \ra \ul L R \ra \ol L R \ra \ol R \qquad\quad\; \text {(3c$_1$)}\\
\ul R \ra \ol R \ra \ul R \ra \ol R, \, \Delta d \mustbe 0\quad \text {(3c$_0$)}
\end{cases}\\
\D n &= +1 \ra -1 \ra \Delta d \ra -1 \ra +1:
\begin{cases}
\ul R \ra \ul L R \ra \ol R L \ra \ol L \qquad\quad\; \text {(3e$_1$)}\\
\ul R \ra \ol R \ra \ul L \ra \ol L, \, \Delta d \mustbe 0\quad \text {(3e$_0$)},
\end{cases}
\end{align*}}
where every second case is only possible if there are no step-vector differences, $\Delta d = 0$, between the second and third occupation difference and reference to Table~\ref{tab:types-of-doubles}. The equivalence of these generator combinations can be seen in Fig.~\ref{fig:two-body-order}. However, even with no $\Delta d$ difference between the second and third occupation difference both generator combinations still contribute to the Hamiltonian matrix element.

With this method the distinct excitation types, listed in Table~\ref{tab:types-of-doubles} of the main text, can be efficiently identified. We would like to mention that this type of excitation identification does not take into account the cumulative $b_k$ restrictions, $\abs{b_k - b'_k} \leq 1$ for single- and $\abs{b_k - b'_k} \leq 2$ for double-excitations, for non-zero matrix elements, which currently has to be directly accounted for in the subsequent matrix element calculation. However, we plan to make use of this property in future optimizations of our implementation, as it severely limits the number of nonzero coupling coefficients.

\section{\label{app:mat-eles}Detailed Matrix Element Evaluation in the GUGA}

In this section we explain the efficient matrix element calculation in a spin-pure CSF basis based on the GUGA approach in more detail. 

\subsection{\label{app:diag}Diagonal matrix elements}
The diagonal matrix element for a GT state, $\ket m$ of the spin-free Hamiltonian, given in the main text, is given by the sum of the \emph{one-body} matrix elements, $\braopket{m}{\hat H_0}{m}$, with $\hat H_0 = \sum_{ij} t_{ij}\,\hat E_{ij}$, and the two-body contribution, $\braopket{m}{\hat H_1}{m}$, with $\hat H_1 = \sum_{ijkl} V_{ijkl}\,\hat e_{ij,kl}$. The matrix elements of the \emph{weight} generators, $\hat E_{ii}$, are simply just the occupancy of orbital $i$ in state $\ket m$ 
\begin{equation}\label{eq:weight-mateles}
\bra{m'}\hat  E_{ii} \ket m = n_i(m) \delta_{m',m}.
\end{equation}
So the one-body contribution is given by
\begin{align}
 \bra m \hat H_0 \ket m =& \sum\limits_i t_{ii} \bra m \hat E_{ii} \ket m = \sum\limits_i t_{ii} \,n(d_i),\\
 &\text{with }  
n(d_i) = \begin{cases}
           0 & \text{for } d_i = 0 \\
           1 & \text{for } d_i = {1,2} \\
           2 & \text{for } d_i = 3
          \end{cases},
\end{align}
where $d_i$ is step-value of spatial orbital $i$ in $\ket m$. 
The \emph{two-body} contributions are a bit more involved. Let's consider the different cases:

\underline{$i=j=k=l$:} reduces to the sum of doubly occupied orbitals
\begin{equation}
 \bra m \hat H_1 \ket m = \frac{1}{2}\sum\limits_i V_{iiii} \bra m 
\hat E_{ii}^2 - \hat E_{ii} \ket m  = \sum\limits_i V_{iiii} \,
\delta_{d_i,3},
\end{equation}
since $\hat E_{ii}^2 = \hat E_{ii}$ for $d_i = {1,2}$.

\underline{$i=j\neq k=l$:} only the $\hat E_{ii}\hat E_{jj}$ part of $\hat e_{ij,kl} = \hat E_{ij}\hat E_{kl} - \delta_{jk}\hat E_{il}$ remains, which reduces 
to a product of occupation numbers
\begin{align}
 \frac{1}{2}\sum\limits_{i\neq j} V_{iijj} \bra d \hat E_{ii}\hat E_{jj} \ket m &= 
\frac{1}{2}\sum\limits_{i\neq j} V_{iijj} 
\sum\limits_{m'} \bra m \hat E_{ii}\ket{m'}\bra{m'}\hat E_{jj}\ket m = \\ 
\frac{1}{2}\sum\limits_{i\neq j} V_{iijj} \bra m \hat E_{ii}\ket m \bra m \hat E_{jj} \ket 
m &= \frac{1}{2}\sum\limits_{i\neq j} 
V_{iijj} \, n(d_i) n(d_j) = \sum\limits_{i<j}V_{iijj}\, n(d_i)n(d_j).
\end{align}

The last relation comes from the fact that the sums are invariant under $i$, $j$ exchange.

\underline{$i = l \neq j = k$:} Also the exchange integral terms $V_{ijji}$ contribute to the diagonal matrix elements, if the excitation $\bra m \hat E_{ij}\hat E_{ji} \ket m$ leads to the same CSF. The calculation of these matrix elements depends on the step-values between $i$ and $j$ and are obtained by Shavitt's graph rules \cite{Shavitt1981}. The matrix elements between two CSFs for a double excitation are given by the product 
\begin{equation}
\label{eq:mat-prod-supp}
 \bra{m'} \hat e_{ij,kl} \ket m = \prod_{p\in S_2} W(Q_p;d_p, d'_p,\Delta b_p, b_p)  
 \times \sum_{x=0,1} \prod_{p\in S_1} W_x(Q_p; d_p, d'_p, \Delta b_p, b_p) 
\end{equation}
with $S_2 = (i,j) \cup (k,l) - S_2$ being the non-overlap range and $S_1 = (i,j) \cap (k,l)$ being the overlap region of the indices of the involved generator $\hat e_{ij,kl}$. The one-body segment values $W(Q_p;d_p, d'_p,\Delta b_p, b_p)$ can be found in Table~\ref{tab:single-mateles} in the main text and the two-body segment values $W_x(Q_p; d_p, d'_p, \Delta b_p, b_p)$ in Ref.~[\citen{Shavitt1981}]. In the case relevant for diagonal terms the matrix elements depending on the beginning $d_i$ and end $d_j$ step-values are given in Table~\ref{tab:diag-start} with
\begin{equation}
 A(b,x,y) = \sqrt{\frac{b+x}{b+y}} \quad \text {and} \quad 
 f(b,d) = 
 \begin{cases}
  1 & \text{for } d = {0,3} \\
  A(b,2,0)A(b,-1,1) & \text{for } d = 1 \\
  A(b,0,2)A(b,3,1) & \text{for } d = 2
 \end{cases}
\end{equation}
\begin{table*}
\centering
\caption{\label{tab:diag-start}Relevant terms of the exchange contribution to diagonal matrix elements.}
\resizebox{\textwidth}{!}{
{\renewcommand{\arraystretch}{1.2}
\begin{tabular}{c|ccccc}
\toprule
$d_i \vert d_j$ & & 0 & 1 & 2 & 3\\
\hline
0 &&  0 & 0 & 0 & 0\\
1 & &0 & $-\frac{1}{2}\left(1 +A(b_i,2,0)A(b_j,-1,1) \prod_k f(b_k,d_k)\right)$ & $-\frac{1}{2}\left(1 - A(b_i,2,0)A(b_j,3,1)\prod_k f(b_k,d_k)\right)$ & -1\\
2 && 0 & $-\frac{1}{2}\left(1 - A(b_i,0,2)A(b_j,-1,1) \prod_k f(b_k,d_k)\right)$ & $-\frac{1}{2}\left(1 + A(b_i,0,2)A(b_j,3,1)\prod_k f(b_k,d_k)\right)$ & -1\\
3 && 0 & -1 & -1 & -2 \\
\botrule
\end{tabular}
}
}
\end{table*}
Unfortunately this requires the consideration of all step-vector and b-values between $i$ and $j$ to calculate the diagonal matrix element.
\begin{equation}
 \sum\limits_{i\neq j} \frac{V_{ijji}}{2} \bra m \hat e_{ij,ji} \ket m = 
 -\sum\limits_{i<j}  \frac{V_{ijji}}{2}\Big(n(d_i)n(d_j) + X(i,j)\Big)
\end{equation}
The first term $\frac{1}{2}n(d_i)n(d_j)$ accounts for the first singlet coupled $x=0$ matrix elements in the Table~\ref{tab:diag-start} and $X(i,j)$ accounts for the triplet coupled $x=1$ matrix elements. And only yields a contribution if both $d_i$ and $d_j$ 
are either 1 or 2
\begin{equation}
\small
 X(i,j) = 
 \begin{cases}
  \phantom{-}A(b_i,2,0)\prod_k f(b_k,d_k) A(b_j,-1,1) & d_i = 1, d_j = 1 \\
  -A(b_i,2,0)\prod_k f(b_k,d_k) A(b_j,3,1) & d_i = 1, d_j = 2 \\
  -A(b_i,0,2)\prod_k f(b_k,d_k) A(b_j,-1,1) & d_i = 2, d_j = 1 \\
  \phantom{-}A(b_i,0,2) \prod_k f(b_k,d_k) A(b_j,3,1) & d_i = 2, d_j = 2 \\
  \phantom{-}0 & \text{otherwise}
 \end{cases}
\end{equation}
and the rest gets accounted by the product of occupation numbers $n(d_i)n(d_j)$. In total the diagonal Hamilton matrix element for a CSF $m$ is given by
\begin{equation}
 \bra m \hat H \ket m = \sum_i \Bigg\{ t_{ii}\, n(d_i) + V_{iiii}\, \delta_{d_i,3} 
  + \sum_{j>i} \left[ V_{iijj} \, n(d_i) n(d_j) - \frac{1}{2} V_{ijji}\,\Big(n(d_i)n(d_j) + X(i,j)\Big)\right]\Bigg\}.
\end{equation}

We would like to mention that the $f(b,d)$ and $X(i,j)$ functions are based on and can be expressed as the $D(b,p)$ and $RL$ segment shape function found in Ref.~\citen{Shavitt1981}. However, we chose the above formulation for conciseness of this manuscript. 

\subsection{\label{app:off-diag}Off-diagonal matrix elements}
The off-diagonal matrix element between two CSFs $\ket m$ and $\ket{m'}$ is given by
\begin{equation}
 \bra{m'} \hat H \ket{m} = \sum_{ij} t_{ij} \bra{m'}\hat E_{ij}\ket{m} + \frac{1}{2}
 \sum_{ijkl} V_{ijkl}\bra{m'}\hat e_{ij,kl}\ket m
\end{equation}
Similar to the Slater-Condon rules \cite{slater1929, condon1930} for matrix element calculation between Slater determinants, we need to identify the involved orbital indices $(i,j,k,l)$ connecting $m'$ and $m$ and by comparing the \emph{orbital occupation differences} between the two CSFs, $\Delta n_i = n(d_i) - n(d_i')$, already mentioned in Sec.~\ref{app:excit_info}.
There are the following possibilities for $\Delta n_i$, which yield a possible non-zero matrix element between $\ket{m'}$ and $\ket m$:

$\underline{\Delta n_i = 0}$ for all orbitals, but $\ket{m'}$ and $\ket m$ differing for some orbitals, implies a full-start $\underline{RL}$ into full-stop $\overline{RL}$ double excitation, type (2c) in Table~\ref{tab:types-of-doubles}, with only changes in the open-shell orbitals. The matrix element can be expressed as
\begin{equation}
\bra{m'}\hat H \ket m = \frac{1}{2}\sum_{i\neq j} V_{ij,ji} \bra{m'} \hat e_{ij,ji}
\ket m + V_{ji,ij}\bra{m'}\hat e_{ji,ij}\ket m 
= \sum_{i\neq j} V_{ij,ji} \bra{m'}\hat   e_{ij,ji}\ket m.
\end{equation}
Because these full-start into full-stop excitations are symmetric concerning conjugation of the generator indices and the molecular two-body integrals also, it reduces to
\begin{equation}
\bra{m'}\hat H \ket m = 2\sum_{i<j} V_{ij,ji} \bra{m'}\hat e_{ij,ji}\ket m.
\end{equation}
To yield a non-zero matrix element between $m'$ and $m$ the indices $i$ and $j$ have to engulf all the differing orbitals, yielding a maximum lower index $I$, and a minimum upper index $J$. Because full-start into full-stop excitations have the possibility to leave $d_i$ unchanged, basically all combination $i \le I$ and $j \ge J$ in the summation have to be considered. There has to be at least two differences between $m$ and $m'$, or otherwise it would just be a diagonal matrix element. The singlet coupled $x_0$ matrix element branch can be discarded, as a change in $m$ implies $\Delta b = \pm2$ at least at one orbital. Furthermore the integral, $F(I,J)$, between the region of the first to the last change in $m$ ($I\rightarrow J$) is the same for all matrix elements. The remaining product terms are given by the triplet coupled $x_1$ elements for non changing $d_i$ value from orbital $i$ to $I$ and $J$ to $j$ given as
\begin{equation}
\bra{m'}\hat H \ket m = 2\sum_{\substack{i<I\\j>J}} F(I,J)\, V_{ij,ji} \prod_{k=i}^{I-1} RL(d_k) \prod_{k'=J+1}^{j} RL(d_j) \quad \text{with} \quad F(I,J) = \prod_{k=I}^{J}RL(d_k), 
\end{equation}
where $RL(d_i)$ indicates the triplet-coupled $W_1(Q_i;d_i,d'_i,\Delta b_i,b_i)$ matrix elements for a mixed $RL$ generator combination, depending on $d$ and $b$, which can be found in Ref.~[\citen{Shavitt1981}]. Since we calculate the matrix elements on-the-fly in the excitation generation step of the FCIQMC method it is useful to formulate the matrix elements in terms of already calculated terms, i.e.\ $F(I,J)$, to reduce the computational effort of the spin-adapted FCIQMC implementation. 

$\underline{\Delta n_k = \pm 1:}$ for \emph{two} spatial orbitals $i,j$. This implies a one-body contribution, as well as two-body contributions with two indices being identical, over which has to be summed. However, there is the additional constraint that the double excitation also has to lead to the same orbital occupancy difference $\Delta n$, which only leaves following terms:
\begin{align}
\label{eq:mat-1}
\bra{m'} \hat H \ket m = \; t_{ij} \bra{m'}\hat E_{ij}\ket m 
+ \frac{1}{2} \sum_k \,\Big( &V_{ij,kk} \bra{m'} \hat e_{ij,kk} \ket m + V_{kk,ij} \bra{m'} \hat e_{kk,ij} \ket m \nonumber \\
+ &V_{ik,kj} \bra{m'} \hat e_{ik,kj} \ket m + V_{kj,ik} \bra{m'}\hat  e_{kj,ik} \ket m   \Big),
 \end{align}
where the second line, involving weight generators, due to $V_{ijkk} = V_{kkij}$, reduces to
\begin{equation}
\bra{m'}\hat E_{ij}\ket m \sum_{k\neq i,j} V_{ij,kk} \, n(d_k).
\end{equation}
For $k \neq (i,j)$ both the remaining terms yield (without the two-particle integrals for clarity and $\hat e_{ij,kl} = \hat E_{ij}\hat E_{kl} - \delta_{jk} \hat E_{il}$)
\begin{equation}
\bra{m'}\hat E_{ij}\hat E_{ii}\ket m + \bra{m'}\hat E_{ij}\hat E_{jj}\ket m - \bra{m'}\hat E_{ij}\ket m 
=  \bra{m'}\hat E_{ij}\ket m \big(n(d_i) + n(d_j) - 1\big),
 \end{equation}
 which in total yields
 \begin{equation}
  \bra{m'}\hat E_{ij}\ket{m} \left(\sum_k V_{ij,kk} \, n(d_k) - V_{ij,jj} \right).
 \end{equation}
 The third line in Eq.~\eqref{eq:mat-1}, due to $V_{ij,kl} = V_{kl,ij}$ and $\hat e_{ij,kl} = \hat e_{kl,ij}$,  reduces to:
 \begin{equation}
  \sum_k V_{ik,kj} \, \bra{m'} \hat e_{ik,kj} \ket m 
 \end{equation}
 and is a bit more involved to calculate. Depending on the relation of the index $k$ to $(i,j)$, the two-body integral corresponds to certain sequences of generator combinations (assuming $i<j$ for now, which is easily generalized):
 
$\underline{k < i:} \qquad \underline{LR} \rightarrow \overline{L}R \rightarrow \overline{R}$: type (1j) excitations in Table~\ref{tab:types-of-doubles}, \\
without a change in the spin-coupling in the overlap region $(k,i)$ between the two CSFs $m$ and $m'$ . The $\Delta b = 0$ branch matrix elements of the mixed generator $RL$ contribute multiplicatively $-t^2 n(d_k)$ terms, see Ref.~[\citen{Shavitt1981}], and the $x_1$ matrix element contributions are $x_1 = \prod_{l=k}^{i-1} RL(l)$, where $RL(l)$ are just the normal mixed generator $x_1$ product elements. The product only goes until index $i-1$ to still be able to formulate it in terms of the single excitations $\hat E_{ij}$. To formulate it multiplicatively, special factors depending on the step-vector $d(i)$ have to determined, so the semi-stop $x_1$ elements of $\ol R L(i)/\ol L R(i)$ have the same elements as the single starts $\ul R/\ul L$. 
The formulation in terms of $\hat E_{ij}$ enables us to reuse the already calculated one-body elements in the excitation generation of the FCIQMC method.

The modified values starting $\ul \Delta(i)$ and end $\ol \Delta(j)$ values for $\ul{RL} \ra R \ol L \ra \ol R$ and $\ul{RL} \ra \ol R L \ra \ol L$ type of excitation can be found in Table~\ref{tab:modified-mixed}.
\begin{table*}
\centering
\small
\renewcommand{\arraystretch}{1.2}
\caption{\label{tab:modified-mixed}Modified matrix element contributions $\ul \Delta(i)$ and $\ol \Delta(j)$ necessary for the on-the-fly matrix element calculation during the excitation process in the FCIQMC algorithm.} 
\begin{tabular}{ccccccccc}
\toprule
$d'$ & $d$ & $\ul R$ & $R\ol L$ & $\ul\Delta(i)$ & & $\ol L$ & $\ul R L$ & $\ol\Delta(j)$\\
\hline
1 & 0 & 1 & -$tA(0,2)$ & -$tA(0,2)$ & & 1 & $tA(2,0)$ & $tA(2,0)$\\
2 & 0 & 1 & $tA(2,0)$ & $tA(2,0)$& & 1 & -$tA(0,2)$ & -$tA(0,2)$\\
3 & 1 & $A(1,0)$ & -$tA(-1,0)$ & $tA(-1,1)$& & $A(0,1)$ & -$tA(2,1)$ & -$tA(2,0)$\\
3 & 2 & $A(1,2)$ & -$tA(3,2)$ & -$tA(3,1)$& & $A(2,1)$ & $tA(0,1)$ & tA(0,2)\\
\hline\\
$d'$ & $d$ & $\ul L$ & $\ol R L$ & $\ul\Delta (i)$ & & $\ol R$ & $R\ul L$ & $\ol\Delta(j)$\\
\hline
 0 &  1 &  1 &  -t$A(-1,1)$ & -$tA(-1,1)$ & & 1 & $tA(2,0)$ & $tA(2,0)$\\
 0 &  2 &  1 &  $tA(3,1)$ & $tA(3,1)$ & & 1 & -$tA(0,2)$ & -$tA(0,2)$\\
 1 &  3 &  $A(2,1)$ &  $tA(0,1)$ & $tA(0,2)$ & & $A(0,1)$ & -$tA(2,1)$ & -$tA(2,0)$\\
 2 &  3 &  $A(0,1)$ &  -$tA(2,1)$ & -$tA(2,0)$ & & $A(2,1)$ & $tA(0,1)$ & $tA(0,2)$ \\
 \botrule
\end{tabular}

\end{table*}
The rest of the double excitation overlap matrix elements is the same. So for $k < i$ the two-body matrix elements are given by
\begin{equation}
 \sum_{k=1}^{i-1} \left( -t^2\,n(d_k) + \ul\Delta(i) \prod_{l=k}^{i-1} RL(d_l,d'_l) \right),
\end{equation}
with $\ul \Delta(i)$ from Table~\ref{tab:modified-mixed} for the corresponding generator combination.

$\underline{k = i:} \qquad W\underline{R} \rightarrow \overline{R}$: type (0c) excitations in Table~\ref{tab:types-of-doubles}, \\
which just reduce to $\bra{m'}\hat E_{ij}\ket m (n(d'_i) - 1)$, similarly:\\
$\underline{k = j:}$\\
reduces to $\bra{m'}\hat E_{ij}\ket m (n(d_j) - 1)$.

$\underline{k > j:} \qquad \underline{R} \rightarrow R\underline{L} \rightarrow \overline{RL}$: type (1f) excitations in Table~\ref{tab:types-of-doubles}\\
As in the $k < i$ case the $x=0$ contribution is a multiplicative $-t^2\,n(d_k)$ factor. The non-vanishing $x_1$ overlap matrix elements are again calculated multiplicatively by the use of modified semi-stop segments at $j$ to formulate the matrix element in terms of single excitations for $\underline{R} \rightarrow R\underline{L} \rightarrow \overline{RL}$ and $\underline{L} \rightarrow L\underline{R} \rightarrow \overline{RL}$ generator combinations. The modified terms $\ol \Delta(j)$ terms can also be found in Table~\ref{tab:modified-mixed}.
Here only the step-vector combinations, which lead to the $\Delta b = 0$ branch in the overlap region are allowed, since there is not step-vector difference above $j$. The full matrix elements are given by
\begin{equation}
 \sum_{k>j}^n \left(-t^2\,n(d_k) + \ol\Delta(j)\prod_{l=j+1}^k RL(d_l,d'_l) \right),
\end{equation}
with $\ol \Delta(j)$ from Table~\ref{tab:modified-mixed} for the specific generator combination.

$\underline{i(j) < k < j(i):} \; \underline{R} \rightarrow \underline{R} \overline{R} \rightarrow \overline{R}\;(\uL \ra \oL\uL \ra \oL)$: type (0b) excitations in Table~\ref{tab:types-of-doubles}\\
These types of generator combinations correspond to one-body terms actually. At index $k$ the usual product term for the $\hat E_{ij}$ matrix element calculation takes on a different than usual value. This modification can be applied multiplicatively, but depends on $d_k, d'_k, b_k, \Delta b_k$ and the type of generator($i<j$ or $i>j$). The modified values can be found in Table~\ref{tab:mod-inter}.
\begingroup
\begin{table*}
\centering
\small
\caption{\label{tab:mod-inter}Modified two-body terms at the single overlap site $k$ used to formulate the two-body contribution to the single excitations multiplicatively and allow an on-the-fly matrix element calculation thereof.}
\resizebox{\textwidth}{!}{
  {%
\newcommand{\mc}[3]{\multicolumn{#1}{#2}{#3}}
\renewcommand{\arraystretch}{1.2}
\newcommand{\pd}{\phantom{-}}
\begin{tabular}{ccccccccccccccc}
\toprule
 & \mc{4}{c}{Usual value} & & \mc{4}{c}{Modified Value} & & \mc{4}{c}{Multiplicative Factor $r_k$}\\
 \hline
 &  \mc{2}{c}{$R$} & \mc{2}{c}{$L$} &  & \mc{2}{c}{$\underline{R}\overline{R}$} & \mc{2}{c}{$\underline{L}\overline{L}$} & & \mc{2}{c}{$i < j$} & \mc{2}{c}{$i > j$}\\
 \hline
 $d'd\vert \Delta b$ & -1 & +1 & -1 & +1 & &- 1 & +1 & -1 &+1 & & -1 & +1 &-1 & +1\\
 \hline
00 & \pd 1 & \pd 1 & \pd 1 & \pd 1 &  & 0 & 0 & 0 & 0 &  & \pd 0 &\pd 0 &\pd 0 &\pd 0\\
11 & -1 & $C(b,0)$ & $C(b,1)$ & -1 &  & 1 & 0 & 0 & 1 & & -1 &\pd 0 &\pd 0 & -1\\
12 & -$1/(b+2)$ & $-$& $1/(b+1)$ & $-$ & & 1 & $-$ &1  & $-$ & & -$(b+2)$ & $-$ & $(b+1)$ & $-$\\
21 & $-$ & $1/b$ & $-$ & -$1/(b+1)$ &  & $-$  & 1 & $-$ & 1 & & $-$ &\pd $b$ & $-$ & -$(b+1)$\\
22 & $C(b,2)$ & -1 & -1 & $C(b,1)$ & & 0 & 1 & 1 & 0  & & \pd 0 & -1 & -1 &\pd 0\\
33 & -1 & -1 & -1 & -1 & & 1 & 1 & 1 & 1 & & -1 & -1 & -1 & -1\\
\botrule
  \end{tabular}
}%
}
\end{table*}
\endgroup
By defining $r_k$ as
\begin{equation}
\label{eq:rk-eq}
 r_k(d'_k,d_k,b_k,\Delta b_k) = \begin{cases}
        -t^2\,n(d_k) + \ul\Delta(i)\prod_{l=k}^{i-1} RL(d_l,d'_l) & \text{for } k < min(i,j) \\
        -t^2\,n(d_k) + \ol\Delta(j)\prod_{j+1}^{l=k} RL(d_l,d'_l) & \text{for } k > max(i,j) \\
        n(d'_i) - 1 & \text{for } k = i \\
        n(d_j) - 1 & \text{for } k = j \\
        \text{entries from Table~\ref{tab:mod-inter}} & \text{for } k \in (i,j)
       \end{cases}
\end{equation}
the total matrix element of a single excitation with one-body and two-body contributions can be expressed as
\begin{equation}
 \bra{m'}\hat H \ket m = \bra{m'}\hat E_{ij}\ket m \left( t_{ij} - V_{ij,jj} + \sum_k
 V_{ij,kk}\,n(d_k) + V_{ik,kj}\,r_k(d'_k,d_k,b_k,\Delta b_k)  \right)
\end{equation}
in terms of the one-body coupling coefficient. The evaluation requires the calculation of the single excitation matrix element $\bra{m'}\hat E_{ij}\ket m$ through Shavitt's  graphical rules  \cite{Shavitt1978, Shavitt1981}, and the summation of terms depending on $d'_k$ and $d_k$ entries of the two CSFs (although in a sequential dependence, since $r_k$ depends on $\Delta b_k$).
We can calculate the $r_k$ terms similar to Shavitt's matrix product terms with $r_k$~\eqref{eq:rk-eq} in the excitation range given in Table~\ref{tab:mod-inter} during excitation generation. This requires an $\bigO{N}$ effort in calculation of $r_k$, since only occupied orbitals contribute.

$\underline{\Delta n = \pm 1:}$ at \emph{two orbitals} and \emph{additional step-vector differences} $\Delta d$ below or above the excitation range, correspond to (1e, 1f, 1i) or (1j) excitations of Table~\ref{tab:types-of-doubles}, depending on the ordering of the remaining indices. These are $d=1,d'=2$, and vice versa, step-vector differences outside the range $(i,j)$, corresponding to double excitations with mixed generator full-starts $\ul{RL}$ or full-stops $\ol{RL}$. Similar to the $\Sigma \abs{\Delta n} = 0$ case all possible excitations connecting the two CSFs have to engulf the first step-vector change, $I$, if it occurs before $\min(i,j)$ or the last step-vector change, $J$, if it is after $\max(i,j)$. However, all mixed full-starts before $I$ or after $J$ have to be considered too, since there is the possibility of a $\ul{RL}(d=1,d'=1)$ or $\ul{RL}(d=2,d'=2)$ start with non-zero $x_1$ matrix element \cite{Shavitt1981}. So the matrix element is given by: 
\begin{align}
 \sum_{k<I}^{i-1} \ul\Delta(i) F(I,i) \bra{m'}\hat E_{ij}\ket m \prod_{l=k}^{I} RL(d_l) \quad \text{for } I < \min(i,j) \\
 \sum_{k>J} \ol\Delta(j) F(j,J) \bra{m'}\hat E_{ij}\ket m \prod_{l>J}^n RL(d_l) \quad \text{for } J > \max(i,j),
\end{align}
with $RL(d_l)$ again being the $x_1$ matrix elements, $F(I,i)/F(j,J)$ being the always involved $x_1$ matrix elements engulfing all step-vector changes in the overlap region and $\ul\Delta(k)/\ol \Delta(k)$ being the modifying terms to express it in terms of single excitation matrix elements $\hat E_{ij}$, see Table~\ref{tab:modified-mixed}. 

$\underline{\Delta n_i = \pm2}$ at \emph{two spatial} orbital $i$ and $j$ implies a full-start into full-stop double excitation with two alike generators ($\underline{RR}\rightarrow\overline{RR}$ or $\underline{LL}\rightarrow\overline{LL}$, corresponding to type (2a) and (2b) in Table~\ref{tab:types-of-doubles}). This completely specifies the indices and the full matrix element is just given by
\begin{equation}
 \bra{m'}\hat H \ket m = V_{ij,ij}\bra{m'}\hat e_{ij,ij}\ket m 
\end{equation}
and calculated with Shavitt's graphical rules \cite{Shavitt1981}. The order of the orbitals, where electrons are removed and added, determines the type of generators.

$\Delta n \neq 0 $ at \emph{three} different orbitals  with 
one $\Delta n = \pm 2$ and two $\Delta n = \mp 1$, corresponds to type (1c, 1d, 1g) or (1h) excitations of Table~\ref{tab:types-of-doubles}. This determines all four indices, with two indices being identical, with $\Delta n_k = \pm2$ and $\Delta n_i = \Delta n_j = \mp 1$. Leaving the matrix element to be:
\begin{equation}
 \bra{m'}\hat H \ket m = \begin{cases}
                          V_{ik,jk}\bra{m'}\hat e_{ik,jk}\ket m & \text{if } \Delta n_k = -2\\
                          V_{ki,kj}\bra{m'}\hat e_{ki,kj}\ket m & \text{if } \Delta n_k = +2\\
                         \end{cases}
\end{equation}

$\underline{\Delta n_i \neq 0} $ at \emph{four} different orbitals with two times $\Delta n = 1$ and two times $\Delta n = -1$ values, corresponds to type (3a, 3b, 3c, 3d, 3e) or (3e) excitations of Table~\ref{tab:types-of-doubles}. This also completely determines all four indices of the excitation, but there are four different combinations of these indices which can lead to the same state, where two of them, however, are equivalent. The relation and ordering of these indices determines the type and combinations of generators, with the total matrix element given by
\begin{align}
 \bra{m'}\hat H \ket m =& \frac{1}{2}\bra{m'}\big(V_{li,kj}\,\hat e_{li,kj} + 
 V_{kj,li}\,\hat e_{kj,li} + V_{ki,lj}\,\hat e_{ki,lj} + V_{lj,ki}\,\hat e_{lj,ki}\big)\ket m \nonumber \\
 =& \bra{m'}\left(V_{li,kj}\,\hat e_{li,kj} + V_{ki,lj}\,\hat e_{ki,lj}\right) \ket m,
\end{align}
where for orbitals $l$ and $k$, $\Delta n_l = \Delta n_k = +1$ and for $i$ and $j$, $\Delta n_i = \Delta n_j = -1$.
The relative positions of the $\Delta n = +1$ and $\Delta n = -1$ orbitals determines the generator combinations and type of excitations involved.

For alike generator combinations, e.g. $\uR \ra \uR R \ra \oR R \ra \oR$, we have to take into account the sign flip due to an exchange of operator indices. Since $\hat e_{ik,jl}$ and $\hat e_{il,jk}$ ($i<j<k<l$), both contribute to the same excitation. As already pointed out by Paldus \cite{Paldus1980, BoylePaldus1980}, the Coulomb and exchange type contributions can be expressed in terms of the same $x=0$ and $x=1$ matrix element contributions with 
{\small
\begin{align}
w_0 &= \prod_{k\in S_2} W(Q_k;d'_k,d_k,\Delta b_k,b_k)\prod_{k\in S_1} W_1(Q_k;d'_k,d_k,0,b_k) \label{eq:w0}\\
 w_1 &= \prod_{k\in S_2} W(Q_k;d'_k,d_k,\Delta b_k,b_k) \prod_{k\in S_1} W_1(Q_k;d'_k,d_k,\Delta b_k,b_k)\label{eq:w1},
\end{align}
}
where $w_0 \neq 0$ only if $\Delta b_k = 0, \forall k\in S_1$. By sticking to the convention to use the standard order of operators, as indicated in Table~\ref{tab:types-of-doubles}, the contribution of an exchange of orbital indices in the generator can be expressed as 
\begin{align}\label{eq:coulomb-exchange-mateles}
\bra {m'} \hat e_{jl,ik}\ket m &= w_0 + w_1 \\
\bra {m'} \hat e_{jk,il}\ket m &= w_0 - w_1,  
\end{align}
with a type (3a) excitation from Table~\ref{tab:types-of-doubles} as an example. 
The total matrix element is then given by 
  \begin{equation}
   \bra{m'}\hat H \ket m = w_0\left(V_{jlik} + V_{jkil}\right) + w_1\left(V_{jlik} - V_{jkil}\right),
  \end{equation}
The case of alternating orbital occupancy differences and $\Delta n = \pm 1 \ra \Delta n = \mp 1$ involve no sign change in mixed generator semi-start and semi-stops for the $x_1$ matrix element. These type (3c-3f) excitations of Table~\ref{tab:types-of-doubles} also contain non-overlap generator combinations, indicated by the subscript $_0$. Because the non-overlap double excitations are contained as the $\Delta b_k = 0, \forall k \in S_1$ special case of these mixed generator excitation, we do not treat them explicitly, but stick to the convention to always use the mixed generator combination in the excitation generation. For an excitation, which left the $\Delta b_k = 0$ path at some point in the overlap region, only the $x=1$ matrix element contributes. If $\Delta b_k = 0, \forall k \in S_1$, the Coulomb type contribution can be obtained by the $x = 0$ term of the exchange matrix element \cite{Paldus1980}
\begin{equation}\label{eq:non-overlap-exchange}
\bra {m'} \hat e_{il,kj} \ket m = -\frac{w_0}{2} + w_1, \quad
\bra {m'} \hat e_{ij,kl} \ket m = w_0.
\end{equation}
otherwise $w_0 = 0$. The total matrix element is then given by 
\begin{equation}
\braopket{m'}{\hat H}{m} = w_0\left(-\frac{V_{ilkj}}{2} + V_{ijkl}\right) + w_1\,V_{ilkj}.
\end{equation}

$\underline{\Delta n \neq 0}$ at \emph{more than four} different spatial orbitals or $\underline{\Delta b \neq 0}$ outside of excitation range for $\Sigma \abs{\Delta n} = 4$, yields a zero matrix element, as such excitations cannot be obtained by a single application of the Hamiltonian. 

For brevity of this manuscript the extensively used remaining two-body segment value tables are not listed here;
the interested reader is referred to References~[\citen{Shavitt1978, Shavitt1981}] for a detailed explanation and listing of the GUGA matrix elements by I. Shavitt.

\section{\label{app:pick-orb}Weighted Orbital Choice with GUGA Restrictions}
A note on the weighting of the integral contribution to the orbital picking process: As one can see in Eq.~(\ref{eq:coulomb-exchange-mateles}) and (\ref{eq:non-overlap-exchange}) it is not as easy as in a SD based implementation to weight an integral contribution of orbitals by the exact matrix element, $V_{ikjl}$ for spin-opposite and $V_{ikjl} - V_{iklj}$ for spin parallel excitations, as it is done in the current FCIQMC implementation. Since the relative sign of the $w_0$ and $w_1$ contribution (\ref{eq:w0}, \ref{eq:w1}) depends on the chosen excitation $\ket {m'}$ and can not be easily predetermined and there is no notion of a $m_s$ quantum number in a spin-adapted calculation. Our choice was to weight the integral contribution by the magnitude of the integrals $\abs{V_{ikjl}} + \abs{V_{iklj}}$ to capture the strongest couplings at least. This leads to some inefficiencies in the excitation generation of the CSF based implementation. 

\subsection{\label{app:pick-orb-singles}Restrictions on the Orbital Choice for Single Excitations}
To ensure at least one possible non-zero excitation, $\hat E_{ij} \ket m$, we have to place some additional restriction on the choice of orbital $(i,j)$ compared to a SD-based implementation. The idea is to first pick an electron in an occupied spatial orbital $j$ at random with $p(j) = 1/N$. Depending on the step-value $d_j$ certain restriction on the to-be-picked orbital $i$ are placed. A general restriction is that $i$ must not be doubly occupied $d_i \neq 3$. 

If $d_j = 3$, since both $\Delta b$ branches can end at $d_j$ for raising generators R $(i<j)$ and also both branches can start for a lowering generator L $(i>j)$ there are no additional restrictions on the orbital $i$, except $d_i \neq 3$.

\begin{figure*}
\centering
\includegraphics[width=\textwidth]{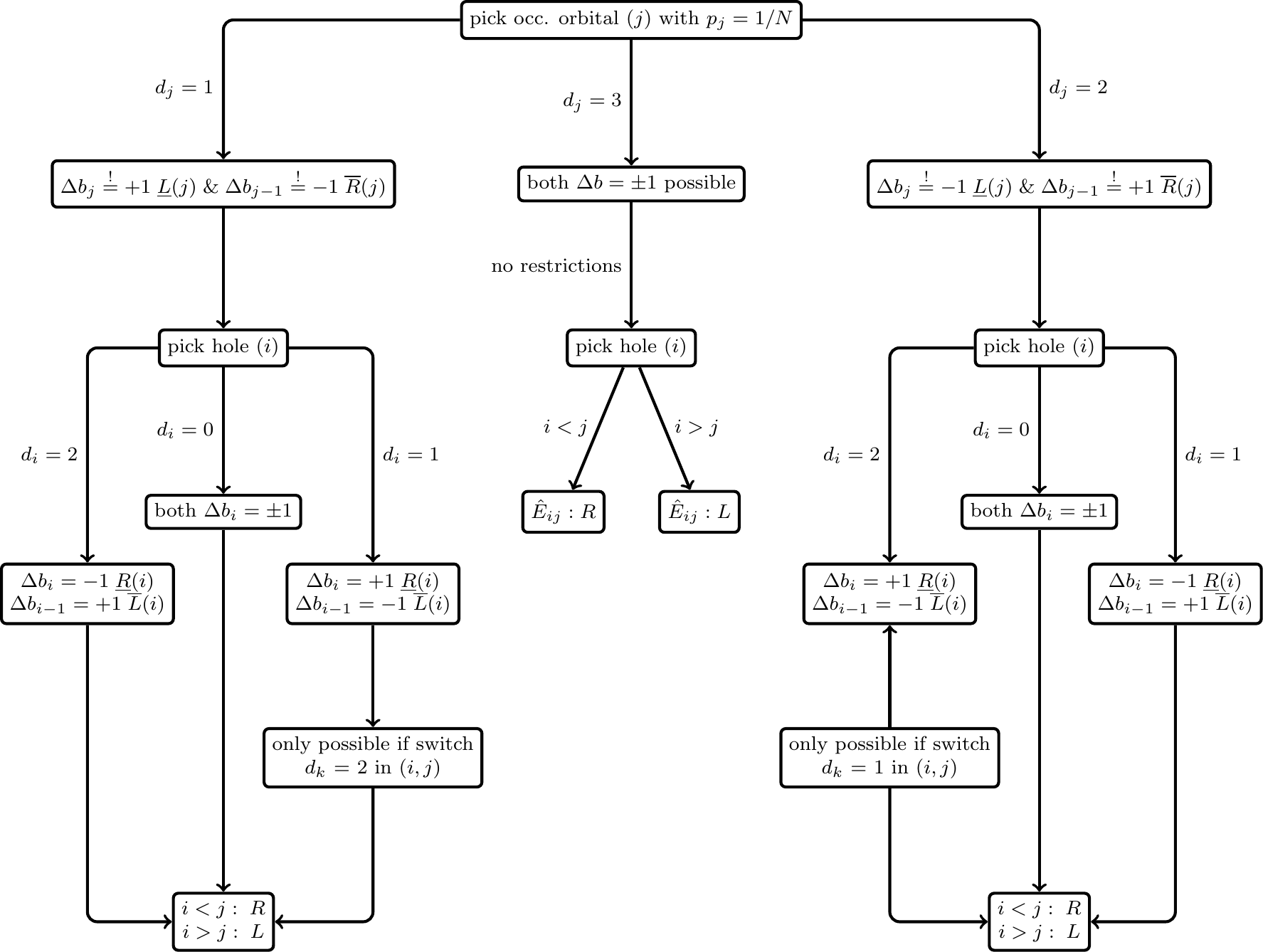}
\caption{\label{fig:singles-flow}Flow-chart of the decision-making process to find a valid index combination $(i,j)$ to ensure at least one non-zero single excitation $\hat E_{ij} \ket m$ and determine the type of generator, $R$ or $L$, depending on the order of $i$ and $j$.}
\end{figure*}

For $d_j = 1$ there is only a $\Delta b_j = +1$ start $\underline{L}$ and a $\Delta b_{j-1} = -1$ end $\overline{R}$ possible. So there is the restriction, that $d_i$ must be $0$, which allows both $\Delta b$ branches to start or to end, or $d_i = 2$, which would lead to the correct $\Delta b_i = -1$ start for $\underline{R}$ or would allow the $\Delta b_{i-1} = +1$ end for $\overline{L}$. A $d_i = 1$ value is only allowed, if there is a valid switch possibility $d_k = 2$ in the range $(i,j)$. 
For a chosen $d_j = 2$ electron orbital the restrictions are similar with $d_i=\{0,1\}$ being valid in general, and $d_i=2$ only if a switch possibility $d_k=1$ for $k\in(i,j)$. The actual restriction is implemented by finding the adjacent opposite spin-coupled orbitals $i_{lower}$ and $i_{upper}$ for a $d_j = \{1,2\}$ and only allowing $d_i = d_j$ to be picked if $i < i_{lower}$ or $i > i_{upper}$. 
A flow-chart of this decision-making process is given in Fig.~\ref{fig:singles-flow}.

If we want to make use of an Abelian point group symmetry, e.g. $D_{2h}$, it is much easier than suggested in the literature \cite{brooks-schaefer-2}, to just restrict the choice of orbital $i$ to the symmetry allowed orbitals $n_j$ corresponding to the picked electron orbital $j$, since for single excitations, in order for $t_{ij}$ to be nonzero, the product of the irreducible representations for orbitals $i$ and $j$ must be totally symmetric.

\subsection{\label{app:pick-orb-doubles}Restriction on the Orbital Choice for Double Excitations}

The excitation generation for doubles is a bit more involved, but due to the product structure of the matrix elements (\ref{eq:mat-prod-supp}). We follow the same approach as for single excitations to pick the four orbitals of $\hat e_{ij,kl}$ in such a manner to have the probability $p(ijkl)$ be related to the integral contribution of the Hamiltonian matrix element and at the same time ensure that at least one valid excitation can be reached. 

We first pick the ordered electron pair $(j < l)$ at random with a probability $p(jl) = 1/N_{pairs}$, where $N_{pairs}$ is the number of electron pairs in the simulation. The first orbital $(i)$ to excite to is then picked out of all, non-doubly occupied $d_i \neq 3$ orbitals, weighted with the Cauchy-Schwarz inequality based approximation \cite{cauchy-schwarz-1, Holmes2016} of the integral contribution $V_{ijkl}$.

The major change, compared to a SD based implementation, now comes only in the choice of the second orbital $(k)$ to excite to. Here we place the restrictions depending on the possible used spatial symmetry and the additional restriction, due to the UGA to obtain non-zero excitations. Additionally, we restrict the picking of orbital $(k)$ in such a way that we do not pick \emph{quasi-}single excitations, which are already taken account of in the single excitation matrix elements. 
The overall restriction $d_k \neq 3$ remains of course. 

To formulate the conditions for a valid orbital index choice, $(i,j,k,l)$, we have to look at the properties of the non-zero two-body segment shapes. The \emph{semi-start} segments, $\uL L,\uR R$, behave similar to single segment shapes concerning the restrictions on the in-coming and out-going $\Delta b$ values of an excitation and are listed in Table~\ref{tab:semi-start-alike}.
And similar to the end of a single excitation there are certain restriction for non-zero two-body elements at the end of the overlap range, depending on the type of the two alike generators, see Table~\ref{tab:semi-end-alike}. As one can see in these tables these segments behave like a single-excitation starts for an in-going $\Delta b_{k-1} = 0$ branch and like a single excitation end segment for the approaching $\Delta b_{k-1} = \pm 2$ branches. 

\begin{table*}
\centering
\small
\renewcommand{\arraystretch}{1.1}
\caption{\label{tab:semi-start-alike}$L\underline{L}, \underline{L}L^*$ and $\uR R, R\uR^*$ contributions at the overlap range start. $^*$ indicates the sign change of the $x=1$ matrix element, depending on the order of operators. $N_k' = N_k - 2$ for two lowering generators and $N'_k = N_k + 2$ for two raising generators. And $RR$ and $LL$ intermediate segments in the overlap region of a double excitation, depending on $\Delta b_{k-1}$.}
\begin{threeparttable}
\begin{tabular}{ccccccccccccccc}
\toprule
 \multicolumn{4}{c}{$L\uL/\uL L^*$} & & \multicolumn{4}{c}{$\uR R/R\uR^*$} & & \multicolumn{5}{c}{$RR\,/\,LL$}\\
 \cline{1-4} \cline{6-9} \cline{11-15}
\multicolumn{2}{c}{$\Delta b_{k-1}:$} & -1 & +1 & & && -1 & +1 & & & & 0 & -2 & +2 \\
\cline{1-4} \cline{6-9} \cline{11-15}
$d'$ & $d$ & \multicolumn{2}{c}{$\Delta b_k$}  &  & $d'$ & $d$ &  \multicolumn{2}{c}{$\Delta b_k$} & & $d'$ & $d$ & \multicolumn{3}{c}{$\Delta b_k$}\\
\hline
0 & 1 & 0  & +2 & & 1 & 0 & -2 & 0 & & 0 & 0 & 0 & -2 & +2\\
0 & 2 & -2 & 0  & & 2 & 0 & 0\tnote{a} & +2\tnote{b} & & 1 & 1 & 0 & -2 & +2 \\
1 & 3 & -2 & 0  & & 3 & 1 & 0 & +2 & & 2 & 1 & +2\tnote{b} & 0\tnote{a} & $-$ \\
2 & 3 & 0\tnote{a} & +2\tnote{b} & & 3 & 2 & -2 & 0 & & 1 & 2 & -2 & $-$ & +2\tnote{b} \\
\cline{1-9}
& & & & & & & & & & 2 & 2 & 0 & -2\tnote{a} & +2\tnote{b} \\
& & & & & & & & & & 3 & 3 & 0 & -2 & +2\\
\cline{11-15}
\end{tabular}
\begin{tablenotes}
\footnotesize
\item [a] No $b_k$ restriction, since $\Delta b_{k-1} = -1$ or $\Delta b_{k-1} = -2$.
\item [b] This path is only possible if $b_k >1$, otherwise $S_k <0$.
\end{tablenotes}
\end{threeparttable}
\end{table*}

\begin{table}
\centering
\small
\caption{\label{tab:semi-end-alike}Segment value restriction for the end of the overlap range for two lowering generators $\overline{L}L,L\overline{L}^*$,  and two raising generators $R\oR,\oR R^*$, depending on $\Delta b_{k-1}$ value. $N_k' = N_k \pm 1$ depending on the generator type. $^*$ indicated that the $x=1$ matrix element contribution has opposite sign for exchanged order of generators.}
\begin{threeparttable}
{
\newcommand{\mr}[1]{\multirow{2}{*}{#1}}
\renewcommand{\arraystretch}{1.1}
\begin{tabular}{ccccccccc}
\toprule
\multicolumn{4}{c}{$\oL L\,/\,L\oL^*$} & & \multicolumn{4}{c}{$R\oR\,/\,\oR R^*$}\\
\cline{1-4} \cline{6-9}
$d'$ & $d$ & $\Delta b_{k-1}$ & $\Delta b_k$ & & $d'$ & $d$ & $\Delta b_{k-1}$ & $\Delta b_k$ \\
\cline{1-4} \cline{6-9}
\mr{1} & \mr{0} & 0 & -1 & & \mr 0 & \mr 1 & 0 & +1 \\
 &  & +2 & +1 & &  & & -2 & -1 \\
 \cline{1-4} \cline{6-9}
\mr 2 & \mr 0 & 0 & +1\tnote{a} & & \mr 0 & \mr 2 & 0 & -1 \\
 & & -2 & -1 & &  & & +2 & +1 \\
 \cline{1-4} \cline{6-9}
\mr 3 & \mr 1 & 0 & +1 & & \mr 1 & \mr 3 & 0 & -1 \\
& & -2 & -1 & & & & +2 & +1 \\
\cline{1-4} \cline{6-9}
\mr 3 & \mr 2 & 0 & -1 & & \mr 2 & \mr 3 & 0 & +1\tnote{a} \\
& & +2 & +1 & & & & -2 & -1 \\
\botrule
\end{tabular}
}
\begin{tablenotes}
\footnotesize
\item [a] Only possible if $b_k > 0$, otherwise $S_k < 0$.
\end{tablenotes}
\end{threeparttable}
\end{table}

For two-body generators with identical starting indices, $\hat e_{ij,ik}$, only the $\Delta b_k = 0$ branch contributes, due to a zero $x=1$ matrix element in the overlap region \cite{Shavitt1981}. In addition only a $\Delta b_k = 0$ branch leads to a non-zero matrix element with two coinciding upper indices, $\hat e_{ij,kj}$. This means that for the $\underline{R}R$ or $\underline{L}L$ segments only the $\Delta b_k = 0$ branch can be chosen. So these type of double excitations can be treated very similar to single excitations, since the $x=0$ contribution is very easy to compute, ($-1^{n_2}$, with $n_1$ being the number of singly occupied orbitals in the overlap region) and are only non-zero if $d_k' = d_k$ in the overlap range. So no switch decisions have to be made in the excitation generation.

The case of mixed generators $R+L$ is a bit more involved. A simultaneous start $e_{ij,ki}$ acts similar to an usual double intermediate segment value, except that a $d_i = 0$ value leads to a zero matrix element, see  Tables~\ref{tab:mixed-start-end} and Ref.~[\citen{Shavitt1981}]. And similar to intermediate segment values of alike generators in the overlap region, the $x = 0$ matrix elements are zero for the $\Delta b_k = \pm 2$ branches. 

\begin{table}
\centering
\small
\caption{\label{tab:mixed-start-end}$\underline{RL}$ starting segments, where there is no change in the orbital occupation number $N_k' = N_k$. And $\overline{RL}$ end segment restrictions, depending on the $\Delta b_{k-1}$ value.}
\begin{threeparttable}

\begin{tabular}{ccccc}
\toprule
& & $\uR \uL$ & & $\oR \oL$ \\
\hline
$d'$ & $d$ & $\Delta b_{k}$ & & $\Delta b_{k-1}$\\
\hline
1 & 1 & 0 & & 0 \\
2 & 1 & +2\tnote{a} & & -2 \\
1 & 2 & -2 & & +2 \\
2 & 2 & 0 & & 0 \\
3 & 3 & 0\tnote{b} & & 0\tnote{b}\\
\botrule
\end{tabular}
\begin{tablenotes}
\footnotesize
\item [a] Only for $b_k > 1$.
\item [b] $x=1$ matrix element is zero.
\end{tablenotes}
\end{threeparttable}
\end{table}

The $x=1$ contribution of the mixed two-body segment values, see Ref.~[\citen{Shavitt1981}], is zero for $d_i = 3$, but not for the $\Delta b_k = 0$ branches of $d_i = \{1,2\}$. This leads to a major complication in the implementation of CSFs in the FCIQMC algorithm through the GUGA approach. These contributions with no change in step-value with a non-zero matrix element correspond to an exchange type contribution to double excitations. 
Since in the FCIQMC excitation generation it is necessary to uniquely assign a definite probability $p(m'|m)$, different starting orbitals $i' < I$, with $I$ indicating the first step-value change $\Delta d_I$, can contribute to an excitation with a $\underline{RL}$ start. The matrix element influence was mentioned above, but also the probabilities, $p(i')$, for all possible other starting orbitals $i' < I$ have to be accounted for. 
Similarly for an $\overline{RL}$ end, all other possible $j' > J$, with $J$ indicating the last step-value change $\Delta d_J$, ending orbitals have to taken into account. And for a pure exchange type excitation (type 2c in Table~\ref{tab:types-of-doubles}) $\underline{RL}\rightarrow\overline{RL}$ all combinations $(i'<I,j'>J)$ of possibly contributing orbitals have to be considered in the matrix element and generation probability computation. 

Otherwise a $R\underline{L}$ segment behaves similar to an $L\underline{L}$ and $\underline{R}L$ to an $\underline{R}R$ in terms of $d_k', b_k$ and $\Delta b_{k-1}$ restrictions, except the number of electrons in $N_k' = N_k$. Also the intermediate $RL$ segments behave as $LL$ and $RR$ and the $R\overline{L}$ is equivalent to $L\overline{L}$ and $L\overline{R}$ to $R\overline{R}$ respectively. Except the electron number difference becomes the corresponding value $N_k' = N_k \pm 1$ of the ongoing excitation ($R$ in the case of $R\overline{L}$ and $L$ for $L\overline{R}$).
It should also be noted, that a $\Delta b_k = 0$ branch can end at any $d_j \neq 0$ value, whereas $\Delta b_k = -2$ is restricted to $d_j = 1$ and $\Delta b_k = +2$ to $d_j = 2$, to be able to align the $S_k$ value of $\ket {m'}$ and $\ket m$, so they coincide outside of the range of the generator $\hat e_{ij,jl}$. Since $\Delta b_k = 0$ already indicates $\ket {m'} = \ket m$ in the overlap range, this issue is no problem in a direct CI calculation with CSFs, but is burdensome to implement in FCIQMC, since we want to be able to get one out of all possible excitations for a given CSF $\ket m$ and assign a unique generation probability to it. 
So we also have to take into account all other possible index combinations, which would be able to lead to this excitation and sum their matrix elements of course, but also recompute the probability.

Due to the uniqueness for most type of excitations, this is no problem, except in the case of these exchange type excitations with coinciding indices and a $RL$ generator combination. Unfortunately we have not yet found a more elegant way to treat these cases, except implement it in the most efficient way. With a heavy re-usage of terms to avoid an $\bigO{N^2}$ or even $\bigO{n^2}$ computational cost of these excitations.

\subsection{Orbital Picking and Excitation Identification}

\begin{figure*}
\centering
\includegraphics[width=\textwidth]{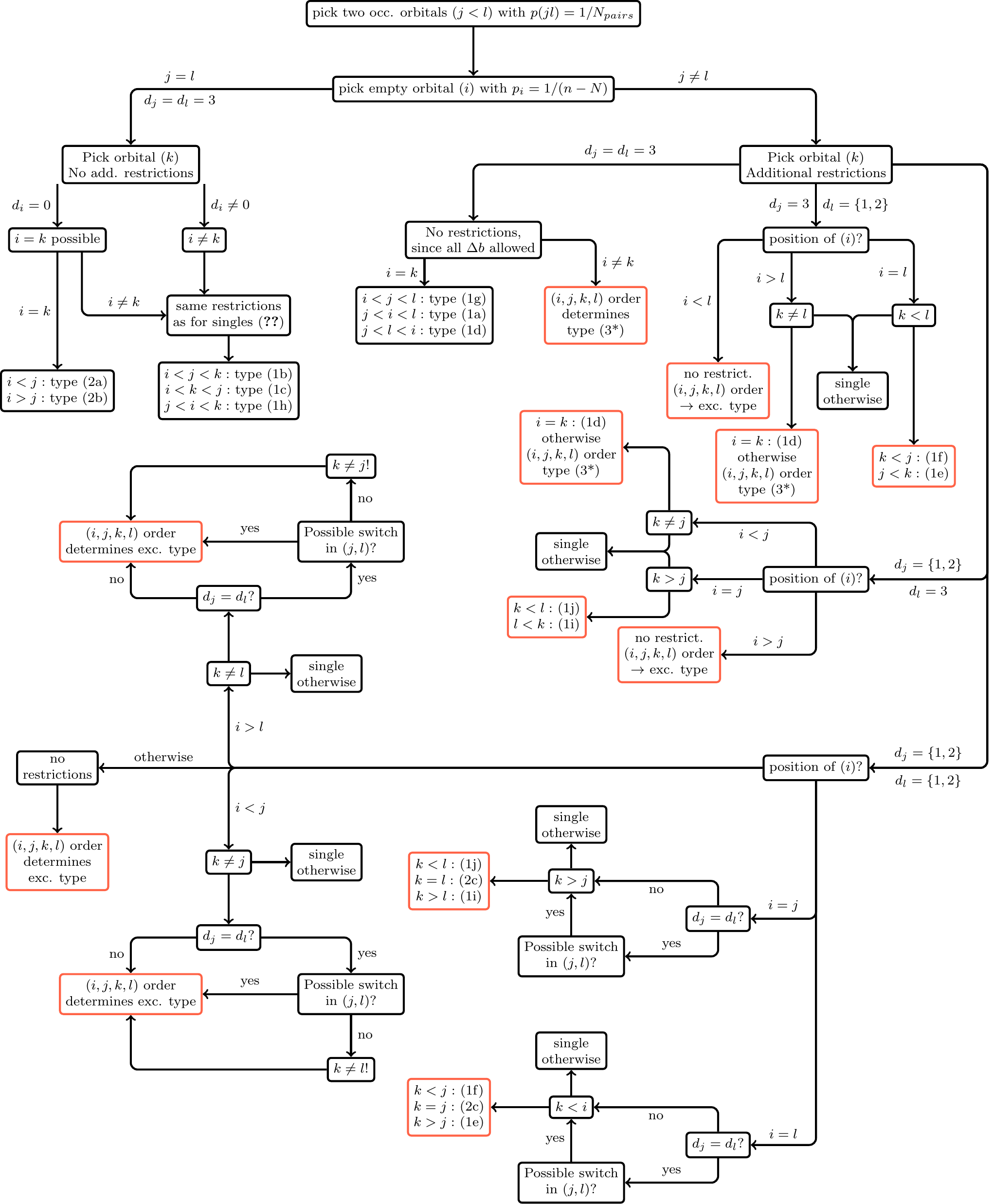}
\caption{\label{fig:doubles-flow}Flow-chart of the decision-making process to find a valid index combination $(i,j,k,l)$ to ensure at least one non-zero double excitation $\hat e_{ij,kl}\ket m$ and identify the excitation type based on these indices.}
\end{figure*}

In the following, the work flow of picking a valid index combination $(i,j,k,l)$ for a non-zero double excitation of a CSF $\ket m$ in the FCIQMC method is presented. A flow-chart of the decision-making process is shown in Fig.~\ref{fig:doubles-flow}. 

Both electron indices $(j < l)$ and the first hole index $(i)$ are picked with uniform or a weighted probability. 
If both picked electrons are in the same spatial orbital $\underline{j = l \rightarrow d_j = 3}$, similar restrictions as for \emph{single excitations} apply, for the remaining orbitals $i$ and $k$. Of course both orbitals $i$ and $k$ have to be non-doubly occupied. This applies in general, independent of the relation of electron orbitals $j$ and $l$ and their step-value $d_j, d_l$. 
And $i = k$ is only possible if $d_k = 0$, since both electrons will be excited to the same spatial orbital. Since both electrons get removed from the same orbital the only possible excitation types are (1b,1c,1d,2a,2b) of Table~\ref{tab:types-of-doubles}. If $i = k$ the type of excitation is (2a) if $i > j$, or (2b) if $i < j$, requiring $d_i = 0$. For $i \neq k$ the same restrictions as for single excitations apply, that $d_i = d_k = 1$ is only possible if a switch possibility $d_m = 2$ in the range $(i,k)$ and vice versa for $d_i = d_k = 2$. The type of excitation only depends on the order of the involved indices. As already mentioned, all these excitations require $\Delta b_m = 0$ in the overlap range. With the trivial case of type (1b) excitation with a single orbital overlap range. Which make the calculation of the excitation very similar to single excitations.

If the electron indices are not equal $j \neq l$, the picking of the remaining orbitals $k$ and $i$ depends more strongly on the step-values of the already chosen orbitals $(j,l)$.\\
If $\ul{d_j = d_l = 3}$ there is no additional restriction on orbital $k$, since in except of $b$ value restrictions on starts of a doubly occupied orbital ($\Delta b = +1$ branch forbidden due to $b = 0$, e.g.) all restrictions mentioned in the previous section can be accounted for, due to the flexibility of the $d_j = d_l = 3$ step-values.

If $i = k$, depending on the order of the indices this leads to excitations of type (1a,1d) or (1g), since $i \neq j,l$ due to $d_j = d_l = 3$. These excitations again can be easily treated, due to the single overlap region (1a) or a necessary $\Delta b_m = 0$ in the overlap region. 

If $i \neq k$ all 4 indices are different, leading to a type (3*) excitation depending on the order of the indices. Where again, the exchange type excitation is chosen by definition and not a possibly non-overlap double excitation (3c$_0$,3d$_0$,3e$_0$,3f$_0$). 

If $\ul{d_j = 3, d_l = \{1,2\}}$: Depending on if the already picked first orbital to excite to $i = l$, orbital $k$ must be restricted to $k < l$. This is because $k > l$ would lead to an exchange contribution to a single excitation, which is already taken into account for in the singles matrix elements. This leads to excitations of type (1e) or (1f) depending on the order of $k$ and $j$. 
If $i > l$ orbital $k \neq l$ since this again would lead to an already accounted exchange contribution to a single excitation. If $i = k$ it is an (1d) excitation otherwise it is one of the type (3*) depending on the relation of the indices. 
If $i < l$ there is no restriction on the indices for $k$ and this can lead to a variety of excitations. 

If $\ul{d_j = \{1,2\}, d_l = 3:}$ There are similar restrictions considering the already picked orbitals. If $i = j$, $k$ must be picked $k > j$ to avoid choosing already accounted for exchange contributions to single excitations. And if $i < j$ $k$ must not be $j$, otherwise there are no additional restrictions. 

If both $\ul{d_j = \{1,2\}}$ and $\ul{d_l = \{1,2\}}$ are singly occupied the most stringent restrictions apply. If $i = l$ there should be a possible switch between the already picked $j$ and $l$ if both have the same step-value $d_j = d_l$, since otherwise it would not be possible to fulfill the $\Delta b$ criteria at the end of an excitation to lead to a non-zero excitation. Additionally, orbital $k$ has to be lower than $i$, otherwise it is again an exchange contribution to a single excitation. Depending on the order of the orbitals, this leads to type (1e,1f,2c) excitation, which, already mentioned, needs additional re-computation of matrix element and generation probability contribution. 

If $i = j$, there also must be a switch possibility for $d_j = d_l$ between $j$ and $l$ and $k > j$. This leads to type (1i,1j,2c) excitations. With the necessity of recalculation of matrix element and generation probability contributions. 

If $i > j$ orbital $k$ must not be $l$ to avoid an exchange contribution to singles and similarly if $i < j$, $k$ must not be $j$. And $k$ can only coincide with $j$ for $i > j$, if there is a switch possibility between $j$ and $l$, if $d_j = d_l$. And similar for $i < j$, $k = l$ is only possible if $d_j \neq d_l$ or there is a switch possibility between $j$ and $l$. 
Otherwise no restrictions are place on the picking of orbital $k$ and the type of excitation depends on the order of the indices and can lead to all sort of excitation types in Table~\ref{tab:types-of-doubles}. 

In the whole picking process, since we allow the empty orbitals to be picked in any order, in addition to $p(i|jkl)$ we also have to recompute $p(k|ijl)$ of having picked the orbitals in the opposite order, since they lead to the same possible excitation. This increases the generation probability by 2 in general, but introduces the effort to recompute. We could, similar to the electron orbitals $j$ and $l$, decide to pick only orbitals $i < l$, which would also make the identification of the excitation type easier. 

To utilize Abelian point group symmetry for double excitations we restrict the last picked empty orbital $k$ to ensure that the product of the four irreducible representations of $(i,j,k,l)$ are totally symmetric and consequently $V_{ijkl}$ is not zero due to symmetry.

\clearpage

\end{document}